\documentclass[preprint,12pt]{elsarticle}
\usepackage[utf8]{inputenc}
\usepackage{lineno,hyperref}
\usepackage{amsmath,fancyhdr,amsthm,amssymb,amsfonts,version}
\usepackage{multirow}
\usepackage{bbm,version}
\usepackage{mathrsfs}
\usepackage{graphicx}
\usepackage[letterpaper,top=1.2in,bottom=1in,left=1.2in,right=1.2in]{geometry}%
\usepackage[section]{placeins}
\usepackage{stmaryrd}
\usepackage{bm}
\usepackage{epstopdf}
\usepackage{float}
\usepackage{subfigure}
\usepackage{appendix}
\usepackage{natbib}
\include{pythonlisting}
\usepackage{enumitem}

\hypersetup{colorlinks=true, linkcolor=blue, citecolor=red}

\makeatletter
  \newcommand\figcaption{\def\@captype{figure}\caption}
  \newcommand\tabcaption{\def\@captype{table}\caption}
\makeatother

\numberwithin{equation}{section}

\journal{}

\begin{document}

\begin{frontmatter}

\title{DeepM\&Mnet for hypersonics: \\
Predicting the coupled flow and finite-rate chemistry behind a normal shock using neural-network approximation of operators}

\author[Brown]{Zhiping Mao}
\author[MIT]{Lu Lu}
\author[Surrey]{Olaf Marxen}
\author[JHU]{Tamer A. Zaki\corref{cor1}}
\cortext[cor1]{Corresponding author}
\ead{t.zaki@jhu.edu}
\author[Brown]{George Em Karniadakis\corref{cor2}}
\cortext[cor2]{Corresponding author}
\ead{george\_karniadakis@brown.edu}

\address[Brown]{Division of Applied Mathematics, Brown University, Providence, RI, 02912, USA}
\address[MIT]{Department of Mathematics, Massachusetts Institute of Technology, Cambridge, MA, 02139, USA}
\address[Surrey]{Department of Mechanical Engineering Sciences, University of Surrey, Guildford GU2 7XH, UK}
\address[JHU]{Department of Mechanical Engineering, Johns Hopkins University, Baltimore, MD 21218, USA}

\date{}

\begin{abstract}
In high-speed flow past a normal shock, the fluid temperature rises rapidly triggering downstream chemical dissociation reactions. The chemical changes lead to appreciable changes in fluid properties, and these coupled multiphysics and the resulting multiscale dynamics are challenging to resolve numerically.  Using conventional computational fluid dynamics (CFD) requires excessive computing cost. Here, we propose a totally new efficient approach, assuming that some sparse measurements of the state variables are available that can be seamlessly integrated in the simulation algorithm.
We employ a special neural network for approximating nonlinear operators, the DeepONet ~\cite{lu2019deeponet}, which is used to predict separately each individual field, given inputs from the rest of the fields of the coupled multiphysics system. 
%
We demonstrate the effectiveness of DeepONet for a benchmark hypersonic flow involving seven field variables. Specifically we predict five species in the non-equilibrium chemistry downstream of a normal shock at high Mach numbers as well as the velocity and temperature fields. We show that upon training, DeepONets can be over five orders of magnitude faster than the CFD solver employed to generate the training data and yield good accuracy for unseen Mach numbers within the range of training.  Outside this range, DeepONet can still predict accurately and fast if a few sparse measurements are available. We then propose a composite supervised neural network, DeepM\&Mnet, that uses multiple pre-trained DeepONets as building blocks and scattered measurements to infer the set of all seven fields in the entire domain of interest. Two DeepM\&Mnet architectures are tested, and we demonstrate the accuracy and capacity for efficient data assimilation. DeepM\&Mnet is simple and general: it can be employed to construct complex multiphysics and multiscale models and assimilate sparse measurements using pre-trained DeepONets in a ``plug-and-play" mode.

\end{abstract}

\end{frontmatter}


\section{Introduction}
\subsection{Motivation}


Simulating the high-speed flow field of a chemically reacting fluid is interesting, challenging and has important applications including hypersonic cruise flights and planetary re-entry. In order to predict such a flow field, a multi-physics and multi-scale approach is essential. The fluid dynamics part of this problem may already feature a large number of effects as well as length and time scales when it features shocks, boundary layers, transition to turbulence to name but a few. In addition to fluid dynamics, also physical chemistry needs to be taken into account, as chemical reactions are likely to occur in the flow, and these need to be modeled accurately.  It is particularly challenging if reactions and flow effects take place on similar spatio-temporal scales because a simplified model for this type of flow field cannot be easily derived. Instead,  the full set of equations describing the physical as well as the chemical model must be solved.

One particularly relevant canonical problem is the high-speed flow downstream of a normal shock. In this case, the temperature of the fluid rises rapidly across the shock, which in turn triggers chemical dissociation reactions. As a result, the flow field changes composition, which influences the energy balance as dissociation reactions are endothermic. A change of composition directly influences several other aspects because it leads to a modification of viscosity as well as heat conduction.

The level of fidelity required to model flows with high-temperature gas effects depends on the typical flow speed and temperature. The rate of chemical reactions is largely influenced by temperature (and to a lesser extent by pressure), and convective mass transport is driven by flow speed. Comparing the typical time scales of these two effects yields three different regimes. In the first, chemical reaction rates are much smaller than the rate of convective transport, and the fluid composition in this regime is considered frozen. If chemical reaction rates are much larger than the rate of convective mass transport, the flow is in chemical equilibrium; reaction rates are therefore assumed infinite  and the gas composition depends on local properties such as temperature and pressure (or density). The most interesting regime, which is considered herein, is referred to as finite-rate chemistry, or non-equilibrium chemistry; it lies in between the other two regimes, when the rates of reactions and transport are commensurate. 

Gas composition strongly effects the relation between temperature and internal energy. For a calorically perfect gas, internal energy is proportional to temperature. A gas composed of a single atomic species typically behaves  calorically perfect, for which the specific heat is constant. A gas composed of a molecular species will experience the excitation of vibrational and electronic modes of the molecules. As a result, the internal energy becomes a non-linear function of temperature and the specific heat will also vary with temperature. Such a gas is denoted as thermally perfect. Vibrational and electronic excitation may not happen infinitely fast, in which case the process of thermal relaxation may need to be taken into account. However, here we will only consider a gas in so-called local thermal equilibrium, i.e.,\,it will be assumed that the vibrational excitation only depends on local properties such as the temperature and not on its (time) history.

In many practical applications, the fluid is a mixture of atomic and diatomic species, such as in air. While air can certainly be modelled as a calorically perfect gas at low temperatures where vibrational and electronic modes are not excited, at high temperatures a thermally perfect gas model would is more appropriate. At even higher temperatures, the threshold for the onset of chemical reactions may be reached and the corresponding changes of gas composition may need to be included in the modelling approach. Such a high temperature may occur as a result of a shock wave, which creates an almost instantaneous increase of temperature. The time scale for a flow passing through a shock is extremely short, owing to the very small thickness of a shock, which is on the order of the mean free path lengths of the species involved. Chemical reactions are much slower than this fast time scale, and hence the composition of the gas mixture does not change across the shock: the gas mixture is frozen. However, downstream of the shock, chemical reactions may set in as a result of the higher temperature, and as the flow speed is reduced significantly, a region of flow in the finite-rate reaction regime is expected, before the flow may reach an equilibrium state far downstream of the shock. As a result of finite-rate reactions, the gas mixture will increasingly change its composition as molecules begin to dissociate, rendering the region immediately downstream of the shock particularly interesting.

Compared to a calorically perfect gas, a gas that undergoes changes in composition is appreciably more complex to model because transport equations must be solved for each species density, with a source term for the creation and destruction of species by chemical reactions. These additional equations complicate the numerical treatment significantly. In particular, the source term can cause numerical stiffness and hence become difficult to integrate.
Several numerical methods for finite-rate chemistry exist, mostly for time-dependent flows involving combustion \cite{hilbert2004impact,najm1998semi,day2000numerical,najm2005modeling,nicoud2000conservative}, but they often rely on a low-Mach-number formulation and are therefore not applicable to hypersonics. At high Mach number, evidence abound regarding the sensitivity of the flow to small distortions \cite{park_zaki_2019,jahanbakhshi_zaki_2019}, and hence accurately capturing non-equilibrium chemistry becomes extremely important. 
A growing number of methods now exist that account for non-equilibrium chemistry while solving the compressible Navier-Stokes or Euler equations at hypersonic speeds \cite{zhong1996additive,kitamura2005shock,matheou2008verification,duan2009procedure,prakash2011high,coussement2012three,zanus2020parabolized,ZhangShu2011JCP}. However, the combination of high Mach numbers and high temperatures not only renders these simulations challenging, but also taxes computational resources heavily.  

\begin{figure}[http]
\begin{center}
\includegraphics[width=0.70\textwidth,angle=0]{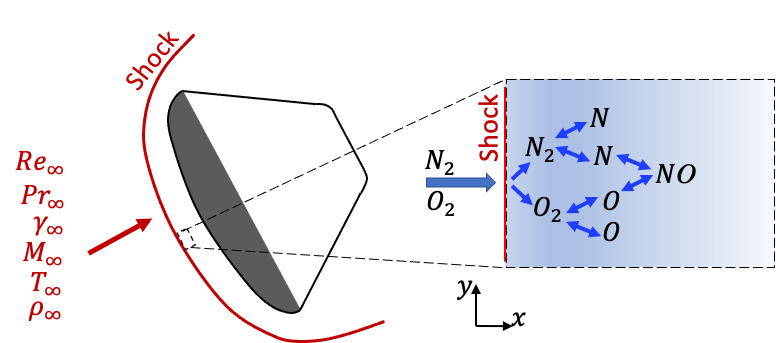}
\end{center}
\caption{Schematic of the non-equilibrium chemistry that takes place downstream of a high-Mach-number shock around a bluff body.  The simulations are performed in the frame of the body.}
\label{fig:Problem_Schematic}
\end{figure}

\subsection{Deep neural networks}

In realistic hypersonic applications, flight data may be comprised of limited measurements, for example of temperature or velocity and perhaps even in special cases of the composition of the species. To integrate such data with the computational approach and in order to simulate the aforementioned multiscale \& multiphysics problems efficiently, we abandon the classical numerical methods and explore in the present work a new approach, namely, a deep neural network (DNN) based approximation of all nonlinear operators.

The machine learning community has made tremendous strides in the past 15 years by capitalizing on the neural network (NN) universal function approximation \cite{cybenko1992approximation}, and building a plethora of innovative networks with good generalization properties for diverse applications. However, it has ignored an even more powerful theoretical result by Chen \& Chen \cite{chen1993approximations, chen1995universal} that states that neural networks can, in fact, approximate functionals and even nonlinear operators with arbitrarily good accuracy. This is an important result with significant implications, especially for modeling and simulation of physical systems, requiring accurate regression and not only approximate classification tasks as in commercial applications. Preliminary results in \cite{del2019learning, lu2019deeponet} have provided a glimpse of the potential breakthroughs in modeling complex engineering problems by encoding different explicit and implicit operators using DNNs. For example, in \cite{del2019learning} Ferrandis et al.\,represented a functional predicting the dynamic motions of a destroyer battleship in extreme sea states, making predictions at a fraction of a second in contrast to one week per simulation using OpenFoam CFD solver. Similarly, in \cite{lu2019deeponet}, Lu et al.\,developed the Deep Operator Network (DeepONet) to approximate integrals, ODEs, PDEs, and even fractional Laplacians by designing a new trunk-branch NN that approximates linear and nonlinear operators, and generalizes well to unseen functions.

Traditional methods, especially high-order discretizations such as WENO \cite{liu1994weighted} and spectral elements \cite{karniadakis2013spectral}, can produce very accurate solutions of multiphysics and multiscale (M\&M) problems but they do not scale well in high dimensions and large domains. Moreover, they cannot be easily combined with data  \cite{WangM_2019,WangQ_2019,mons2019kriging,david_zaki_2019} and are prohibitively expensive for inverse problems. Real-world M\&M problems are typically ill-posed with missing initial or boundary conditions and often only partially known physics, e.g., reactive transport as in the present work. Physics-Informed Neural Networks (PINNs) can tackle such problems given some extra (small) data anywhere in the domain, see \cite{raissi2019physics, raissi2020hidden, mao2020physics, lu2019deepxde}. PINNs are easy to implement for multiphysics problems and particularly effective for inverse problems \cite{nvidiatalk} but not as efficient or accurate for forward multiscale problems. Here, we propose DeepONets to approximate functionals and nonlinear operators as building blocks of a more general M\&M framework that can be used to approximate different nonlinear operators for modeling M\&M problems. Unlike PINNs, we can train DeepONets {\em offline} and 
make predictions for new input functions {\em online} very fast. We refer to this integrated framework that will use both data and DeepONets as DeepM\&Mnet, and, in principle, it can be used for any complex M\&M problem in physics and engineering.
%
Here we consider hypersonic flow downstream of a normal shock, which involves the interaction of seven field variables (see Figure \ref{fig:Problem_Schematic}). 
This formidable M\&M challenge is an excellent testbed to develop the DeepM\&Mnet framework and to demonstrate its effectiveness.  

In the present work, we claim the following contributions:
\begin{itemize}
\item We start with developing DeepONets for the M\&M model, namely, the non-equilibrium chemistry that takes place behind a normal shock at Mach numbers between 8 and 10. We infer the interactions of the flow and five chemical species whose densities span 8 orders of magnitude downstream of the shock. Collectively, these dynamics establish the gas composition and flow velocity, which are governed by the nonlinear Navier-Stokes equations and whose operators will be learned by our DeepONets. The speedup of the trained DeepONets is about 100,000X compared to the CFD solver.
\item Besides the prediction for the case when the Mach number is in the range $[8,10]$, we also test the case when the input is out of the input space, i.e., the Mach number is out of the range $[8, 10]$ (extrapolation). Though the initial prediction is not satisfactory, we obtain very good predictions by combing a few data and the pre-trained DeepONets by developing a supervised NN, which can be efficiently trained.
\item As a preliminary step in developing the multi-physics integrated framework, we employ these pre-trained DeepONets as building blocks to form different types of DeepM\&Mnets. We first develop a {\em parallel} DeepM\&Mnet architecture which, similar to the aforementioned extrapolation algorithm, requires sensor data for all the variables. However, in practice, for data assimilation we may not have access to information regarding the species densities, and may only have sparse data for the flow. Therefore, we develop a {\em series} DeepM\&Mnet architecture that assimilates only a few data for the flow and predicts the entire state. Moreover, we examine the influence of the global mass conservation constraint and demonstrate that not only does it stabilize the training process but it also improves prediction accuracy.
\end{itemize}

The rest of the paper is organized as follows: 
In the next section, we present the M\&M fluid-mechanical model and demonstrate how to generate the training data using a finite difference approach. We then develop in section \ref{sec:deeponet} the DeepONets, which will serve as building blocks for the DeepM\&Mnets discussed in section \ref{sec:deepmmnet}. We conclude with a summary in section \ref{sec:conclusion}.
In the Appendix we present an alternative series type DeepM\&Mnet.

\section{Problem setup and data generation}

In this section, we present in detail the governing equations that model the flow and describe how to obtain the data for training and testing.
The mathematical formulation for fluid motion is given in \S~\ref{sec:fluid_mechanics}, followed by a description of the data generation process including details of the test case in section \S~\ref{sec:data_generation}.

\subsection{Fluid-mechanical model and numerical method}
\label{sec:fluid_mechanics}
The fluid mechanical model used here comprises the Navier-Stokes equations for a compressible fluid, and these equations are advanced in time until a steady state is reached. A detailed description of the model is given in \cite{marxen2013method}. Here, only key elements of this model are described. The governing equations are the principles of conservation of mass, momentum balance and energy conservation, and are formulated for a 5-species mixture of chemically reacting gases in two spatial dimensions ($j=1,2$):
\begin{eqnarray}
\frac{\partial{\rho}}{\partial{t}}+\frac{\partial}{\partial x_j}\left( \rho u_j \right) &=& 0, \label{eq:continuity}\\
\frac{\partial \rho^s}{\partial t}+\frac{\partial}{\partial x_j}\left( \rho^s u_j \right) &=& \textit{\.{w}}^s, \quad s=1 \dots 5, \label{eq:speciescontinuity}\\
\frac{\partial(u_i)}{\partial t}+\frac{\partial}{\partial x_j}\left( \rho u_i u_j +p \delta_{ij} \right) &=& \frac{\partial \sigma_{ij}}{\partial x_j}, \quad i=1,2, \label{eq:momentum}\\
\frac{\partial E}{\partial t}+\frac{\partial}{\partial x_j} \left[ \left( E +p \right) u_j \right] &=& - \frac{\partial q_{j}}{\partial x_j} + \frac{\partial}{\partial x_k} \left( u_j \sigma_{jk} \right). \,\,\quad \label{eq:total_energy}
\end{eqnarray} 
The equations are non-dimensionalized using reference quantities described below. The mixture density is $\rho$ and $\rho^s$ is the species density for species $s=1\dots 5$, and $u_1$, $u_2$ are the velocity components in the streamwise $x=x_1$ and normal $y=x_2$ directions. In eqn.~\eqref{eq:speciescontinuity}, $\textit{\.{w}}^s$ is a source term due to finite-rate reactions that lead to production or consumption of species. It is obtained from the MUTATION library, here used in its version 2.1 \cite{magin2004transport,MaginCaillaultBourdonLaux2006,Wangetal2011}, which has been coupled to the Navier-Stokes solver with the help of an interface code layer.

Cases considered here are based on air with species $N$, $O$, $N_2$, $NO$ and $O_2$. 
The state of the gas mixture is described by the pressure $p$ and temperature $T$, which are related to one another via the following equation of state, and assuming that all species individually behave as ideal gases so that partial pressures sum up to the total pressure of the mixture:
\begin{equation}
\tilde p = \sum_{s} \tilde \rho^s \tilde R^s \tilde T \, . \label{eq:pressure_details} 
\end{equation}
In the equation, $\tilde\bullet$ represents dimensional quantities, and $\tilde R^s = {\mathcal{\tilde R}}/{\tilde M^s}$, where  $\mathcal{\tilde R}$ is the universal gas constant and $\tilde M^s$ is the molar mass of species $s$.
The temperature $T$ is required to calculate the heat flux vector $q_j$, which reads:
\begin{equation}
q_j  = - \frac{1}{Re_\infty Pr_\infty Ec_\infty} \, k \frac{\partial T}{\partial x_j} \,. \qquad \label{eq:heat_flux}
\end{equation}
The temperature does not belong to those quantities governed by transport equations \eqref{eq:continuity} to \eqref{eq:total_energy} above, but it is linked to the internal energy $e$, which in turn contributes to the total energy $E$ together with kinetic energy, in the following way:
\begin{equation}
E = \frac{1}{Ec_\infty} e \rho + \frac{1}{2} \rho u_i u_i \,.  \label{eq:energy_def}
\end{equation}
Computation of the temperature $T$ is based on known values of the internal energy $e$ and the species densities $\rho^s$ so that the total internal energy $e$ can be split into individual contributions $e^s$ from each species. The corresponding iterative solution procedure used to obtain the temperature, again done within he MUTATION library, takes into account the translational energy of atoms and linear molecules, as well as  rotational and vibrational energy and formation enthalpy for molecules.

The right-hand side of both the momentum equation \eqref{eq:momentum} and the energy equation \eqref{eq:total_energy} contains the the viscous stress tensor $\sigma_{ij}$, which is given by:
\begin{eqnarray}
\sigma_{ij} &=& \frac{\mu}{Re_\infty} \left( \frac{\partial u_i}{\partial x_j} + \frac{\partial u_j}{\partial x_i} - \frac{2}{3} \frac{\partial u_k}{\partial x_k} \delta_{ij} \right).
\end{eqnarray}
Unlike in a calorically perfect gas, transport properties such as viscosity $\mu$ and thermal conductivity $k$ are not simple functions of (local) temperature, but also depend on gas composition. These quantities are also computed using the MUTATION library. 
Due to the nature of the flow field considered here, these transport properties do not play a major role. Therefore no further description of how $\mu$, $\tilde k$ are computed is given here. 

Non-dimensionalization is based on inlet conditions marked by $\infty$, resulting in the following non-dimensional quantities. The Reynolds number is defined as $Re_\infty = \frac{\tilde \rho_\infty \tilde a_\infty \tilde L_{ref}}{\tilde \mu_\infty}$, 
where $\tilde L_{ref}$ is an arbitrary length scale and all other dimensional quantities represent the pre-shock conditions, for example $\tilde a_\infty$ is the upstream speed of sound. The Prandtl number is $Pr_\infty = \frac{\tilde \mu_\infty \tilde c_{p,\infty}}{\tilde k_\infty}.$
The reference temperature used for non-dimensionalization is $\tilde T_{ref}=(\gamma_\infty - 1) \tilde T_\infty$, where $\gamma_\infty$=$\tilde c_{p,\infty} / \tilde c_{v,\infty}$ is the specific heat ratio and $\tilde c_{p,\infty}$, $c_{v,\infty}$ are the specific heats at constant pressure and volume, respectively. The Eckert number $Ec_\infty$ in eqns. \eqref{eq:heat_flux} and \eqref{eq:energy_def} assumes a value of 1, as the pre-shock gas is  calorically perfect.

The integration domain extends from immediately downstream of the shock, $x=0$, to $x=0.02$ discretized using $NX=160$ equi-spaced grid points in the streamwise direction. In $y$ direction, $MY=21$ grid points were used within $y \in [0,0.05]$. At the inflow, a Dirchlet condition is prescribed, and corresponding values are described in the next section. Periodic boundary conditions are imposed in $y$ direction.

Spatial derivatives in the transport equations \eqref{eq:continuity} to \eqref{eq:total_energy} are discretized using high-order compact finite differencing, and the solution is advanced in time using a third-order Runge-Kutta method. The discretization is largely identical to that used in Ref.~\cite{Nagarajanetal03}, on which the present code is based. However, discretization at boundaries of the integration domain has been altered to accommodate the boundary conditions applied here.

\subsection{Data generation}
\label{sec:data_generation}
We introduce in this subsection how to generate the training and testing datasets used to develop the DeepONets and DeepM\&Mnets.
We consider ``case S" in reference \cite{marxen2013method}, i.e., the flow field downstream of a normal shock wave, which is set at the origin of the coordinate system. The parameters used are given in Tables \ref{tab:General:parameters}-\ref{tab:postshockcondition}. 
Specifically, the conditions upstream of the shock are given in Table \ref{tab:General:parameters}. The gas composition at the inflow boundary is given in Table \ref{tab:gas_com}. It is assumed that within the short streamwise length of the shock, the composition $\chi^s=\rho^s/\rho$ does not change, and hence the composition upstream and downstream remains the same. However, both temperature and density increase significantly, and the post-shock conditions are used as inflow conditions for the simulations. The initial temperature and total density are presented in Table \ref{tab:postshockcondition}; these have been calculated using Rankine–Hugoniot relations. The flow field is progressed in time until a steady state is reached. 
The same case has been previously considered in \cite{magin2006nonequilibrium}. 
Even though variations only take place along the streamwise direction, the actual computations were performed in two dimensions as stated earlier. 

\begin{table}[h]
    \centering
    \begin{tabular}{c c c c c c}
    \hline
      $Re_\infty$   &  $Pr_\infty$   &$\gamma_\infty$   &$M_\infty$   & $\tilde{T}_\infty~[K]$   & $\tilde{\rho}_\infty~[kg~m^{-3}]$  \\
       $10^{4}$  & $0.69$ & $1.397$ & $[8,10]$& $350$ & $0.3565 \times 10^{-3}$\\
    \hline
    \end{tabular}
    \caption{General parameters upstream of the shock.}
    \label{tab:General:parameters}
\end{table}

\begin{table}[h]
    \centering
    \begin{tabular}{c c c c c}
    \hline
      $N_2$   &  $O_2$   &$N$   &$O$   & $NO$   \\
       $0.767082$  & $0.232918$ & 0 & 0 & 0 \\
    \hline
    \end{tabular}
    \caption{Gas composition $\chi^s$ at the inflow boundary.}
    \label{tab:gas_com}
\end{table}

\begin{table}[h]
    \centering
    \begin{tabular}{c c}
    \hline
      $\tilde{T}~[K]$   & $\tilde{\rho}~[kg~m^{-3}]$  \\
       $5918.87$  & $0.255537 \times 10^{-2}$ \\
    \hline
    \end{tabular}
    \caption{Post shock conditions $(x=0)$.}
    \label{tab:postshockcondition}
\end{table}

\begin{figure}[h]
\begin{center}
\includegraphics[width=0.31\textwidth]{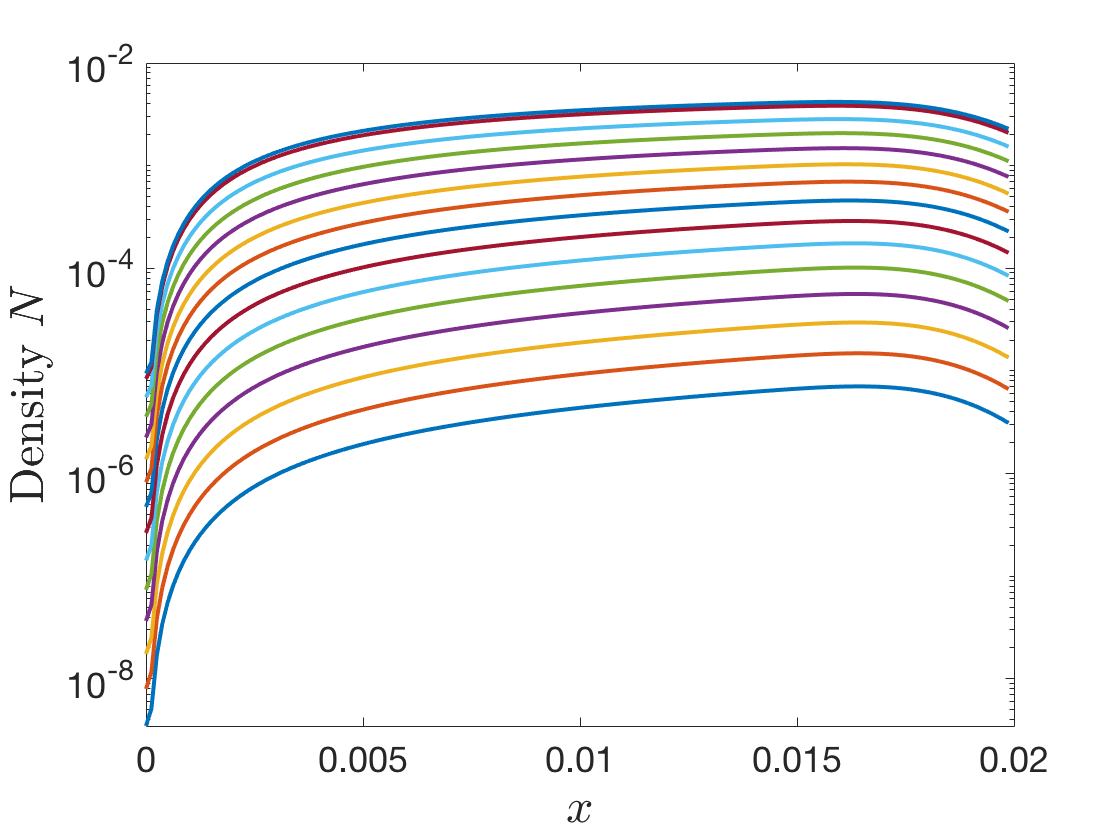}
\includegraphics[width=0.31\textwidth]{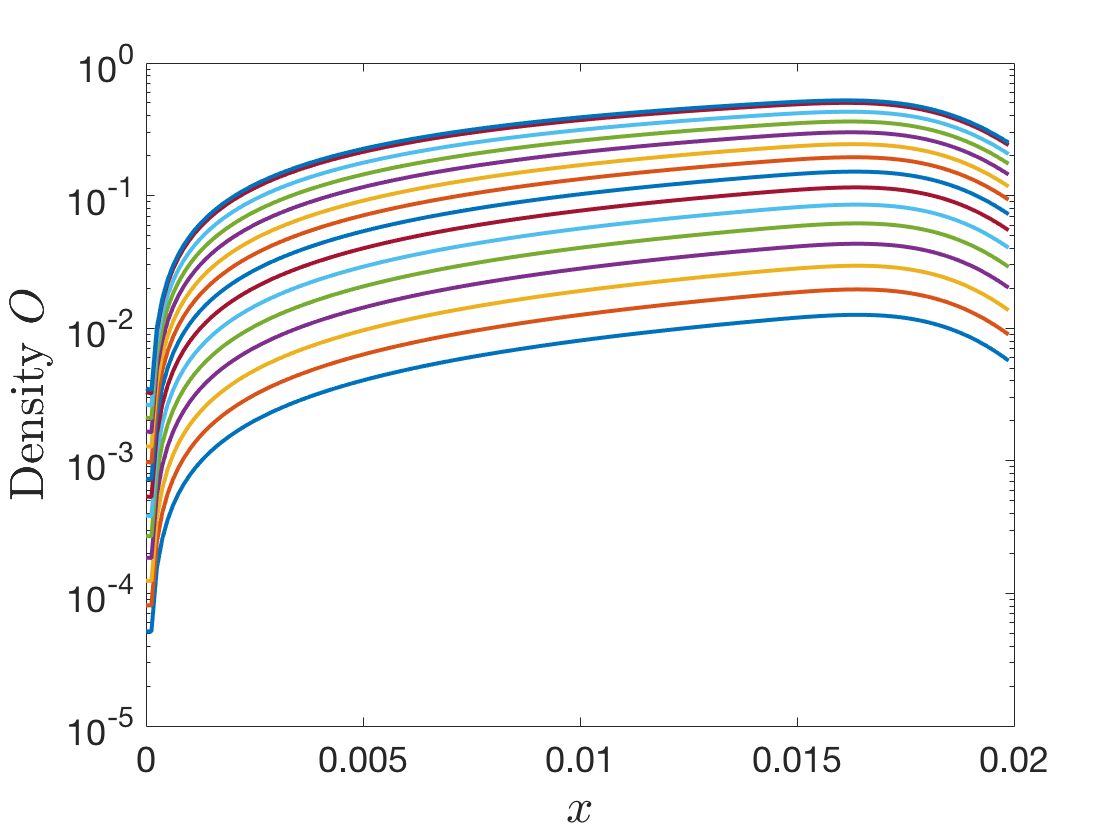}
\includegraphics[width=0.31\textwidth]{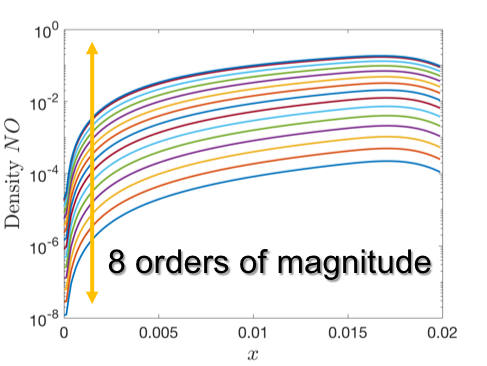}
\includegraphics[width=0.31\textwidth]{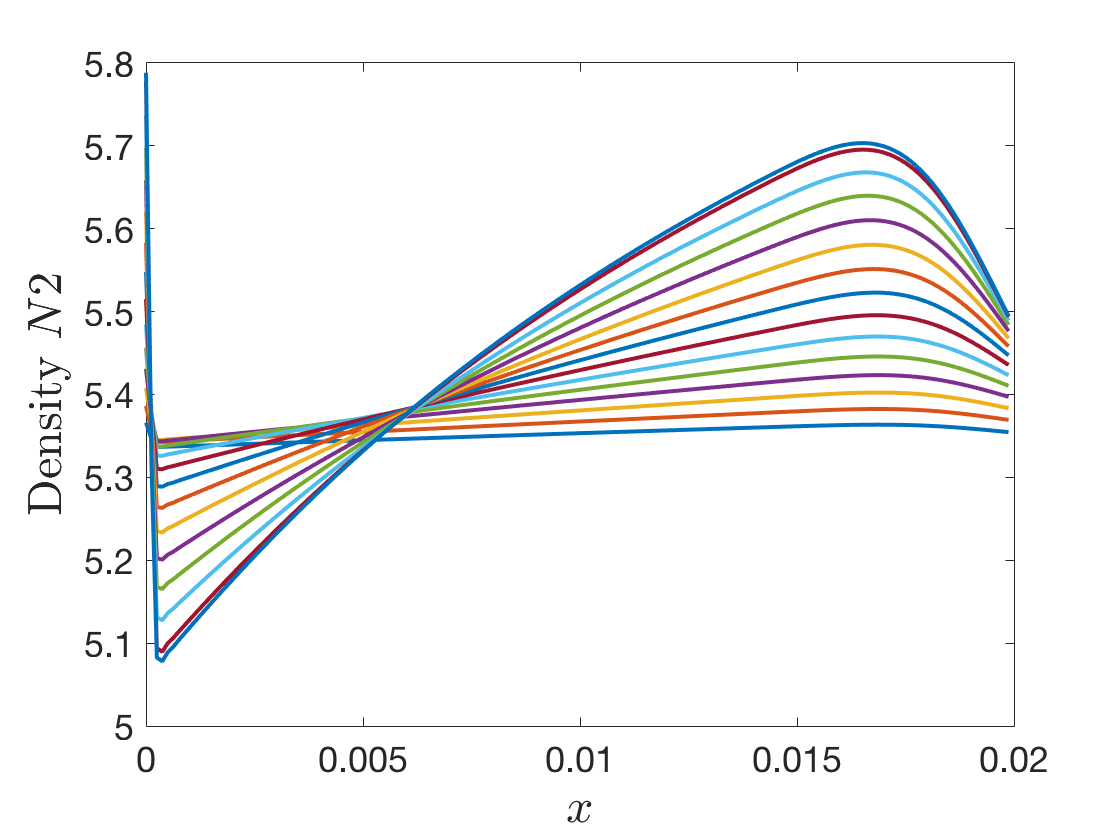}
\includegraphics[width=0.31\textwidth]{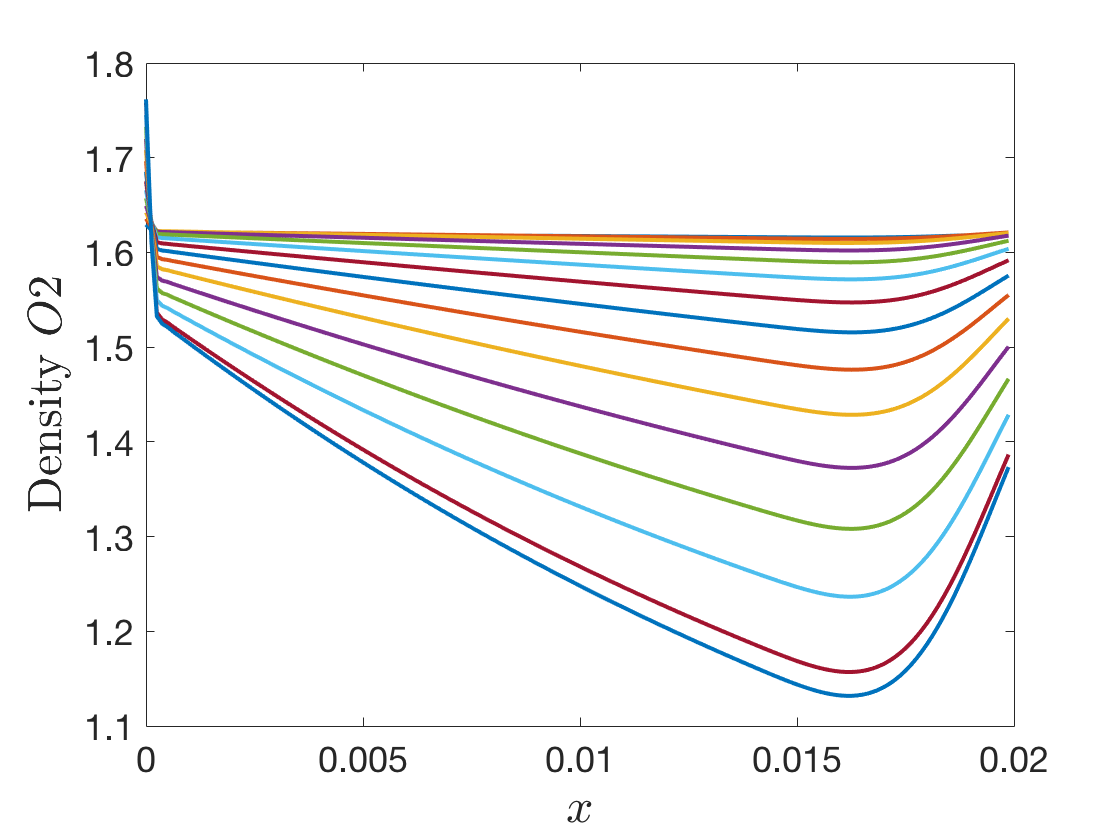}
\\
\includegraphics[width=0.31\textwidth]{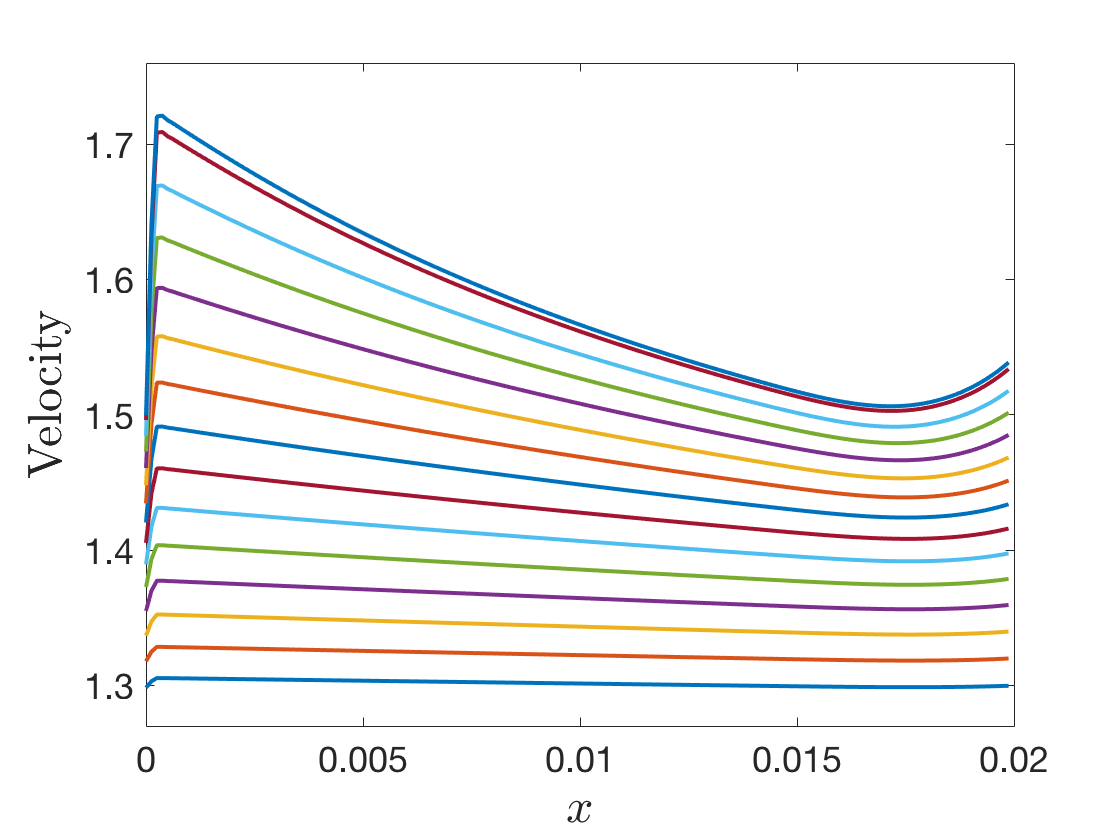}
\includegraphics[width=0.31\textwidth]{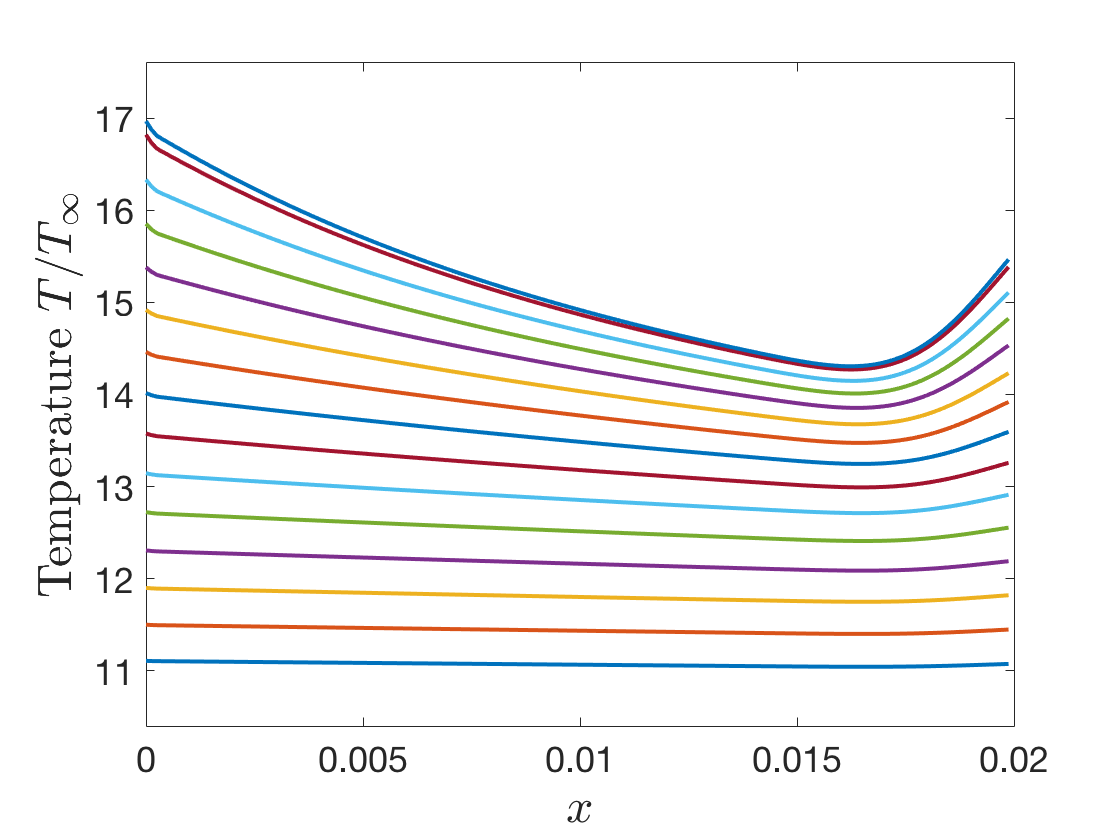}
\end{center}
\caption{Data defining the function spaces for the chemical species, the velocity and the temperature. The range of the function space of $\rho_{NO}$ spans 8 orders of magnitude.}
    \label{fig:Vspace}
\end{figure}

We generated 400 trajectories for $M_\infty \in [8, 10]$ and randomly selected 240 trajectories and 60 trajectories for training and testing, respectively (see figure ~\ref{fig:Vspace}). 
We show the function spaces for the five chemical species and the velocity and the temperature in Figure \ref{fig:Vspace}. The range of the function space of $\rho_{NO}$ spans 8 orders of magnitude across a thin boundary layer, which highlights the challenging of resolving this flow.  


\section{Developing DeepONets as building blocks}\label{sec:deeponet}
We develop in this section the DeepONets, which will serve as building blocks for the DeepM\&Mnet. 
These DeepONets target the coupled dynamics between the flow and the chemical species.  With input from either the flow or the chemical species, the DeepONet predicts all the remaining fields.  These DeepONets will subsequently be used as building blocks for constructing the DeepM\&Mnets.
Note that in \ref{sec:apd:DeepOnet:s}, we provide another example of three simple DeepONets for the chemical species alone, where each uses $\rho_{N_2}$ and $\rho_{O_2}$ as inputs and predicts either $\rho_{N}$, $\rho_{O}$ or $\rho_{NO}$. 

\subsection{DeepONet architecture}
Neural networks are not only universal approximators of continuous functions, but also universal approximators of nonlinear continuous operators~\cite{chen1995universal}. By virtue of the potential applications of the universal approximation theorem of the neural networks in learning nonlinear operators from data, Lu et al. proposed DeepONets to learn operators accurately and efficiently from a relatively small data set~\cite{lu2019deeponet}.

\begin{figure}[http]
\begin{center}
\includegraphics[scale=0.5,angle=0]{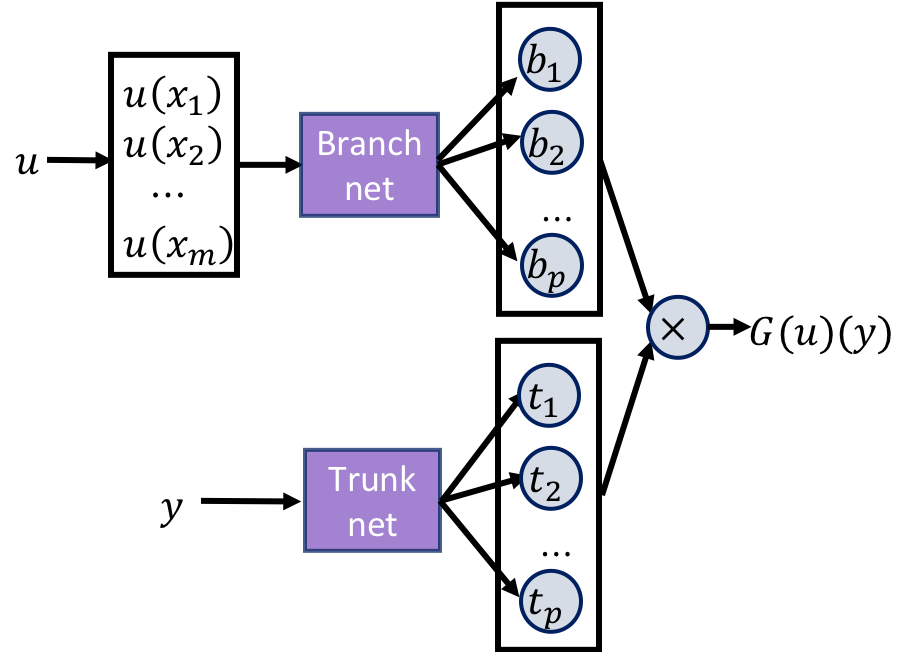}
\end{center}
\caption{Schematic of an unstacked DeepONet, which is the building block of the DeepM\&Mnet. DeepONet learns the operator $G: (u, y) \rightarrow G(u)(y) = \sum_{k=1}^{p} b_k t_k$. Here, $u(x)$ is the input of the branch net observed with $m$ sensors, and $b_i, i=1,2,\dots, p$ are the output of the branch net; $y$ is the input of the trunk net, and $t_i, i=1,2,\dots, p$ is the output of the trunk net. 
}\label{fig:DeepOnet}
\end{figure}

A DeepONet consists of two sub-networks, a branch net and a trunk net. The branch net is for encoding the input function at a fixed number of sensors $x_i , i = 1, \ldots, m$, while the trunk net is for encoding the locations for the output functions. For all the DeepONets presented herein, the number of sensors for the branch net is 75 while the number of sensors for the trunk net is 48. 
In \cite{lu2019deeponet}, two different architectures of DeepONets are proposed, namely, stacked DeepONets and unstacked DeepONets; unstacked DeepONets seem to have better performance. Hence, in this work, we will consider the unstacked DeepONets (Figure \ref{fig:DeepOnet}). A DeepONet learns an operator $G: u\rightarrow G(u)$, where the branch net takes $[u(x_1), u(x_2), \ldots, u(x_m)]$ as the input and outputs $[b_1, b_2, \ldots, b_p]$, and the trunk net takes $y$ as the input and outputs $[t_1, t_2, \ldots, t_p]$. The final output is given by 
\begin{equation*}
    G(u)(y) = \sum_{k=1}^{p} b_k t_k.
\end{equation*}
DeepONets are implemented in DeepXDE \cite{lu2019deepxde}, a user-friendly Python library designed for scientific machine learning.

\subsection{DeepONets for the coupled dynamics between the flow and the chemical species}\label{sec:deeponets:2}
\begin{figure}[http]
\begin{center}
\begin{minipage}{0.495\textwidth}\centering
\includegraphics[scale=0.465,angle=0]{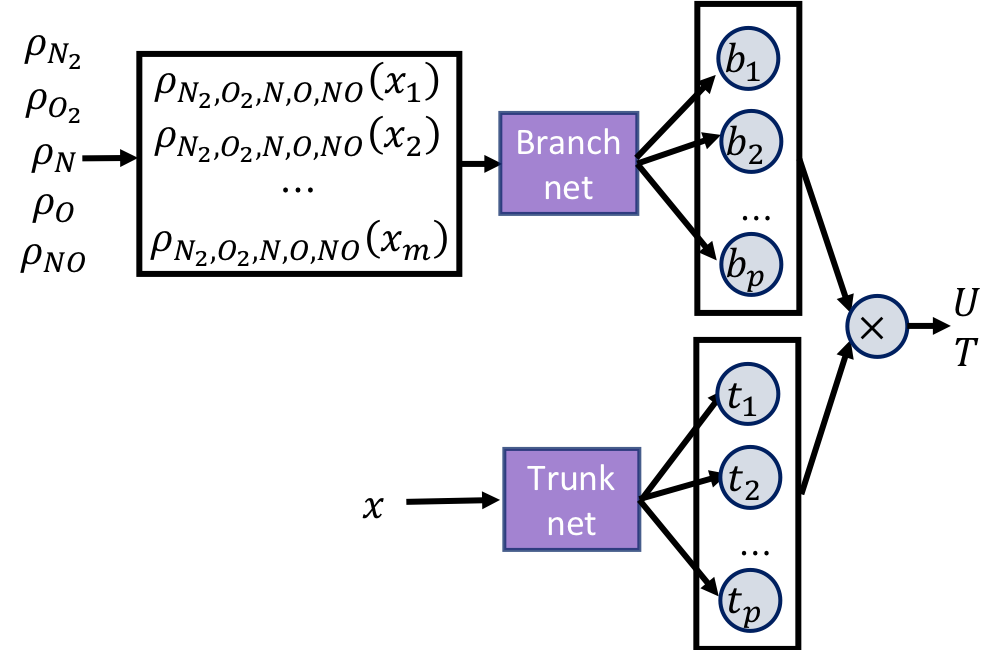} \par{(a)}
\end{minipage}
\begin{minipage}{0.495\textwidth}\centering
\includegraphics[scale=0.465,angle=0]{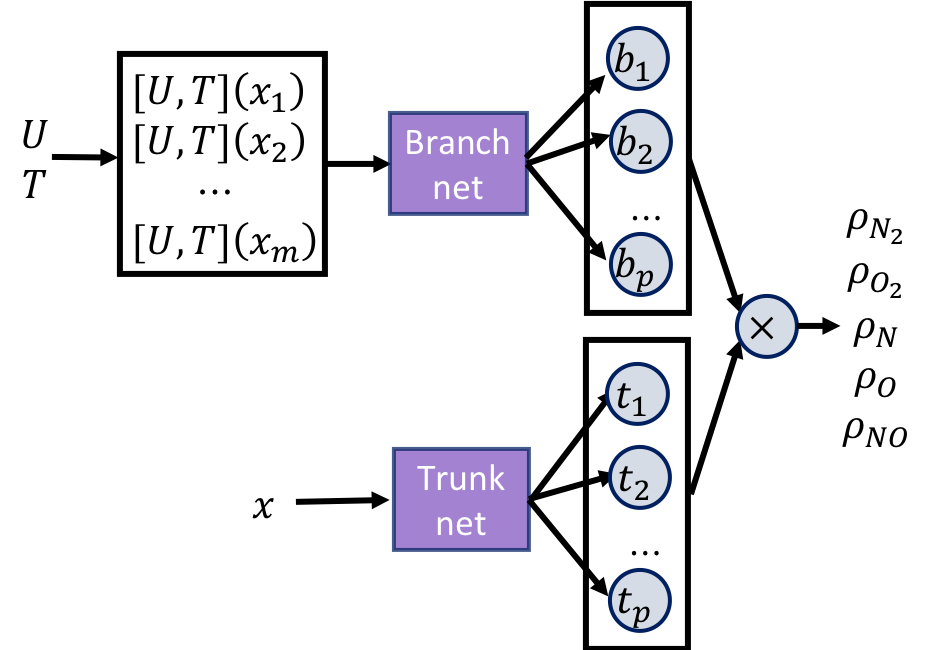} \par{(b)}
\end{minipage}
\end{center}
\caption{DeepONets for the coupled dynamics between the flow and the chemical species. (a): Schematic of using the densities of the five chemical species to predict the velocity and temperature, i.e., the schematic of the DeepONets: $G_{U,T}:~ \rho_{N_2,O_2,N,O,NO}\rightarrow [U,T]$. (b):  Schematic of using the velocity and temperature to predict densities of the five chemical species, i.e., the  DeepONets: $G_{\rho_{N_2,O_2,N,O,NO}}:~ [U,T]\rightarrow \rho_{N2,O2,N,O,NO}$.
}\label{fig:D2UT:Schematic}
\end{figure}

To demonstrate the simplicity and effectiveness of DeepONet, in \ref{sec:apd:DeepOnet:s} we develop three DeepONets to predict the densities $\rho_N$, $\rho_O$ and $\rho_{NO}$ using the densities $\rho_{N_2}$ and $\rho_{O_2}$ as the inputs of the branch nets. Here, however, we focus on the coupled dynamics of the flow an chemical fields.  For this purpose, we develop two DeepONets, which will become building blocks for constructing the DeepM\&Mnets for the entire configuration.  The two DeepONets have the following functionality: 
\begin{enumerate}[label=(\roman*),topsep=0pt,itemsep=0pt,parsep=0pt]
    \item $G_{U,T}:~ \rho_{N_2,O_2,N,O,NO}\rightarrow [U,T]$ uses the densities of the chemical species as the input of the branch net to predict the velocity $U$ and temperature $T$ (Figure \ref{fig:D2UT:Schematic} left).
    \item $G_{\rho_{N_2,O_2,N,O,NO}}:~ [U,T]\rightarrow \rho_{N2,O2,N,O,NO}$ uses the velocity $U$ and the pressure $T$ as the input of the branch net to predict all the densities of the five chemical species (Figure \ref{fig:D2UT:Schematic} right).
\end{enumerate}

We train all the DeepONets independently.
Moreover, we use the logarithms of the original data for all densities while we use the original data for the velocity and the temperature.
The parameters for the neural network are as follows:
\begin{itemize}[topsep=0pt,itemsep=0pt,parsep=0pt]
    \item Hidden layers for both branch and trunk nets: $4\times 100$.
    \item Activation function: adaptive ReLU;
    \item Learning rate: $6\times 10^{-4}$;
    \item Epochs: 120000. \\
\end{itemize}

\begin{figure}[http]
\begin{center}
\includegraphics[scale=0.15,angle=0]{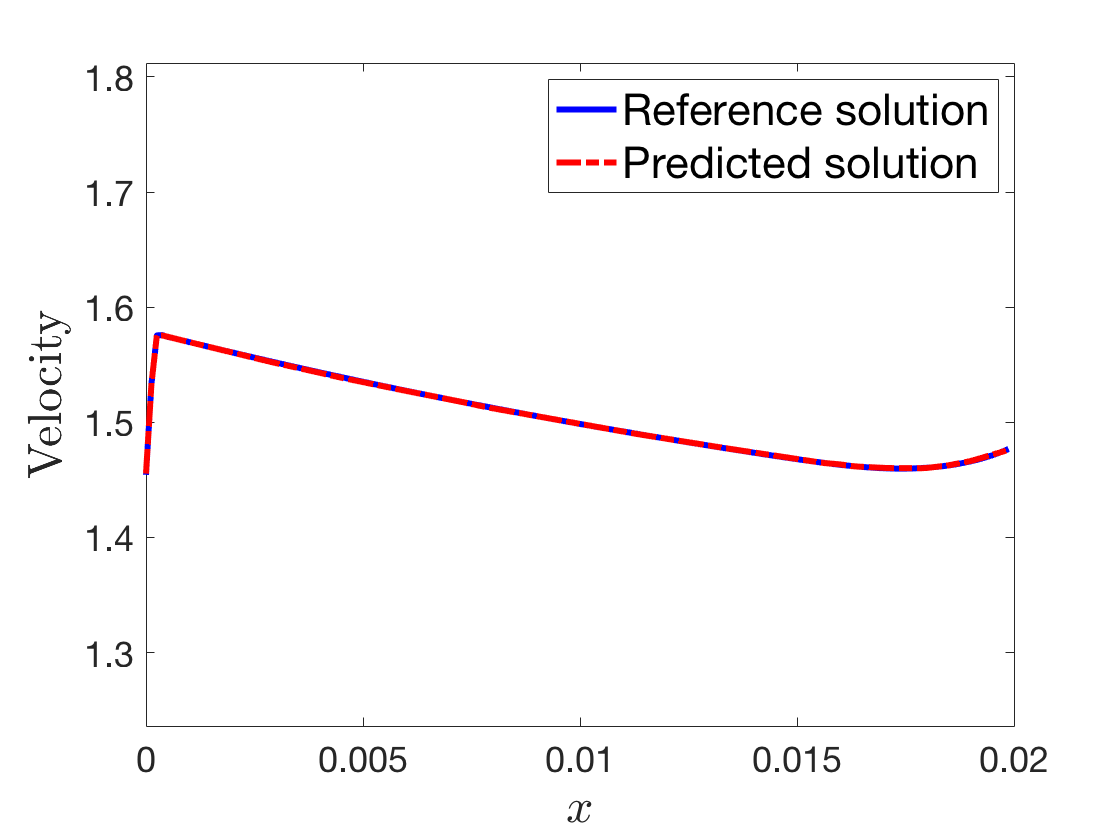}
\includegraphics[scale=0.15,angle=0]{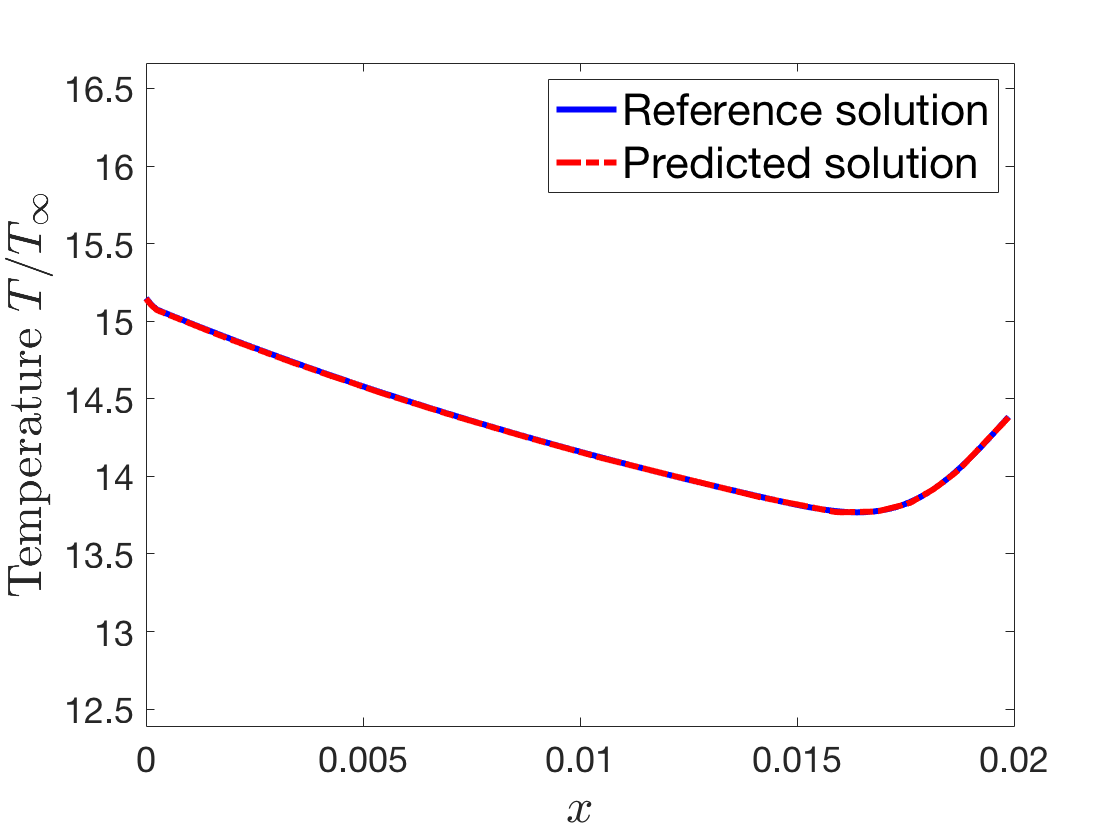}
\\
\includegraphics[scale=0.15,angle=0]{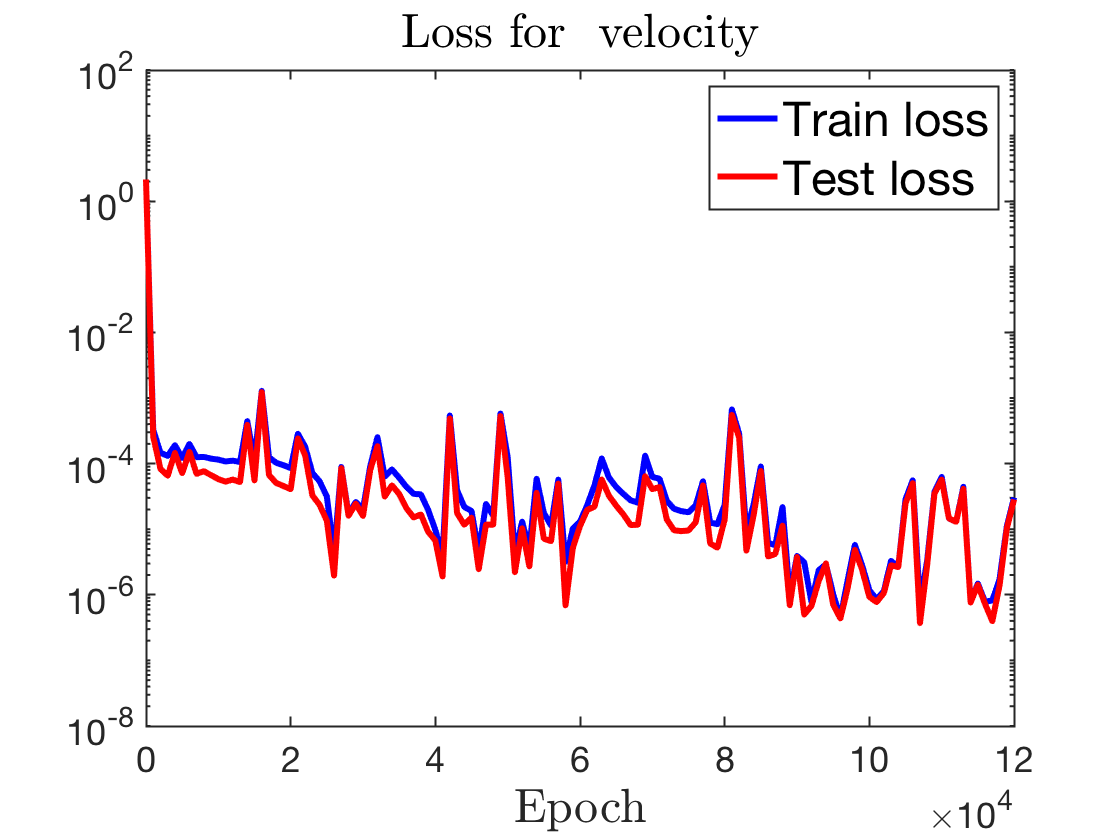}
\includegraphics[scale=0.15,angle=0]{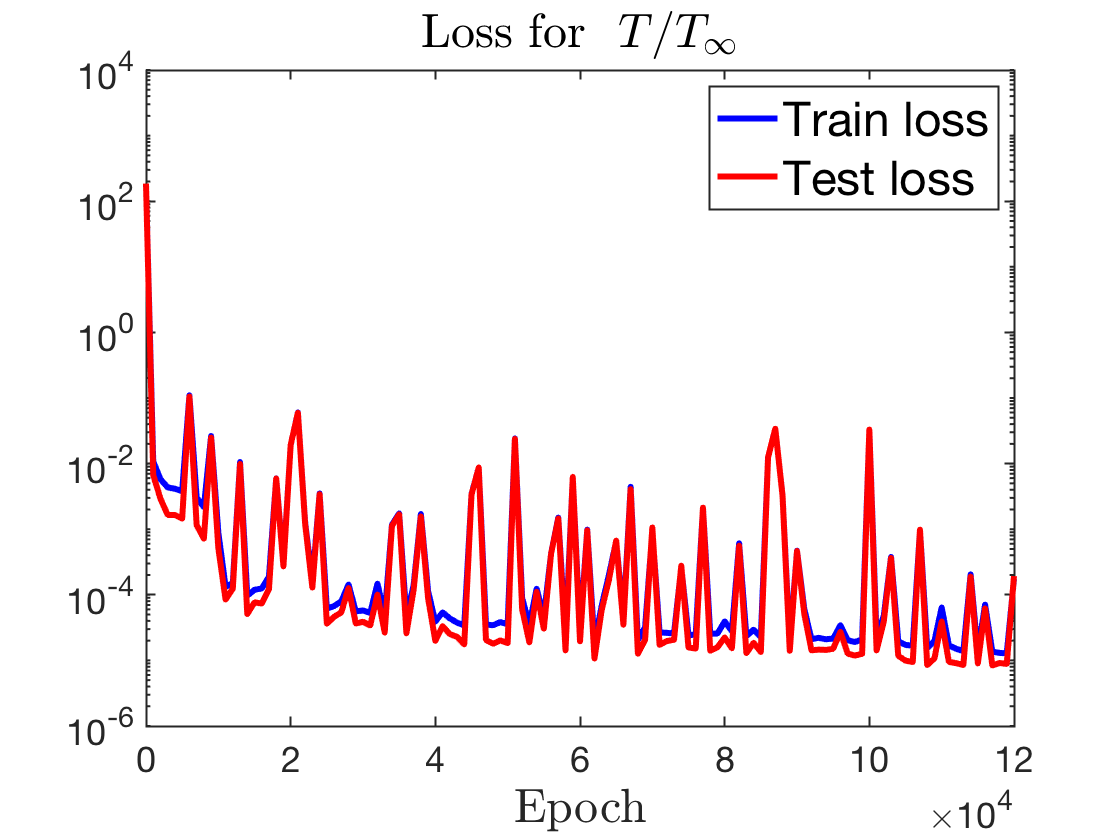}
\end{center}
\caption{Upper: Predictions of the velocity $U$ and the temperature $T/T_\infty$ using the DeepONets $G_{U,T}$. Lower: The training losses and testing losses for the velocity $U$ and the temperature $T/T_\infty$.
}\label{fig:D2UT}
\end{figure}

\begin{figure}[http]
\begin{center}
\includegraphics[scale=0.125,angle=0]{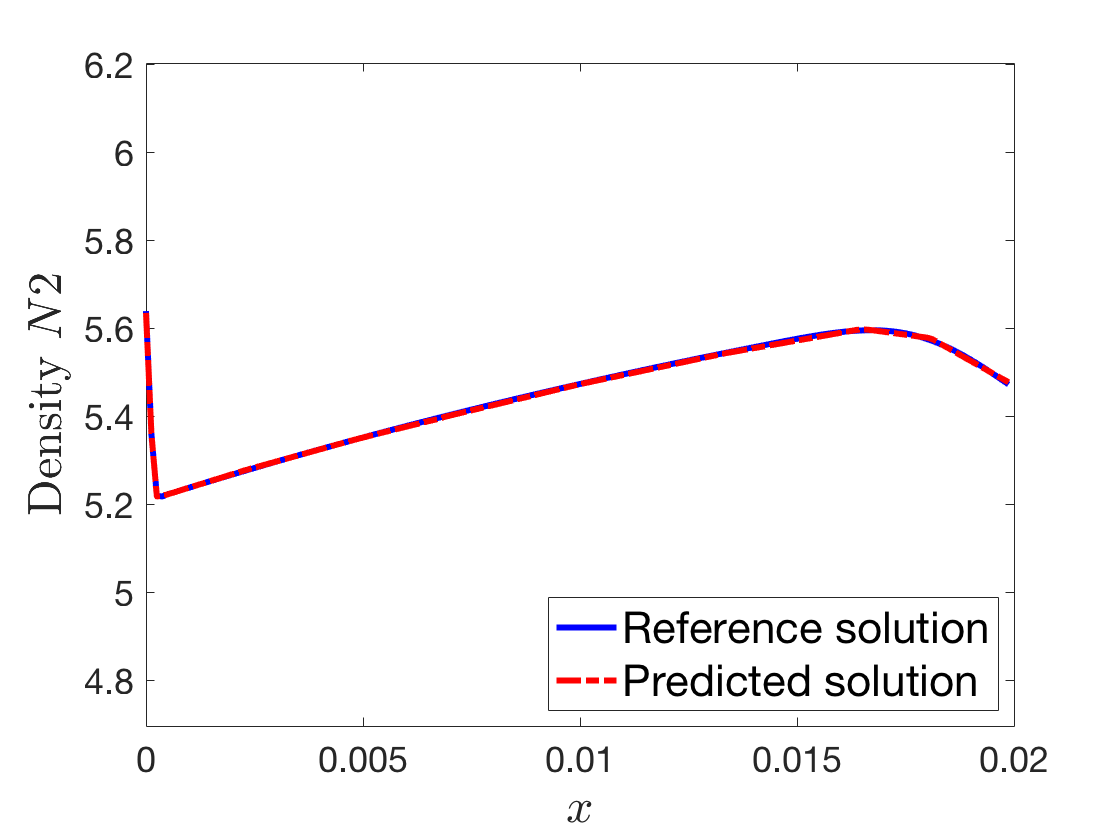} 
\includegraphics[scale=0.125,angle=0]{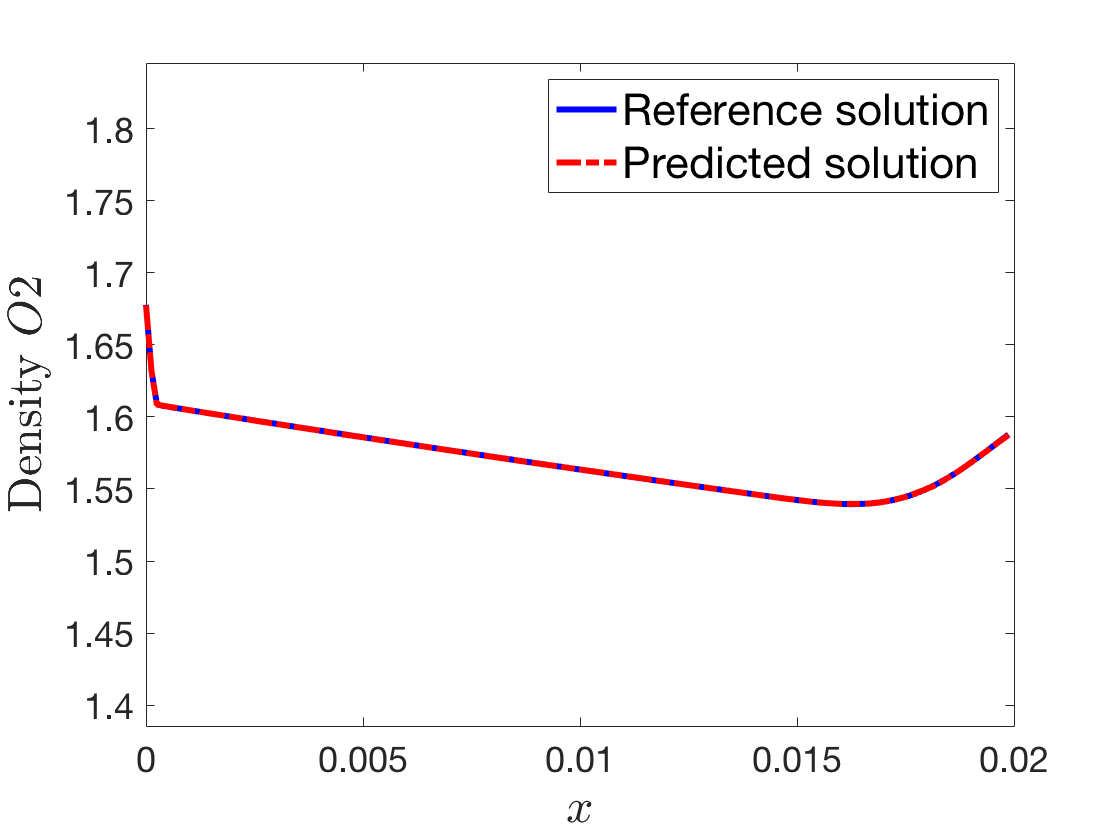} 
\includegraphics[scale=0.125,angle=0]{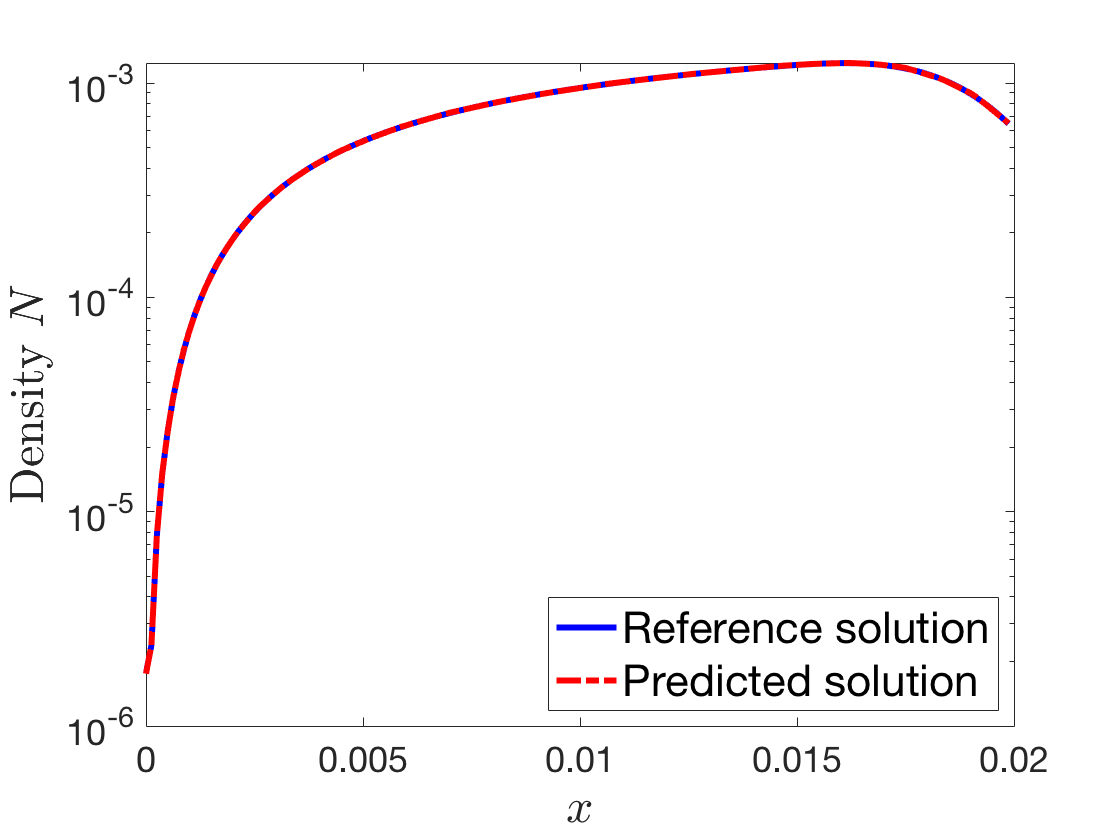} 
\includegraphics[scale=0.125,angle=0]{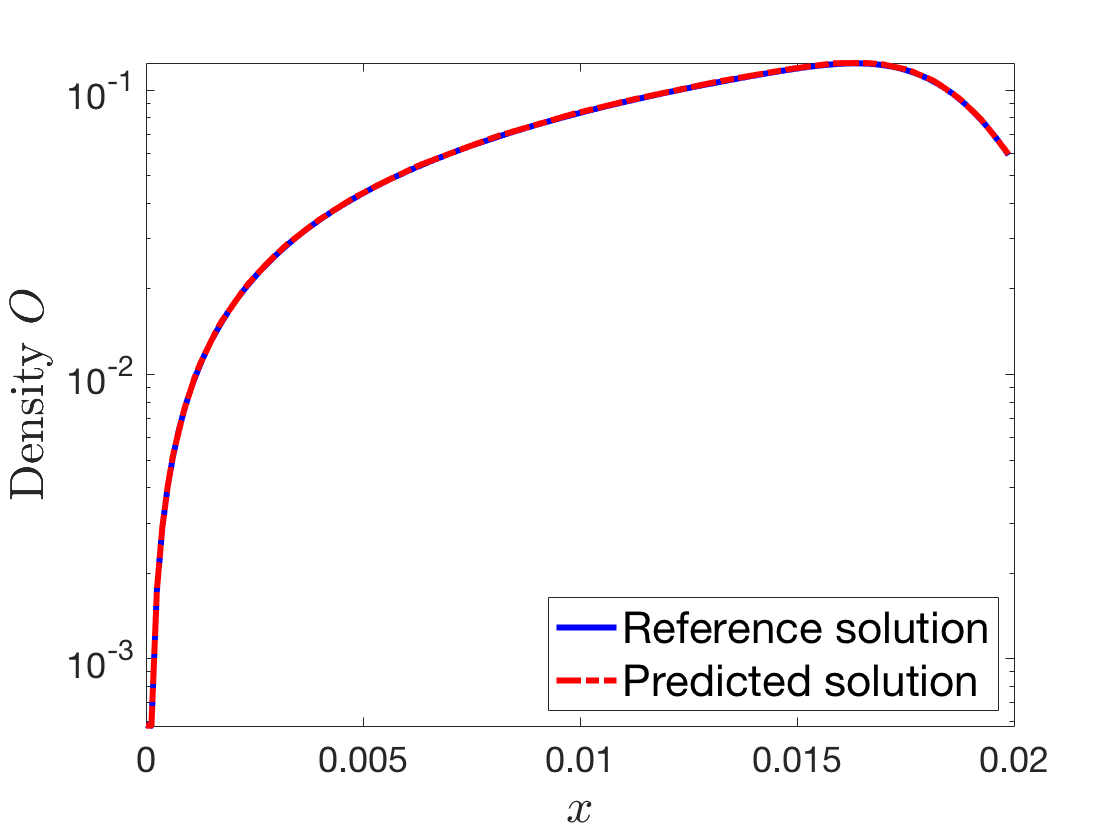} 
\includegraphics[scale=0.125,angle=0]{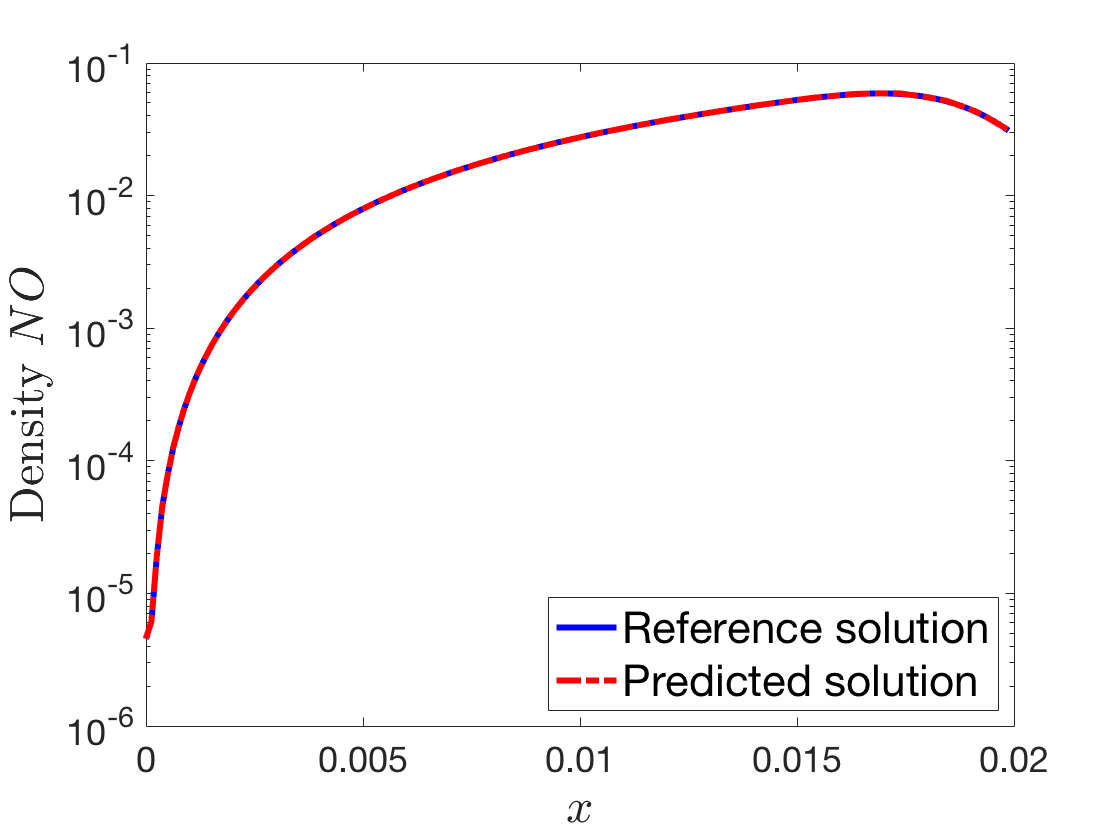} 
\includegraphics[scale=0.125,angle=0]{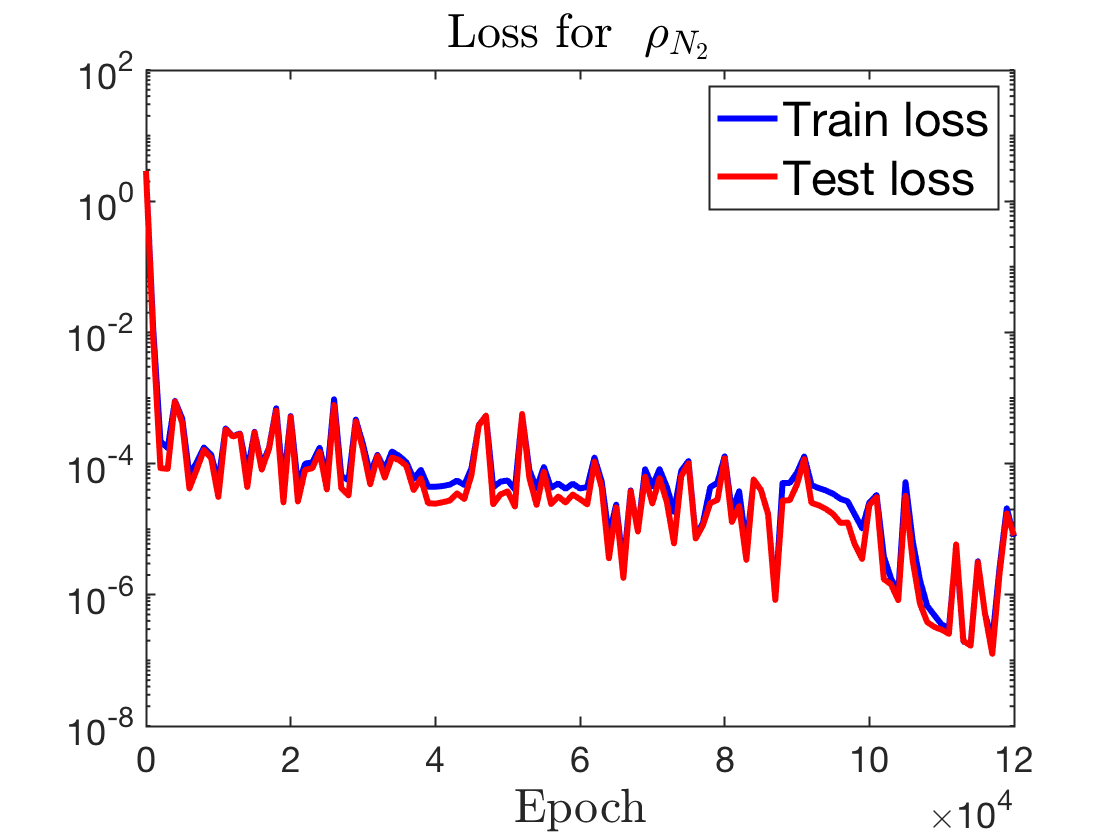} 
\end{center}
\caption{We show in the first five plots the predictions of densities of all the five chemical species, i.e.,  $\rho_{N_2}, ~~\rho_{O_2}, ~~\rho_{N}, ~~\rho_{O}, ~~\rho_{NO},$ using the DeepONets $G_{\rho_{N_2,O_2,N,O,NO}}$. The training losses and testing losses for the density $\rho_{N_2}$ are  shown in the last plot. The training and testing losses are similar for other densities.
}\label{fig:UT2D}
\end{figure}

We show the comparisons between the predictions of the DeepONets $G_{U,T}$ and the independent reference data, and the corresponding losses of the training and testing for $U, T/T_\infty$ in Figure \ref{fig:D2UT}. The agreement demonstrates the capacity of DeepONet to predict the velocity and temperature based on knowledge of the chemical species alone, and the testing errors against independent data are commensurate with the training errors.  We also show the comparisons between the predictions of the DeepONets $G_{\rho_{N_2,O_2,N,O,NO}}$ and the independent reference data as well as the corresponding losses of the training and testing for $\rho_{N_2}$ in Figure \ref{fig:UT2D}.
Again, we observe that the predictions are in excellent agreement with the reference independent data, which were not part of the training. Note that some of the densities vary by orders of magnitude downstream of the shock, and DeepONet predicts these variations accurately.  
An important consideration for achieving this level of accuracy is that the data used for all densities are the logarithms of the original data.  
These results demonstrate that we have successfully trained the DeepONets $G_{U,T}$ and  $G_{\rho_{N_2,O_2,N,O,NO}}$.  
Once trained, the computational cost of DeepONet for predictions of the independent data is much less than CFD simulations. 
With the successfully trained DeepONets $G_{U,T}$ and  $G_{\rho_{N_2,O_2,N,O,NO}}$, we can subsequently predict the interplay of the flow and the chemical species using DeepM\&Mnets (\S\ref{sec:deepmmnet}) and using it for data assimilation.  However, an important pre-requisite for a robust assimilation framework is the ability to handle new data outside the training range, or to extrapolate.  We will therefore first show the prediction when inputs are outside the function space of the training data for the DeepONets.

\subsection{Extrapolation: Predicting inputs outside the function space of the training data}\label{sec:extrapolation}
The DeepONet predictions shown previously are results for Mach numbers in the input training range, i.e., the testing was performed with independent data but with Mach number lies in the interval $[8,10]$, or interpolation. 
Now, we would also like to obtain predictions with the trained DeepONets for Mach numbers that do not belong to the input function space, namely extrapolation. In particular, we want to use the trained DeepONet $G_{U,T}$ to predict the velocity and temperature as well as  $G_{\rho_{N_2,O_2,N,O,NO}}$ to predict the densities of the chemical species for $M_\infty \notin [8,10]$.
For brevity, we only consider the DeepONet $G_{\rho_{N_2,O_2,N,O,NO}}$ and the results can be interpreted as representative of the performance of the method.

\begin{figure}[http]
\begin{center}
\begin{minipage}{0.49\textwidth}\centering
\includegraphics[scale=0.15,angle=0]{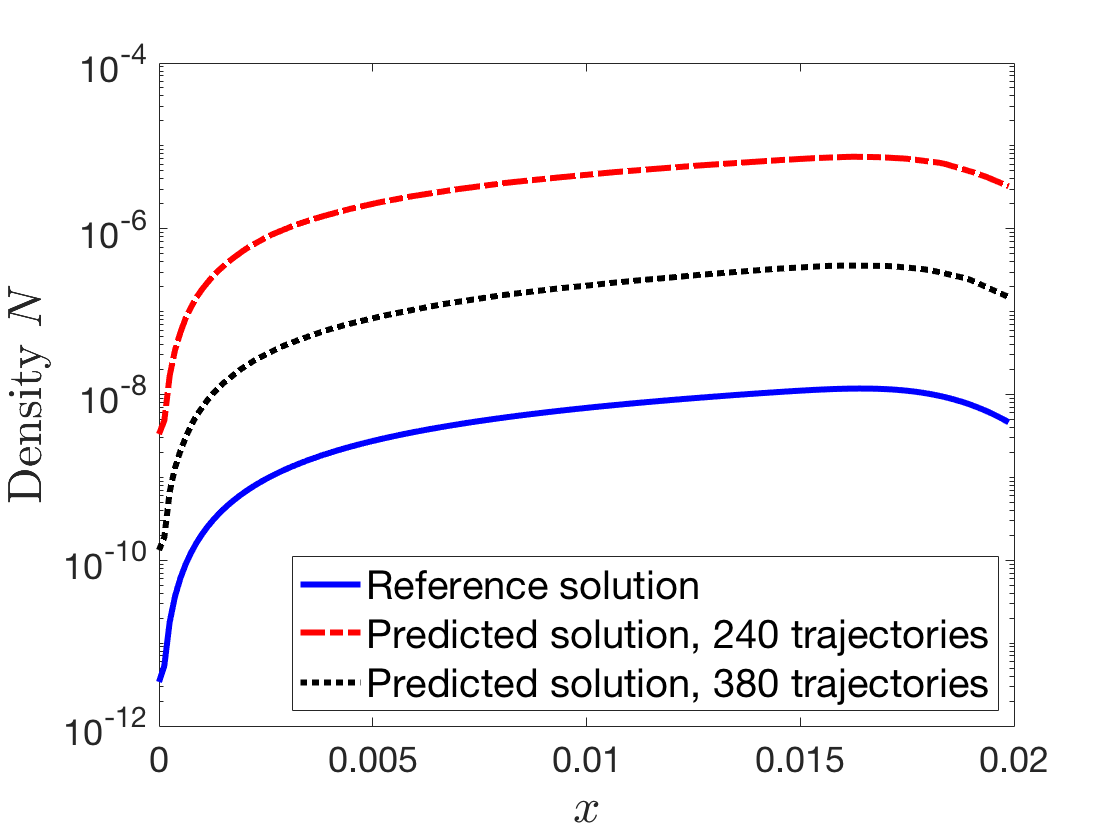} \par{(a) $\rho_N,~Mach = 7.0$}
\end{minipage}
\begin{minipage}{0.49\textwidth}\centering
\includegraphics[scale=0.15,angle=0]{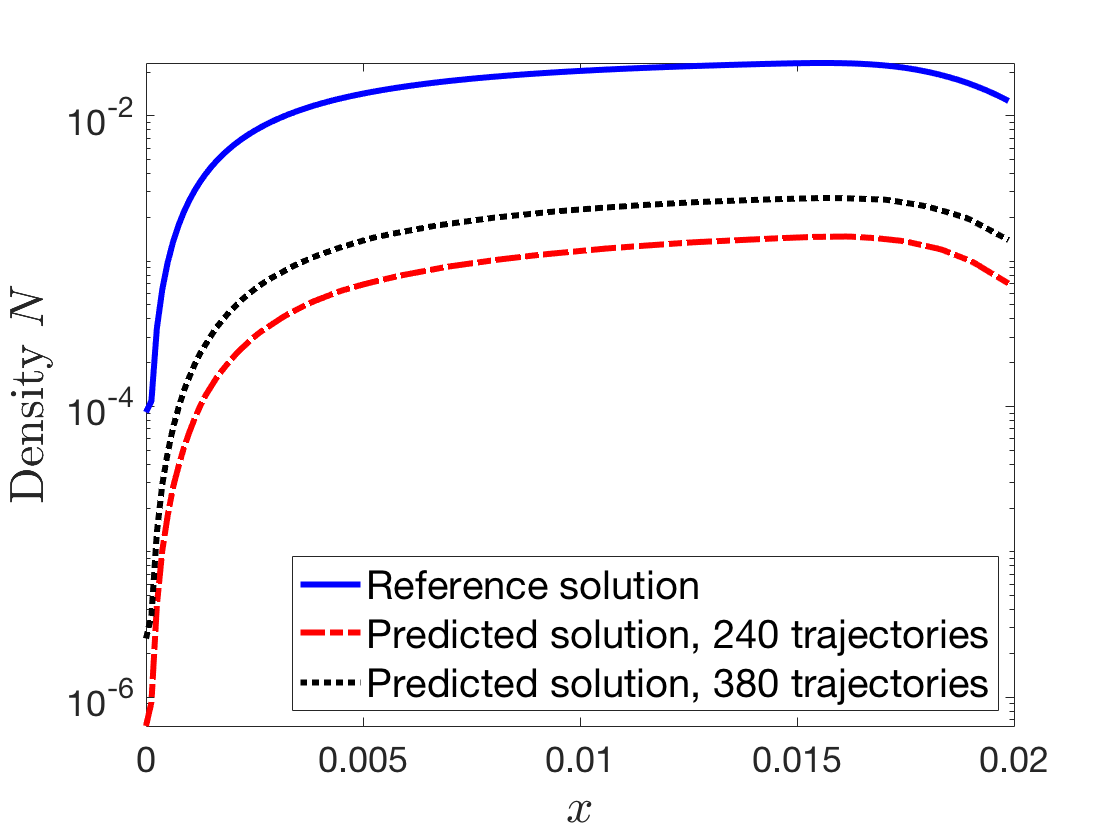} \par{(b) $\rho_N,~Mach = 11.0$}
\end{minipage}
\begin{minipage}{0.49\textwidth}\centering
\includegraphics[scale=0.15,angle=0]{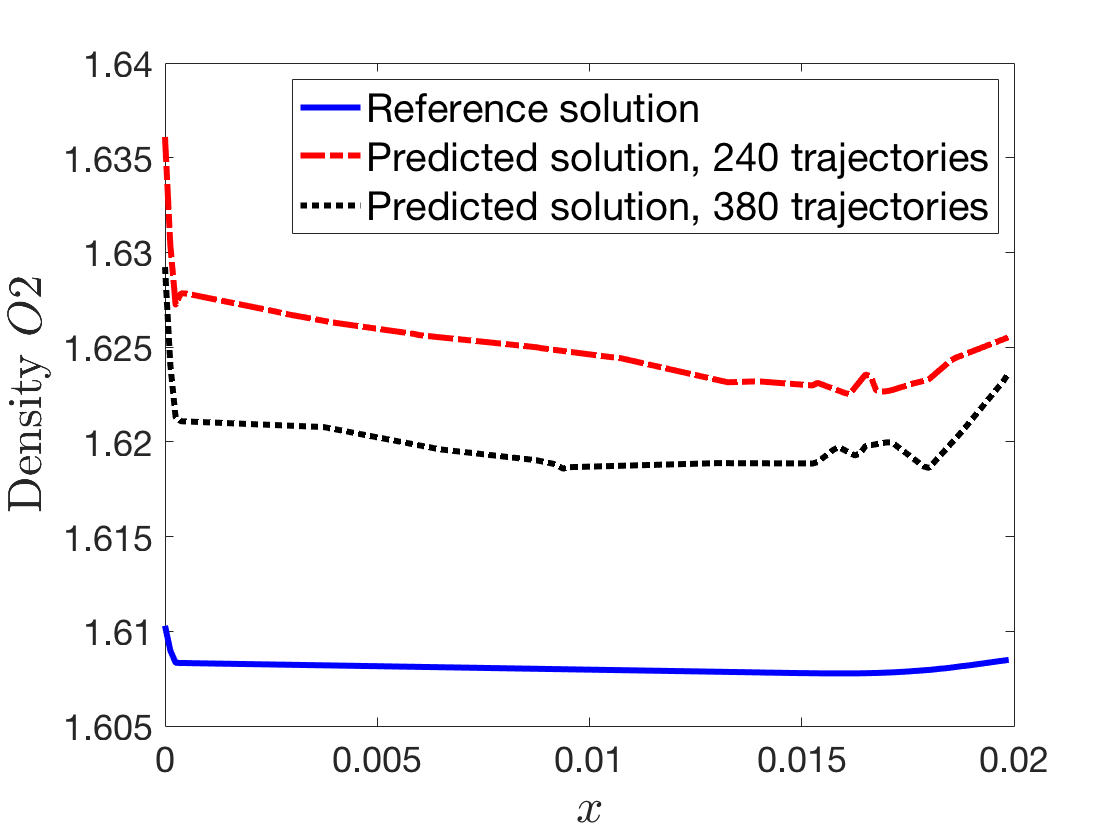} \par{(c) $\rho_{O_2},~Mach = 7.5$}
\end{minipage}
\begin{minipage}{0.49\textwidth}\centering
\includegraphics[scale=0.15,angle=0]{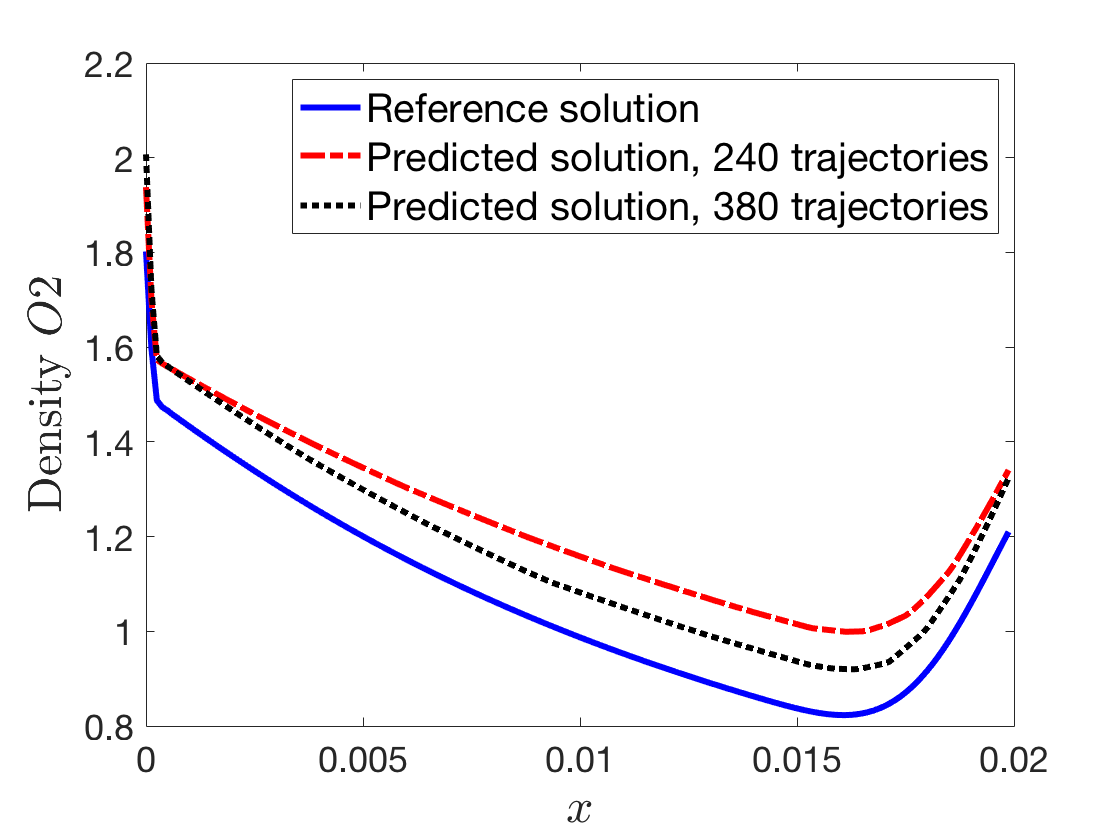} \par{(d) $\rho_{O_2},~Mach = 10.5$}
\end{minipage}
\end{center}
\caption{Extrapolation for the chemical densities with different values of Mach number and different numbers (240 and 380) of training trajectories. The results obtained by using a larger number of training trajectories are closer to the reference solutions compared to the results obtained with a small number of trajectories.
}\label{fig:ext:N:nodata}
\end{figure}

We show the DeepONet predictions for $\rho_N$ and $\rho_{O_2}$ with different values of Mach number and different number of training trajectories in Figure \ref{fig:ext:N:nodata}.
We observe that the predictions do not match the reference data. More precisely, the predictions have a shift from the reference solutions. However, the shapes of the predicted solutions are very similar to the reference solutions. Moreover, we observe that the results obtained by using a larger number of training trajectories are closer to the reference solutions, which implies that the more accurately that 
we represent the input space, the more accurate the extrapolation.  Nonetheless, it is desirable to extrapolate without requiring substantial additional training and to be able to do so robustly `on the fly'.  

\begin{figure}[http]
\begin{center}
\includegraphics[scale=0.95,angle=0]{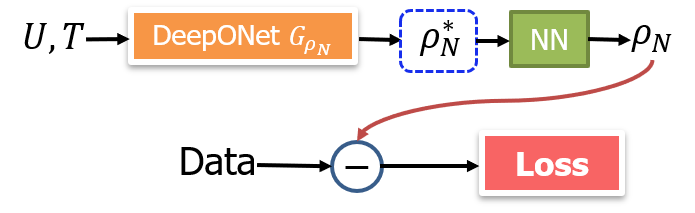}
\end{center}
\caption{Schematic of the architecture for the extrapolation with data for $\rho_N$. The output of the DeepONet $G_{\rho_N}$, i.e., $\rho^*_N$, is used as the input of the neural network ``NN".
}\label{fig:Ext:Schematic:N}
\end{figure}

In the spirit of data assimilation, which we will demonstrate using DeepM\&MNets (\S\ref{sec:deepmmnet}), here too we show how extrapolation in the context of DeepONet can be improved when a few data are available for the output variables, for example from sensors. We propose a simple neural network that takes the DeepONet prediction as its input, and define its loss function to be the mean square error between the scarce available data and its output.  We show a schematic of the proposed neural network for $\rho_N$ in figure \ref{fig:Ext:Schematic:N}, where we use the output of the DeepONet $G_{\rho_N}$, i.e., $\rho_N^*$, as the input of the neural network to be trained. The loss function is given by 
\begin{equation*}
    \mathcal{L}  = \frac{1}{n_{D}}\sum_{i=1}^{n_{D}} \left\|\rho_N(x_i) - \rho_{N, data}^i \right\|^2,
\end{equation*}
where $\rho_N$ is the output of the neural network to be trained, and $\rho_{N, data}^i, i=1,2,\ldots, n_{D}$ are given data.
In this configuration, it is important that the additional network does not require exhaustive training, and is simply required to improve the prediction of DeepONet to match the sensor data.  

\begin{figure}[http]
\begin{center}
\includegraphics[scale=0.25,angle=0]{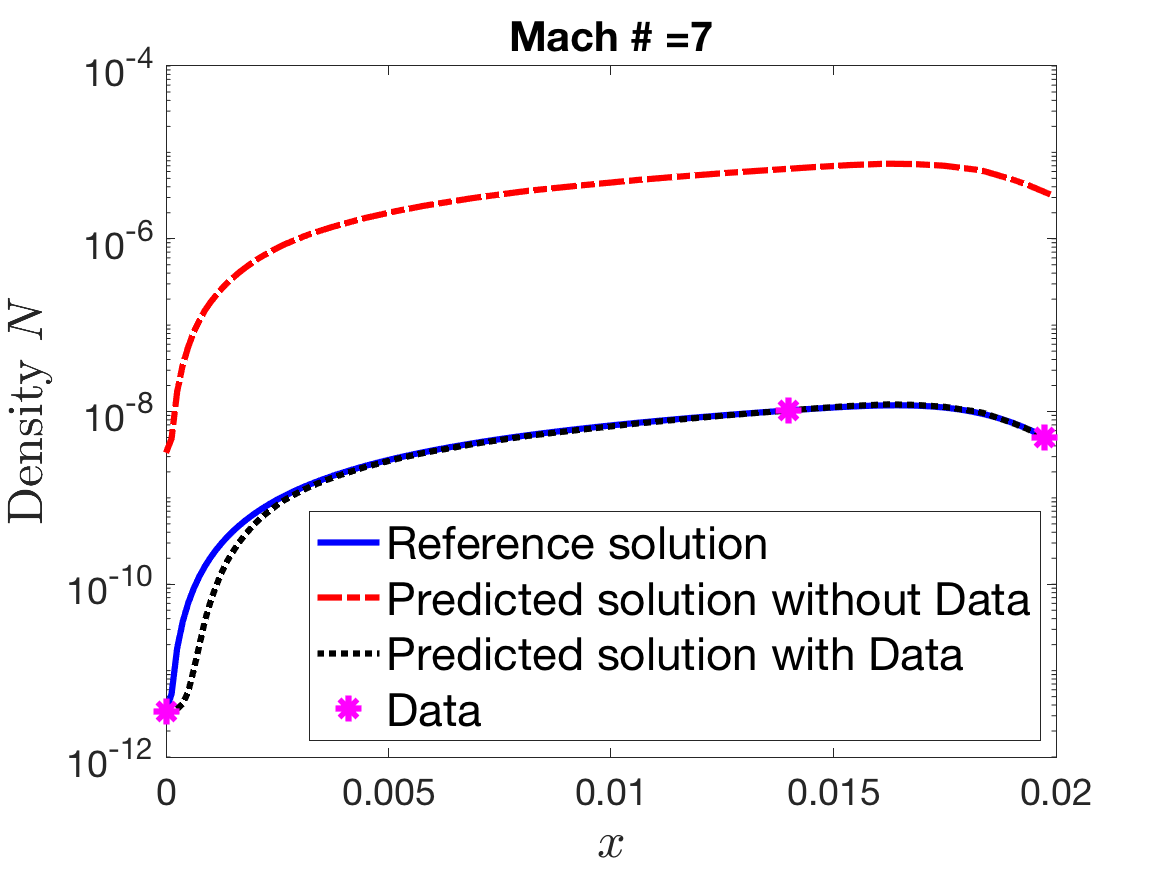}
\includegraphics[scale=0.25,angle=0]{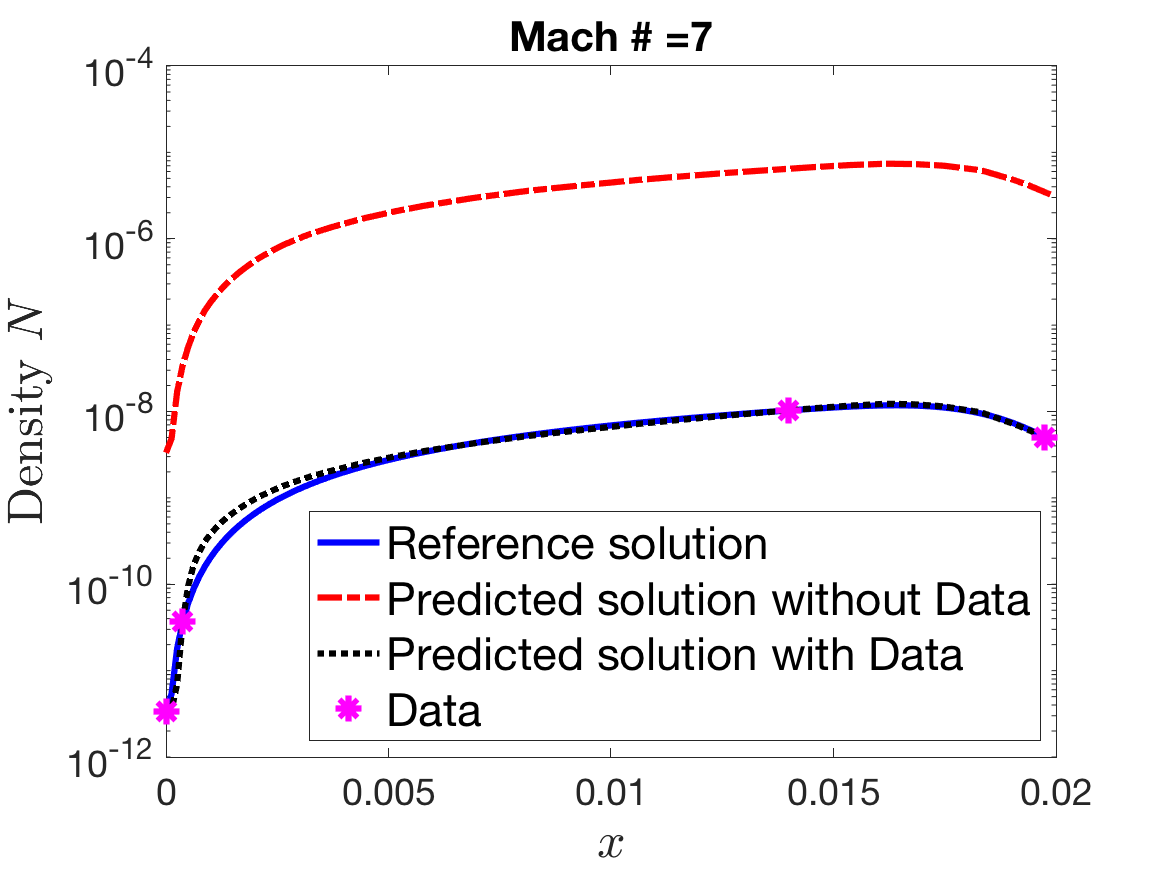}
\includegraphics[scale=0.25,angle=0]{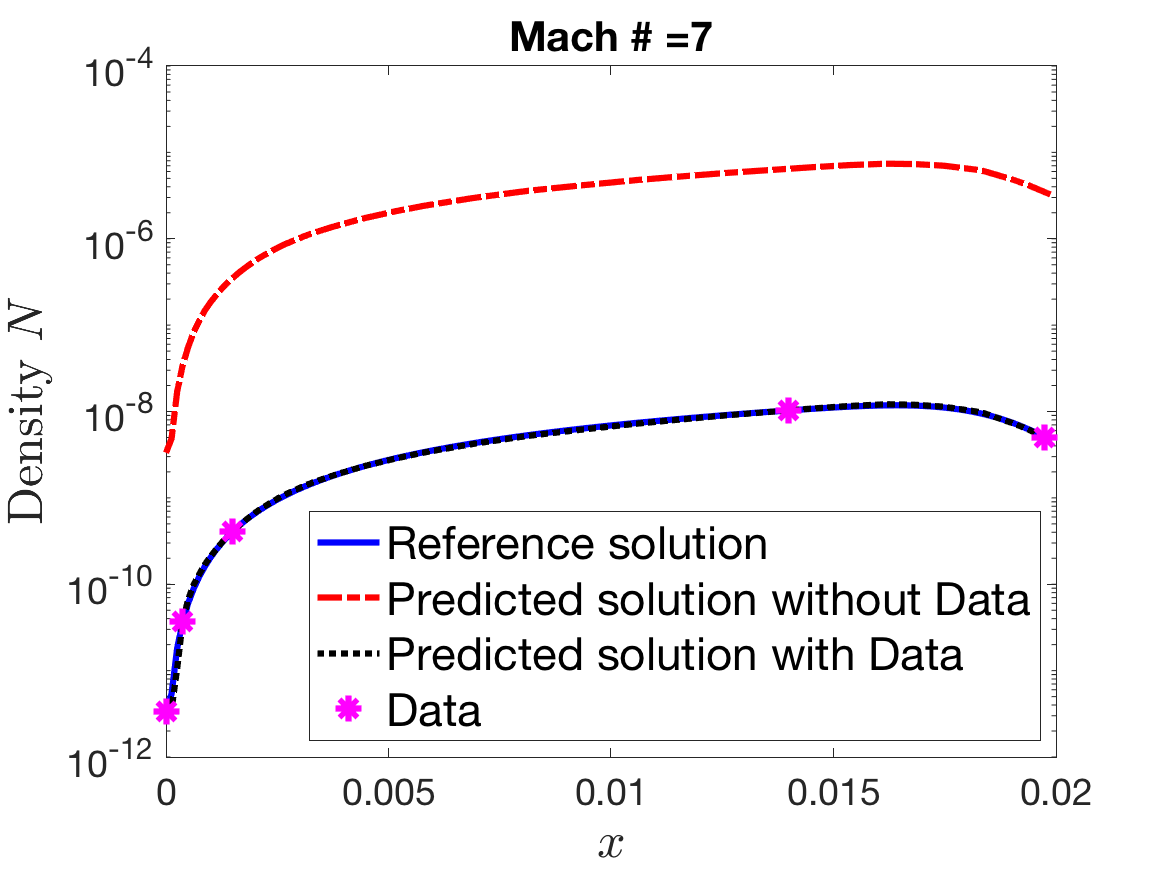}
%
%
\includegraphics[scale=0.25,angle=0]{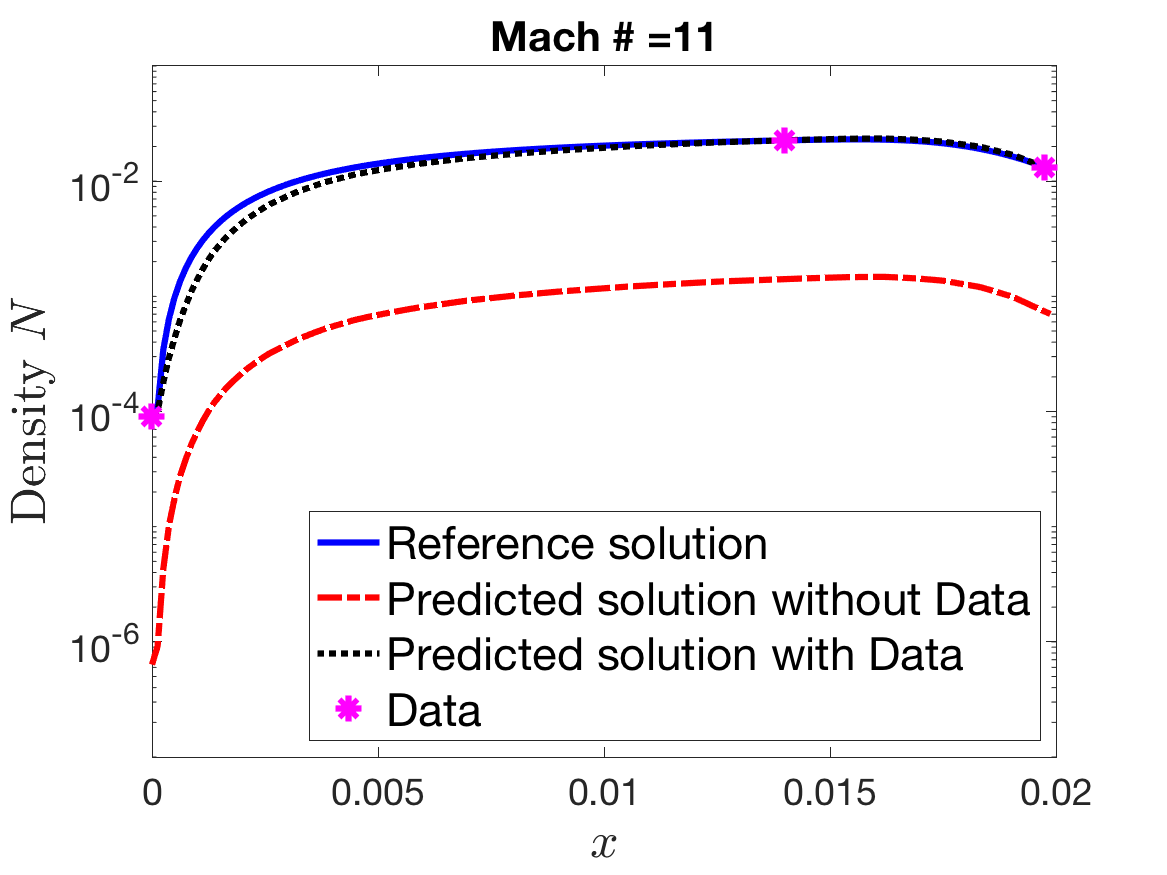}
\includegraphics[scale=0.25,angle=0]{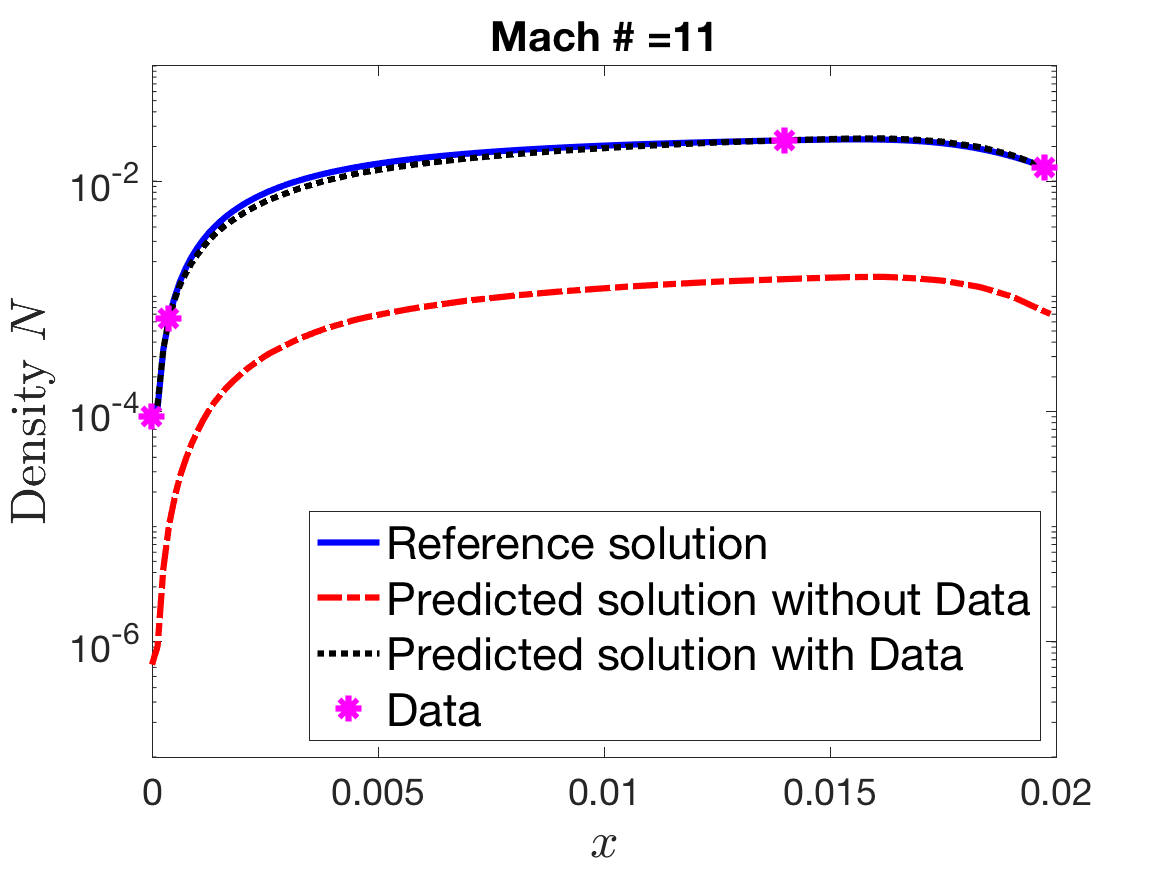}
\includegraphics[scale=0.25,angle=0]{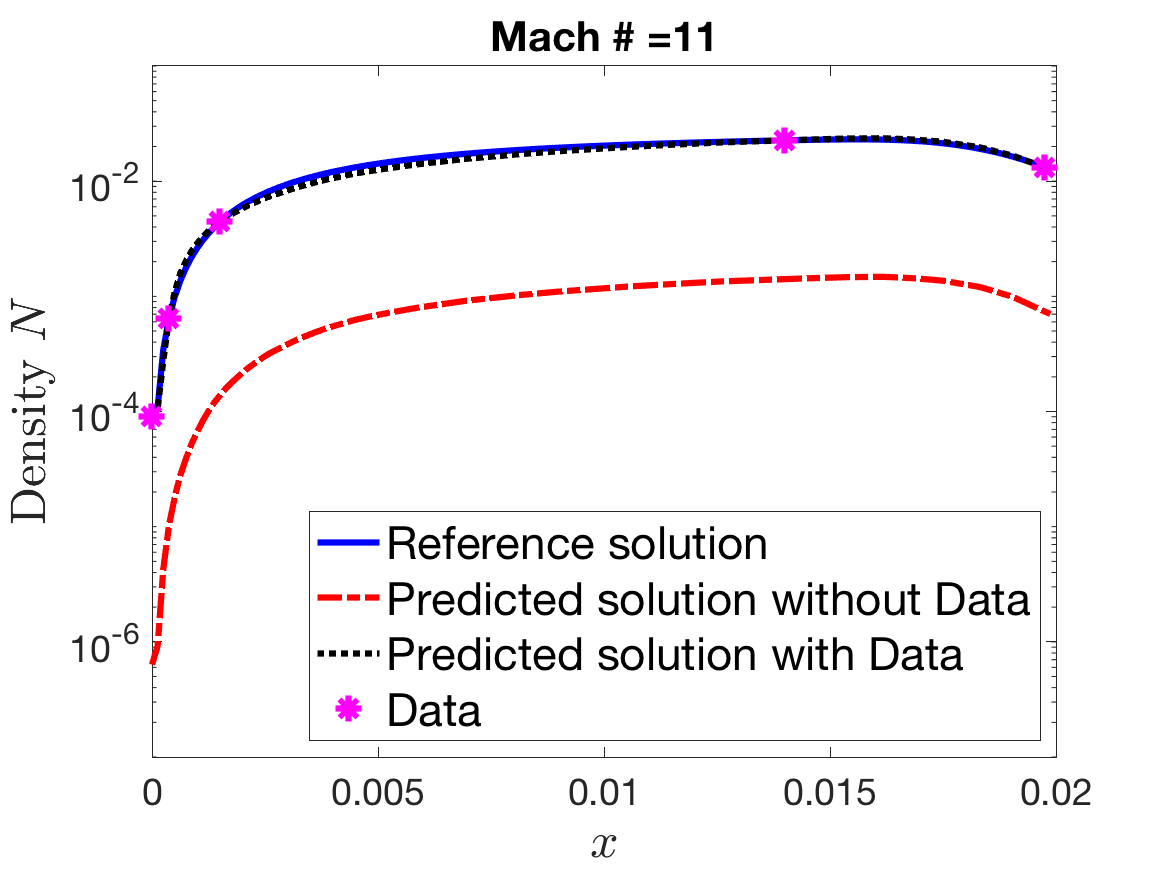}
\end{center}
\caption{Extrapolations of $\rho_N$ for $Mach = 7$ and $Mach = 11$ without and with data. For each Mach number, the number of data is increased from 3 to 4 then 5 points. We obtain good results with just a few data for the extrapolation. The red dot-dash lines are the outputs of the DeepONet $G_{\rho_N}$ without data while the black dash lines are the predictions with data. 
}\label{fig:ext:N}
\end{figure}

\begin{figure}[http]
\begin{center}
\includegraphics[scale=0.25,angle=0]{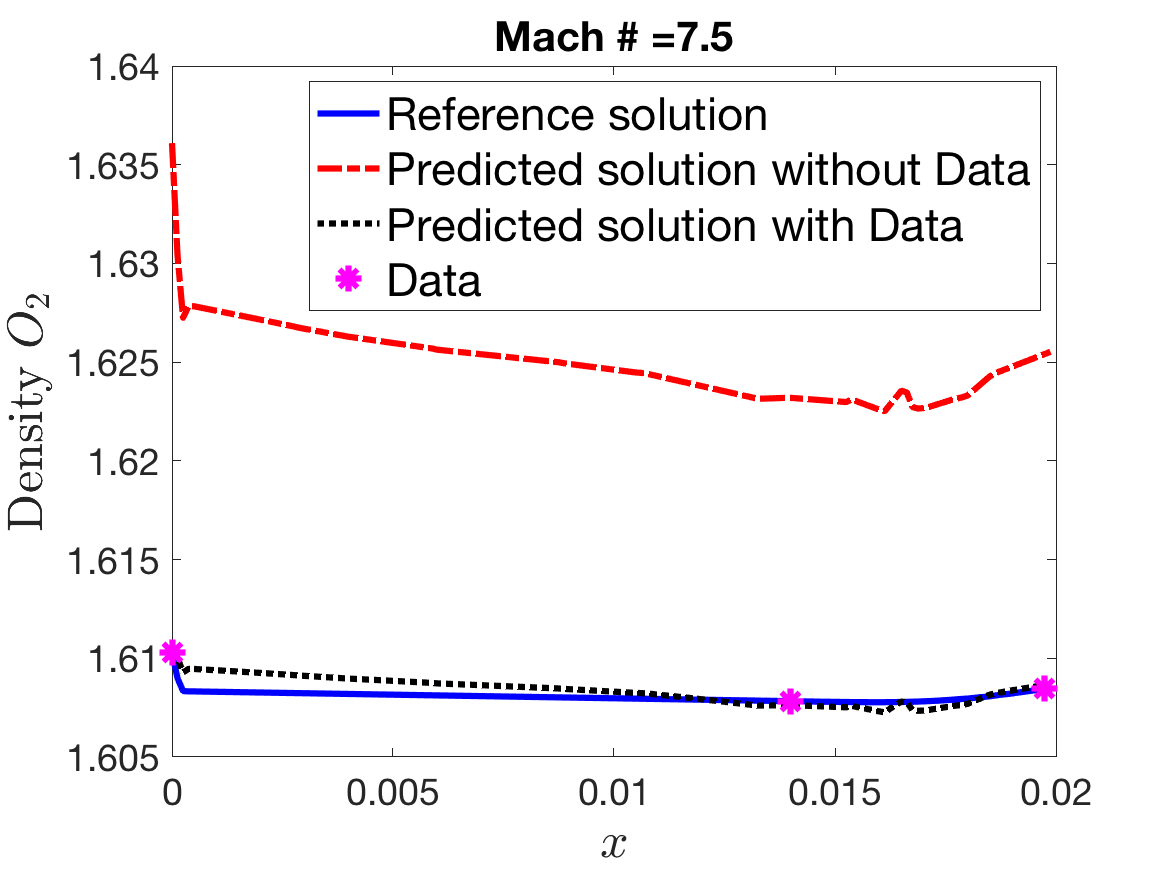}
\includegraphics[scale=0.25,angle=0]{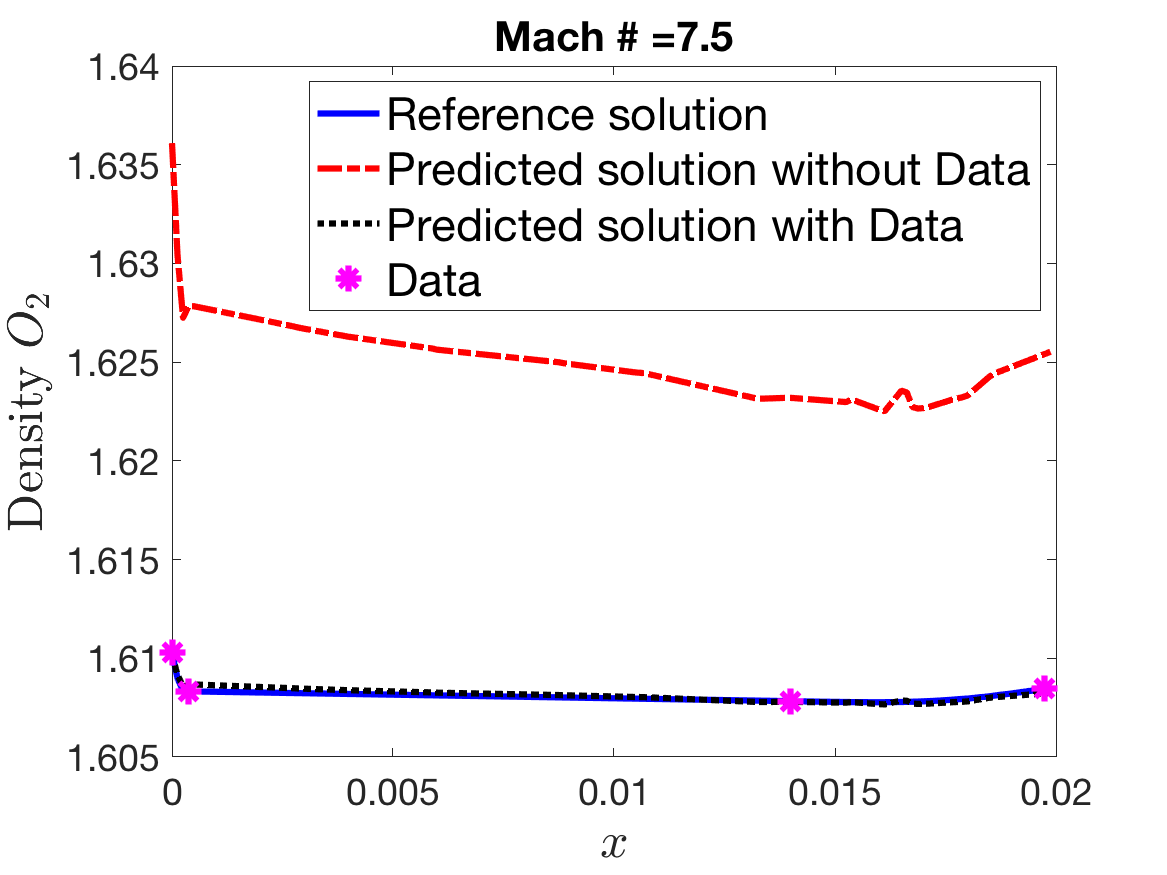}
\includegraphics[scale=0.25,angle=0]{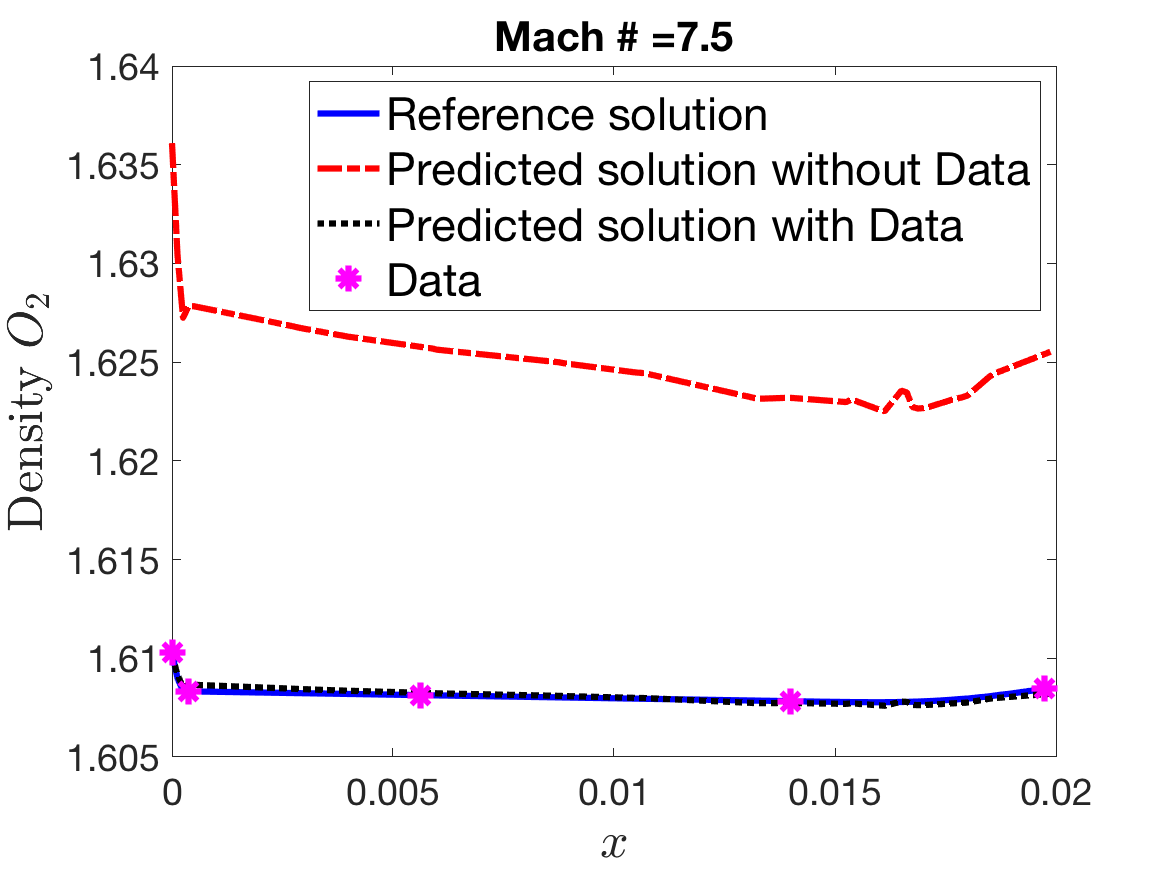}
\includegraphics[scale=0.25,angle=0]{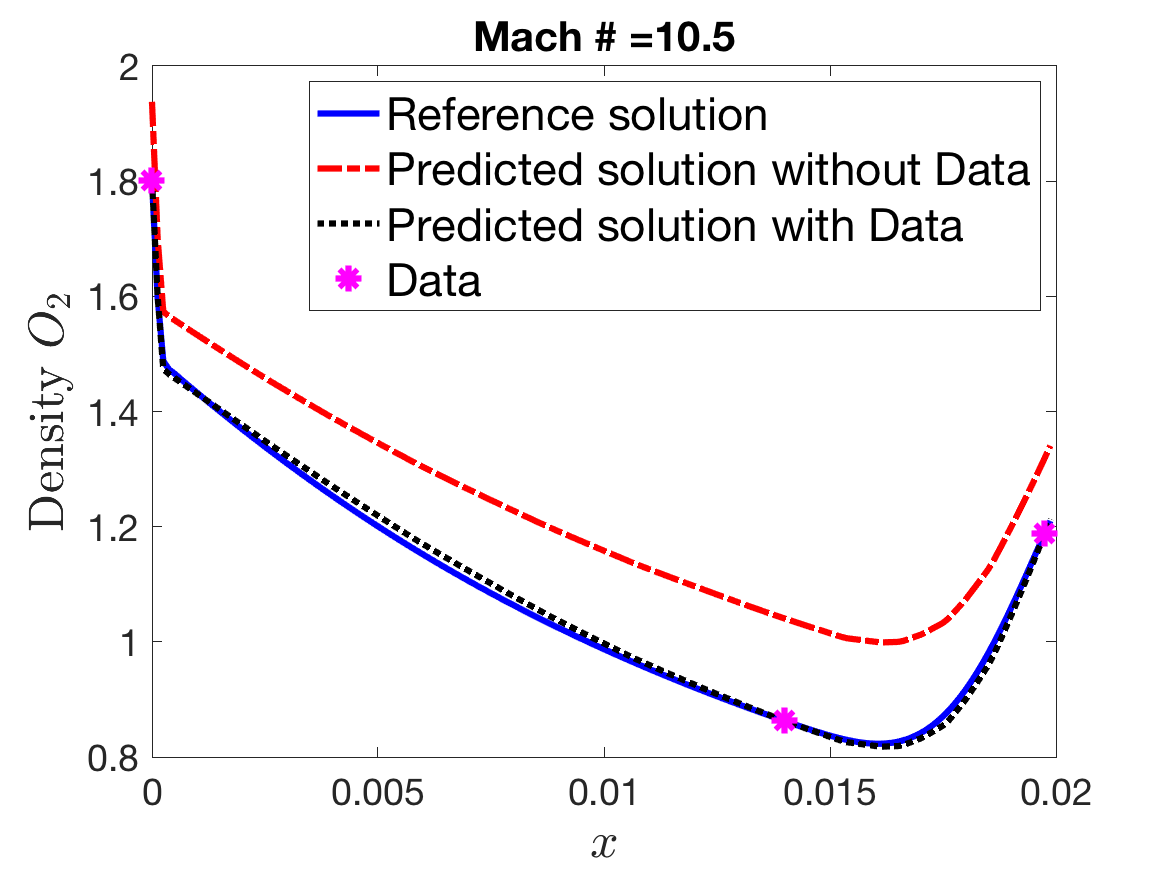}
\includegraphics[scale=0.25,angle=0]{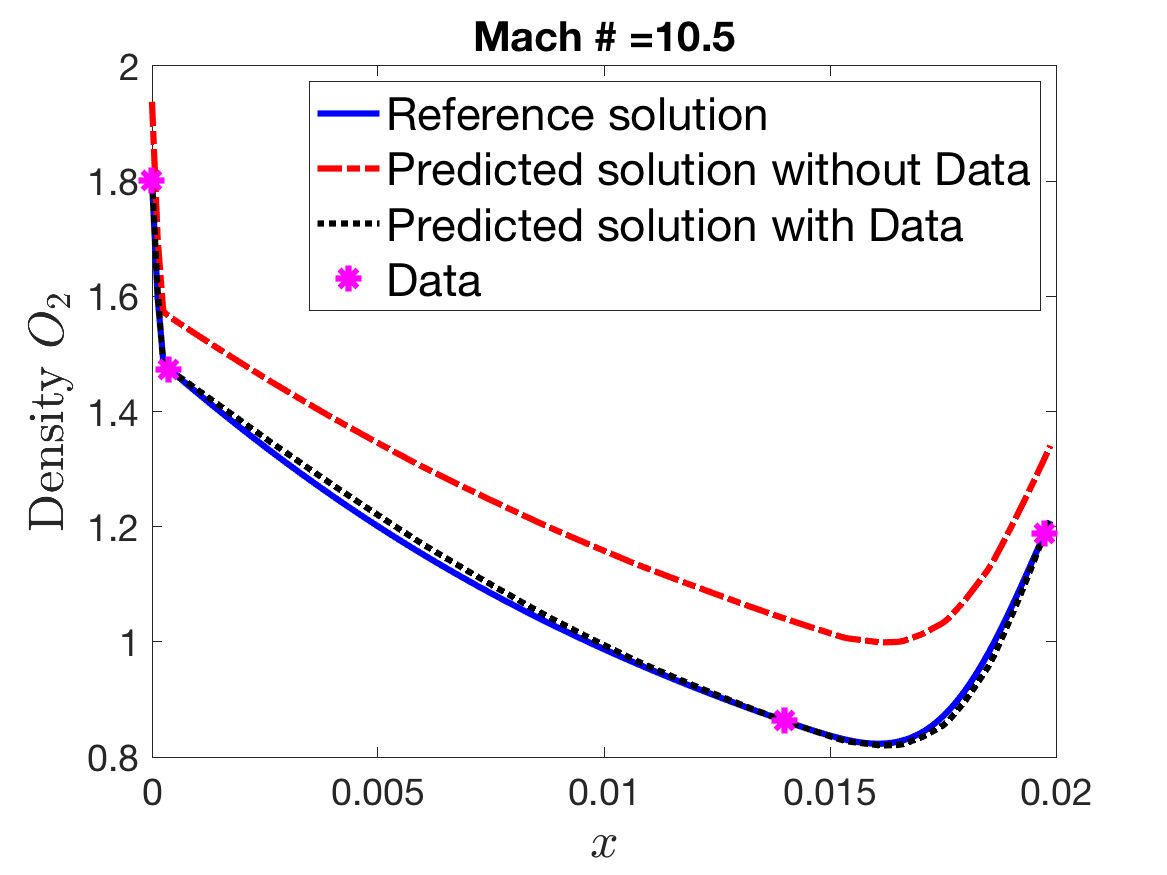}
\includegraphics[scale=0.25,angle=0]{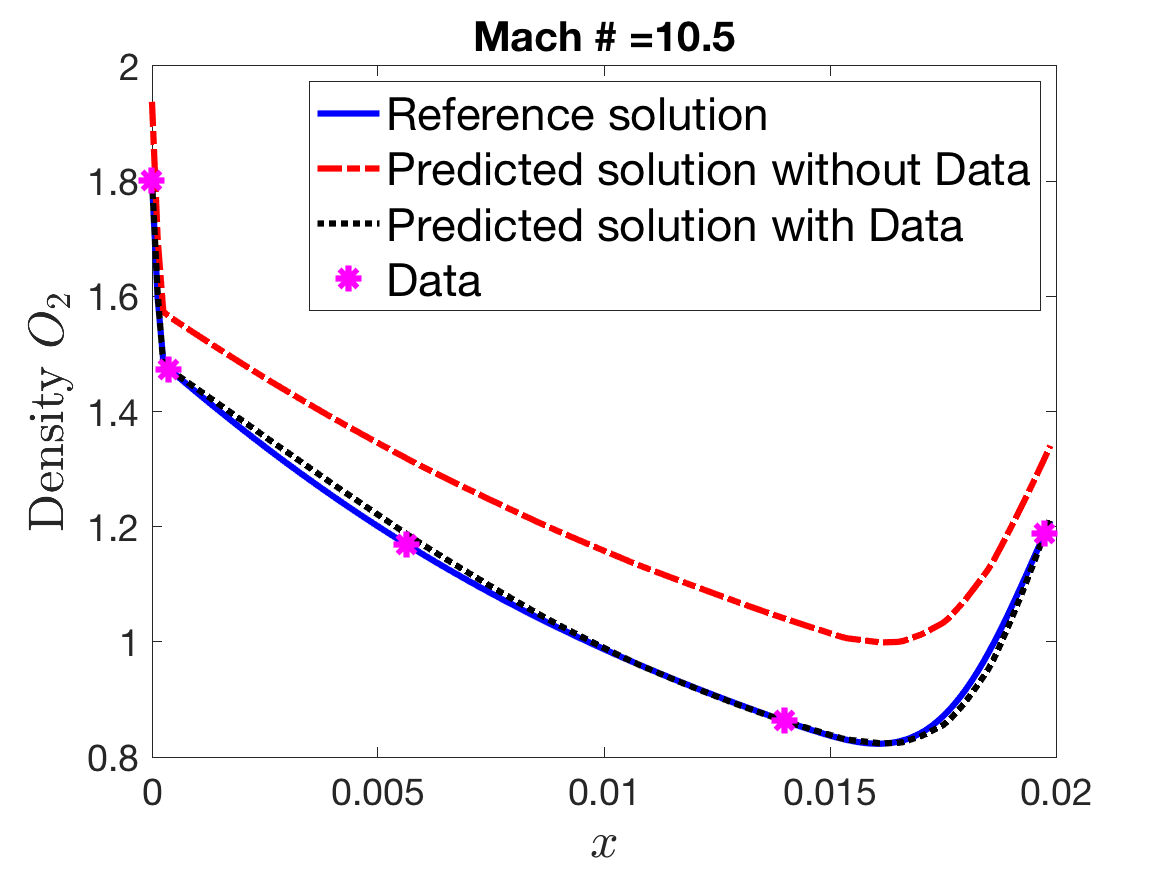}
\end{center}
\caption{Extrapolation of $\rho_{O_2}$ for $Mach = 7.5$ and $Mach = 10.5$ without and with data. For each Mach number, the number of data is increased from 3 to 4 then 5 points.  We obtain good results with a few data for the extrapolation. The red dot-dash lines are the outputs of the DeepONet $G_{\rho_{O_2}}$ without data while the black dash lines are the predictions with data. 
}\label{fig:ext:O2}
\end{figure}

For this add-on network, we use the following parameters:
\begin{itemize}[topsep=0pt,itemsep=0pt,parsep=0pt]
    \item Hidden layers for the neural network ``NN": $6\times 40$.
    \item Activation function: $\tanh$;
    \item Learning rate: $8\times 10^{-4}$;
    \item Epochs: 20000.
\end{itemize}
We note that the training for the above neural network is very fast (usually takes one or two minutes).
The results for the densities of $N$ and $O_2$ with different number of data are shown in Figure \ref{fig:ext:N} and \ref{fig:ext:O2}, respectively. We observe that even with three data points we obtain satisfactory predictions. When we use more data, we obtain very accurate solutions compared to the reference data. We do not show the results of $\rho_{N_2}, ~\rho_O, ~\rho_{NO}$ since they are similar as those for $\rho_{O_2}$ and $\rho_N$.

The results from DeepONnet demonstrate that this architecture can efficiently represent the operators for the behavior of non-equilibrium and the velocity and temperature downstream of the shock.  Within the training range, the network performs exceptionally well in predicting these fields.  For extrapolation, in addition to knowledge of the input function we assume knowledge of only few data points of the output, which may be available from sensor data.  These few sensor data enable accurate predictions with further efficient training.  The above elements are the building blocks for our design of DeepM\&Mets which couple the pre-trained DeepONets, and whose architectures are targeting data assimilation where we do not have knowledge of the full input function but rather just a few data points.

\section{DeepM\&Mnet framework: architectures and results}\label{sec:deepmmnet}
In this section, we propose the DeepM\&Mnet framework for hypersonics multi-physics and multiscale problems, by coupling the pre-trained DeepONets $G_{U,T}$ and  $G_{\rho_{N_2,O_2,N,O,NO}}$ developed in subsection \ref{sec:deeponets:2}.
Unlike the building blocks DeepONets that require input functions and make predictions, we now relax this requirement.  We only assume that we have some sensor data for the inputs.  
For all the tests performed in this section, we randomly select a value of the Mach number in the interval $[8,10]$.

\subsection{Parallel DeepM\&Mnet}
We begin by proposing the parallel DeepM\&Mnet.
Assume that we have some data for all the variables, i.e., $\rho_{k,data}^j, k = N_2, O_2, N,O,NO$ and $U_{data}^j, T_{data}^j, ~j =1,2,\ldots, n_D$, where $n_D$ is the number of data.
We design the parallel DeepM\&Mnet as follows: 
\begin{enumerate}
    \item We construct a neural network ``NN" (to be trained) that takes $x$ as the input, and the velocity and temperature  $U,~ T$ as well as densities of all species, i.e., $\rho_{N_2}$, ~$\rho_{O_2}$, ~$\rho_{N}$, ~$\rho_{O}$, ~$\rho_{NO}$, as the outputs.
    \item We then feed $U,~ T$ as the input of the {\it pre-trained} DeepONets $G_{\rho_{N_2}, \rho_{O_2}, \rho_{N}, \rho_{O}, \rho_{NO}}$ and output the densities of the five species $\rho_{N_2}^*$, ~$\rho_{O_2}^*$, ~$\rho_{N}^*$, ~$\rho_{O}^*$, ~$\rho_{NO}^*$, while we feed $\rho_{N_2}$, $~\rho_{O_2}$, ~$\rho_{N}$, ~$\rho_{O}$, ~$\rho_{NO}$ as the inputs to the {\it pre-trained} DeepONets $G_{U,T}$ and output the velocity $U^*$ and temperature $T^*$.
    \item Then, we define the total loss by combining the mean square errors between $\rho_{N_2}$, ~$\rho_{O_2}$, ~$\rho_{N}$, ~$\rho_{O}$, ~$\rho_{NO}$,~ $U$, ~$T$ and $\rho_{N_2}^*$, ~$\rho_{O_2}^*$, ~$\rho_{N}^*$, ~$\rho_{O}^*$, ~$\rho_{NO}^*$, ~$U^*$, ~$T^*$, and the mean square errors between the data and the outputs of the neural network ``NN". 
\end{enumerate}
\begin{figure}[http]
\begin{center}
\includegraphics[scale=0.65,angle=0]{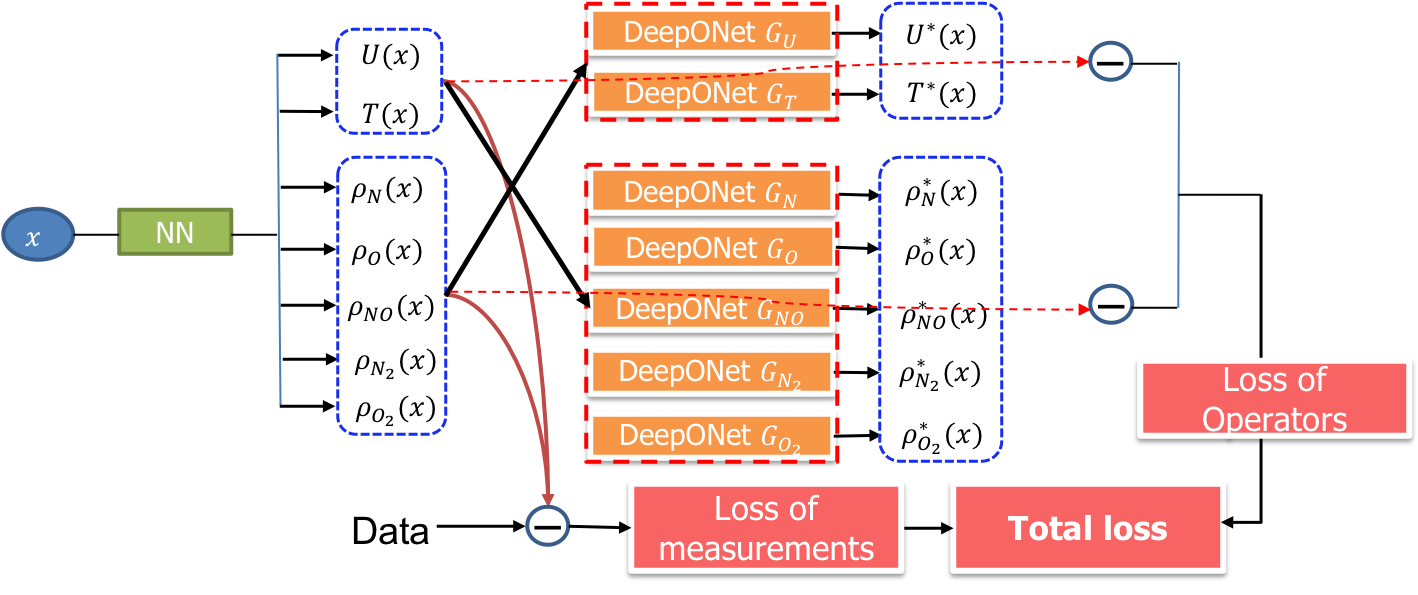}
\end{center}
\caption{Schematic of the parallel DeepM\&Mnet. Here we assume we have some data for all seven state variables.
}\label{fig:DeepMM:Schematic:P}
\end{figure}

We show the schematic of the parallel DeepM\&Mnet in Figure \ref{fig:DeepMM:Schematic:P}.
To stabilize the training process, we add a $L_2$ regularization term in the loss function. 
Moreover, to obtain a more robust training process we also add one more term related to the global mass conservation, i.e., the condition at steady state:
$$\frac{\partial (\rho U)}{\partial x} = 0,$$ 
or equivalently, 
$$\rho U(x) \equiv Const,$$
where $\rho$ is the total density.
Furthermore, we assign each term of the loss function a weight. 
Therefore, the total loss is given as follows:
\begin{equation}\label{eq:loss:DeepMM}
    \mathcal{L} =  \omega_{D}\frac{1}{n_{d}}\mathcal{L}_{data} + \omega_{O}\frac{1}{n_{O}}\mathcal{L}_{op} + \omega_{R}\mathcal{L}_{reg} + \omega_{G}\frac{1}{n_{G}}\mathcal{L}_{G},
\end{equation}
where $\mathcal{L}_{reg} = \|\theta\|_2^2$, $\theta$ is the set of training parameters of the neural network to be trained, and 
\begin{equation*}
    \begin{aligned}
 &\mathcal{L}_{data} = \sum_{j= 1}^{n_{D}} \sum_{k\in \{N_2,O_2,N,O,NO\}}  \|\rho_{k,data}^j - \rho_{k}(x_j)\|^2 + \sum_{j= 1}^{n_{D}} \|U_{data}^j - U(x_j)\|^2 + \sum_{j= 1}^{n_{D}} \|T_{data}^j - T(x_j)\|^2,\\
 &\mathcal{L}_{op} = \sum_{j= 1}^{n_{O}} \sum_{k\in \{N_2,O_2,N,O,NO\}}  \|\rho_{k}^*(x_j) - \rho_{k}(x_j)\|^2 + \sum_{j= 1}^{n_{O}} \|U^*(x_j) - U(x_j)\|^2 + \sum_{j= 1}^{n_{O}} \|T^*(x_j) - T(x_j)\|^2, \\
 &\mathcal{L}_{G} = \sum_{j = 1}^{n_G} \|\rho U (x_j) - Const\|^2,
    \end{aligned}
\end{equation*}
where $\rho = \rho_{N_2} + \rho_{O_2}+ \rho_{N}+ \rho_{O}+ \rho_{NO}$, $n_D$ is the number of data, $n_{O}$ is the number of points for the variables, and $n_{G}$ is the number of points for the global conservation. Here we take the average of $\rho U$ of the CFD data as the data for the  the constant ``Const".

An important consideration is the operator loss, $\mathcal{L}_{op}$, which is a surrogate for the governing equations in the context of PINNs.  Note that using DeepONets and $\mathcal{L}_{op}$ does not preclude including the equations as additional constraints;  in fact we have added mass conservation in the loss.  However, using DeepONets and their loss $\mathcal{L}_{op}$ is an efficient way of incorporating the physics because the DeepONets are pre-trained.

Let $\omega_{D} = 1.0, ~~ \omega_{O} = 1.0$. For each variable, we use 5 data and the following hyperparameters to train the parallel DeepM\&Mnet:
\begin{itemize}[topsep=0pt,itemsep=0pt,parsep=0pt]
    \item Hidden layers for the neural network ``NN": $6\times 50$.
    \item Activation function: $\tanh$;
    \item Learning rate: $5\times 10^{-4}$;
    \item Epochs: 300000.
\end{itemize}

\begin{figure}[http]
    \centering
    \subfigure[]{\label{DeepMM:P:D_N2}
    \includegraphics[width=0.31\textwidth]{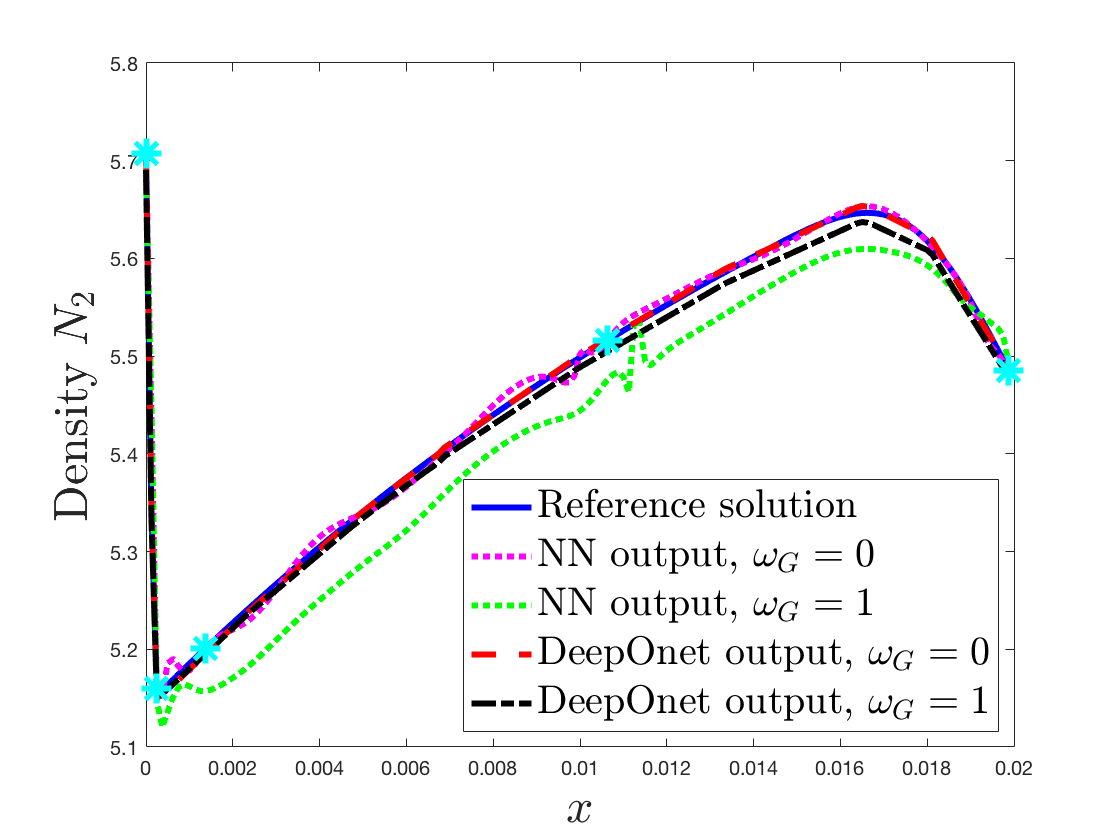}}
    \subfigure[]{\label{DeepMM:P:D_O2}
    \includegraphics[width=0.31\textwidth]{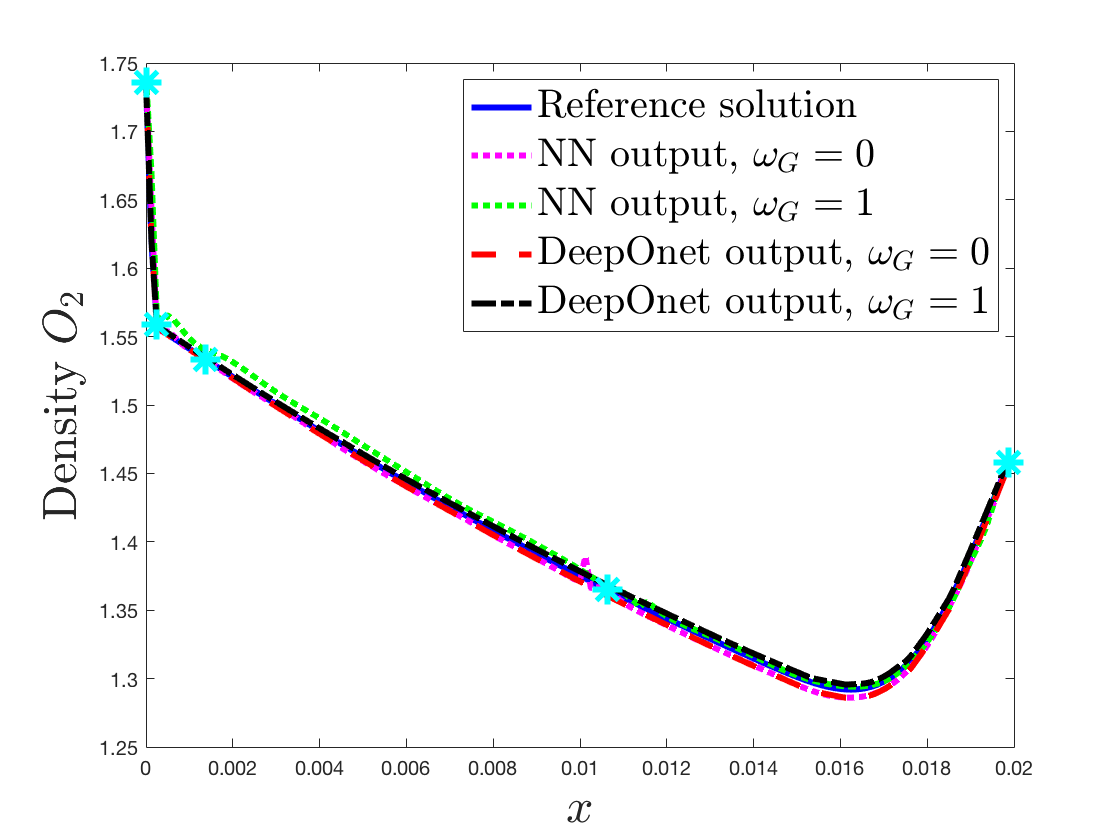}}
    \subfigure[]{\label{DeepMM:P:D_N}
    \includegraphics[width=0.31\textwidth]{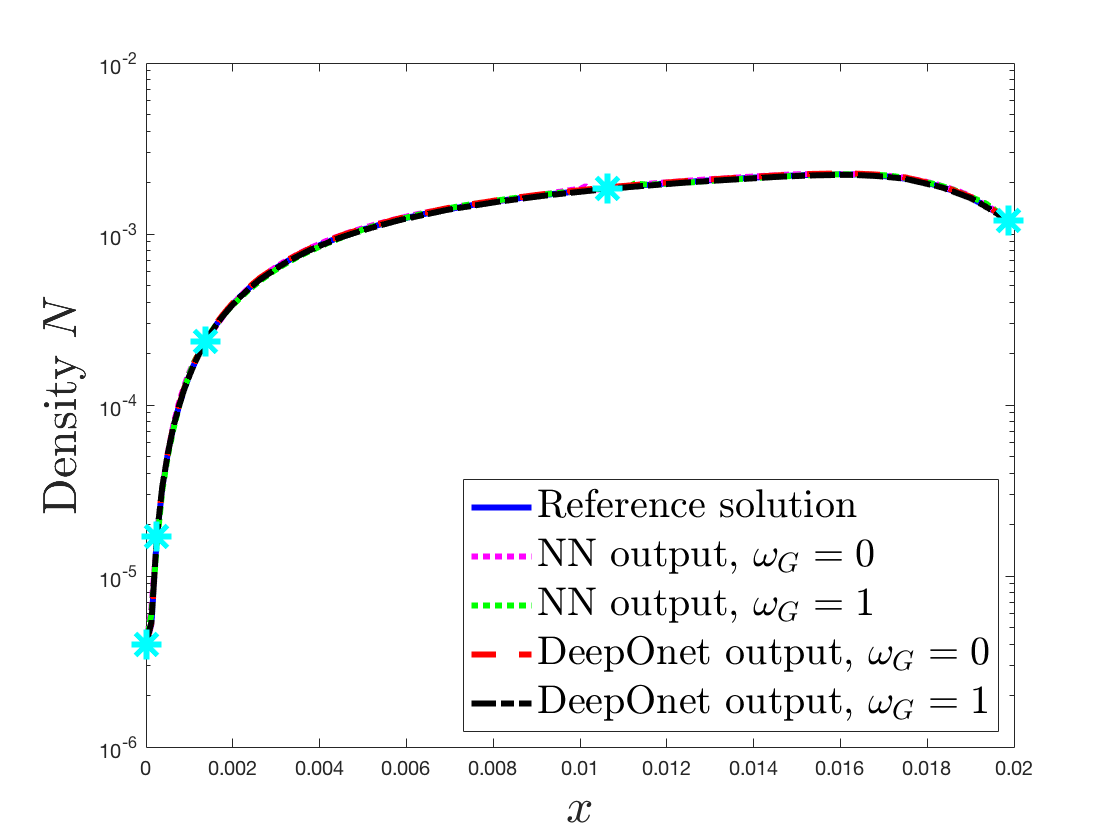}}
    \subfigure[]{\label{DeepMM:P:D_O}
    \includegraphics[width=0.31\textwidth]{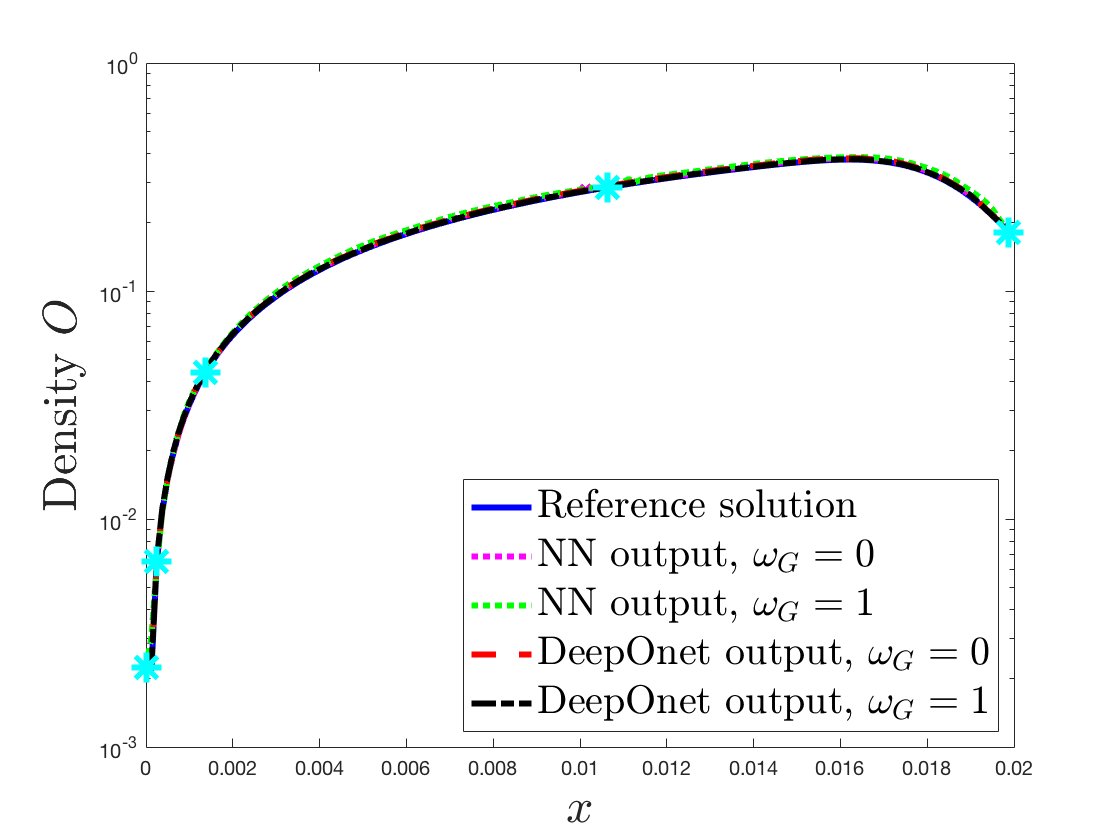}}
    \subfigure[]{\label{DeepMM:P:D_NO}
    \includegraphics[width=0.31\textwidth]{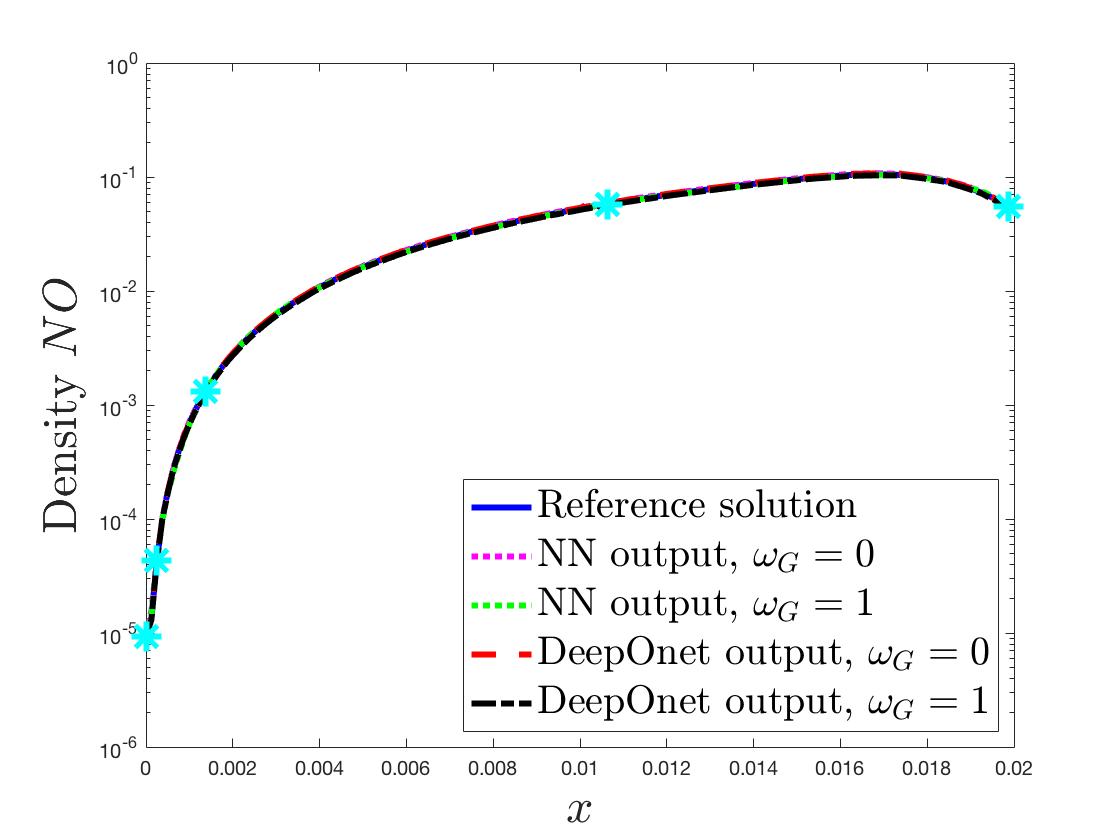}}
    \subfigure[]{\label{DeepMM:P:U}
    \includegraphics[width=0.31\textwidth]{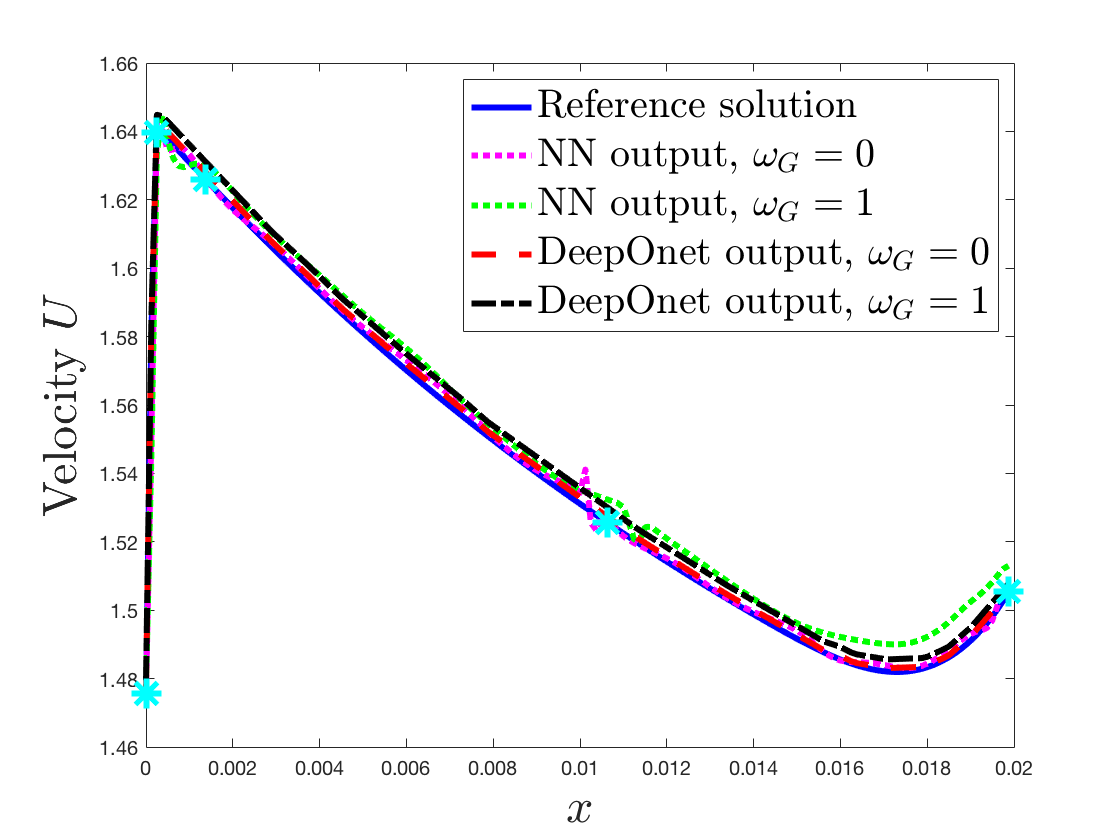}}
    \subfigure[]{\label{DeepMM:P:T}
    \includegraphics[width=0.31\textwidth]{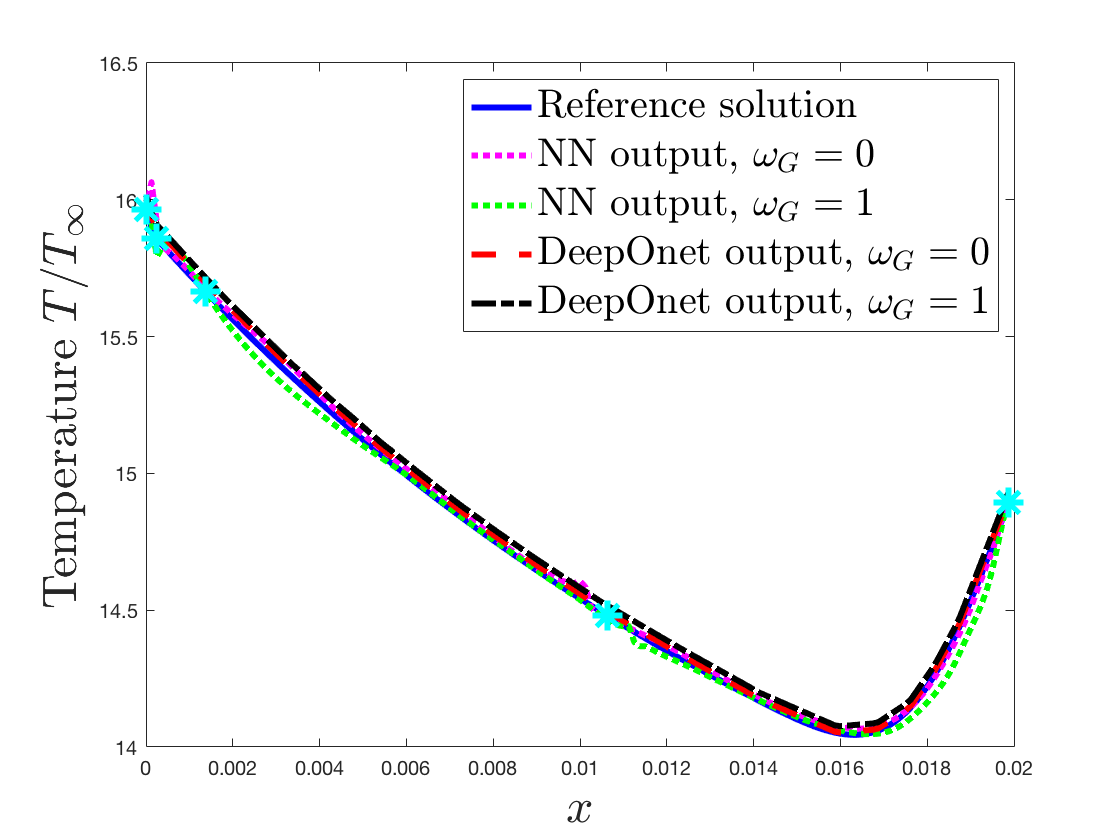}}
    \subfigure[]{\label{DeepMM:P:loss}
    \includegraphics[width=0.31\textwidth]{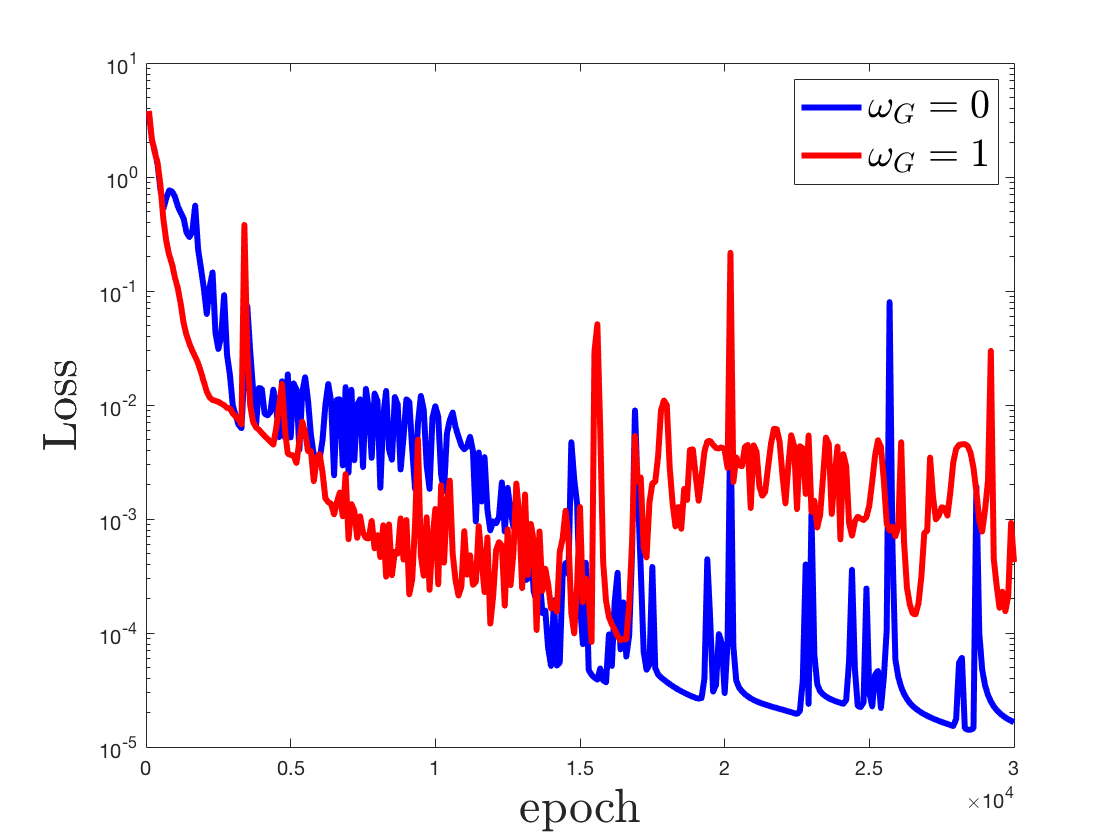}}
    \caption{(a)-(g): Predictions of all the variables (densities of the chemical species, the velocity and the temperature) using the parallel DeepM\&Mnet with ($\omega_G = 1$) or without ($\omega_G = 0$) global conservation. $\omega_{R} = 0$. (h): Loss versus epoch.}
    \label{fig:DeepMM:P:prediction}
\end{figure}

For the first example, we do not use the regularization term ($\omega_{R} = 0$).  In figure \ref{fig:DeepMM:P:prediction}, we report the results with and without the global conservation. Observe that we obtain very accurate predictions. We note that the mild oscillations in teh NN outputs $\rho_{N_2}$, $\rho_{O_2}$, $\rho_{N}$, $\rho_{O}$, $\rho_{NO}$ and $U, ~T$.  In addition, the NN outputs are not very accurate in some cases. However, the DeepONet outputs, $\rho_{N_2}^*$, $\rho_{O_2}^*$, $\rho_{N}^*$, $\rho_{O}^*$, $\rho_{NO}^*$ and $U^*, ~T^*$ are always smooth and accurate, even though we have not used any regularization.  The pre-trained DeepONets that encode the physics therefore deliver an effective regularization in these tests.

\begin{table}[http]
\footnotesize\selectfont
\centering
\begin{tabular}{|c|c|c|c|c|c|c|c|}
\hline
  &  $\rho_{N_2}$  &  $\rho_{O_2}$  & $\rho_{N}$  & $\rho_{O}$  & $\rho_{NO}$ & $U$  & $T$ \\
\hline
    $\omega_{G} = 0, \omega_{R} = 0$  & 1.84e-04 & 7.99e-04 & 3.91e-04 & 5.06e-07 & 7.51e-06 & 2.45e-06 & 4.49e-06 \\
    $\omega_{G} = 1, \omega_{R} = 0$  & 4.33e-04 & 1.36e-05 & 1.24e-04 & 5.72e-07 & 8.27e-07 & 1.46e-06 & 2.04e-06 \\
    $\omega_{G} = 0, \omega_{R} = 10^{-4}$  & 9.13e-03 & 3.25e-02 & 1.71e-03 & 4.21e-05 & 1.69e-04 & 3.57e-05 & 5.34e-05 \\
    $\omega_{G} = 1, \omega_{R} = 10^{-4}$  & 1.14e-03 & 8.04e-03 & 3.08e-04 & 1.06e-05 & 4.99e-05 & 5.80e-06 & 6.83e-06 \\
\hline
\end{tabular}
    \caption{The values of $\|Ref - Pred\|^2$ for the parallel DeepM\&Mnet using 5 data for each variable, where $Ref$  denotes the reference data and $Pred$ denotes the prediction of the DeepONet output in the DeepM\&Mnet architecture. $\omega_{G}$ is the weight for the global conservation term in the loss function. $\omega_R$ is the weight for the regularization term in the loss function.}
    \label{tab:deepMM:P:nui5}
\end{table}

We report the mean square errors between the predictions of the DeepONet output and the reference data with and without  $L_2$ regularization or global conservation in Table \ref{tab:deepMM:P:nui5}. Without $L^2$ regularization, we do not have an appreciably improved accuracy for the results obtained with the global conservation compared with the ones without global conservation. On the other hand, in the case of using the $L^2$ regularization, we obtain significant improvement of the accuracy by enforcing global conservation for all variables.  The results may suggest that regularization is not needed, although it is in fact important as we explain below.

It is important to recognize the performance of the DeepM\&Mnet.  Without complete knowledge of any of the fields, and only starting with a few data for the inputs, we are able to efficiently predict the entire coupled non-equilibrium chemistry and flow.  The predicted solution, reliant on pre-trained DeepONet, respects the operators that govern the physics of the problem.  Compared to conventional data assimilation approaches that attempt to construct the full fields from limited data, our approach is extremely efficient.  But similar to any solution of inverse problem where the physics are nonlinear, convergence is not guaranteed.  For this reason, the regularization term may be needed for the NN in instances where convergence is difficult to achieve. It is also important to ensure that the output of the NN does not suffer from overfitting since it serves as input to all the DeepONets.  An example that demonstrates how regularization can be beneficial is provided in \ref{sec:DeepMMnet:S2}.

\subsection{Series DeepM\&Mnets}\label{sec:DeepMMnet:S}
In practice, sensor data are not generally available for all the field variables.  In the present configuration, it is for example less likely that we will have data for the densities of all the chemical species or perhaps for any of them.  Although it may be seem reasonable to expect access to some sparse data for the velocity and temperature. Therefore, we now consider a different architecture of DeepM\&Mnet, which we term series architecture that only requires the data of the velocity and temperature, i.e., $U_{data}^j, T_{data}^j, j =1,2,\ldots, n_D$. We design the series DeepM\&Mnet as follows:
    \begin{enumerate}
    \item We construct a neural network ``NN" (to be trained) that takes $x$ as the input, and the velocity and temperature  $U,~ T$ as the outputs, which would be fed as the input to the {\it trained} DeepONets $G_{\rho_{N_2}, \rho_{O_2}, \rho_{N}, \rho_{O}, \rho_{NO}}$.
    \item We now feed the outputs of the DeepONets $G_{\rho_{N_2}, \rho_{O_2}, \rho_{N}, \rho_{O}, \rho_{NO}}$, i.e., the densities of the chemical species, i.e., $\rho_{N_2}^*, ~\rho_{O_2}^*, ~\rho_{N}^*, ~\rho_{O}^*, ~\rho_{NO}^*$, as the inputs to the {\it trained} DeepONets $G_{U,T}$ and output the velocity $U^*$ and temperature $T^*$.
    \item Then, we define the total loss as the sum of the mean square error between the outputs of the neural network ``NN", i.e., $U,~ T$ and the outputs of the DeepONets $G_{U,T}$, i.e., $U^*,~ T^*$, and the loss of the measurements, i.e., the mean square error between the data and the outputs of the neural network ``NN", i.e., $U,~ T$. 
    \end{enumerate}

\begin{figure}[http]
\begin{center}
\includegraphics[scale=0.65,angle=0]{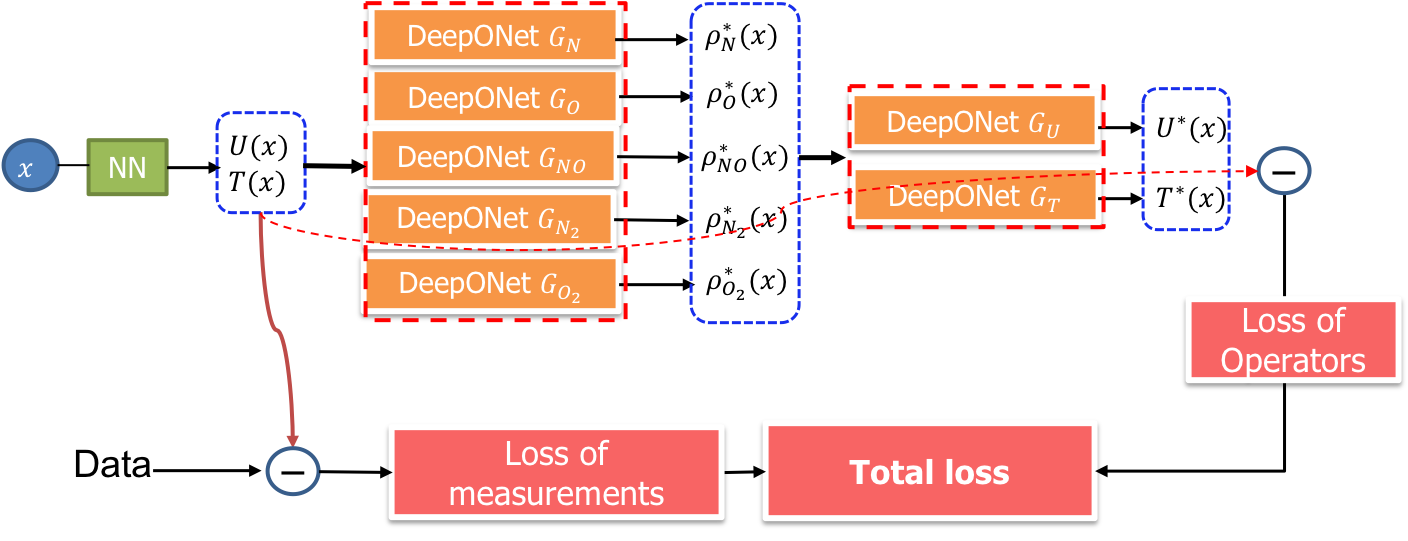}
\end{center}
\caption{Schematic of the series DeepM\&Mnet architecture that employs the data of the velocity and the temperature.
}\label{fig:DeepMM:Series:UT}
\end{figure}

We show the schematic of the series DeepM\&Mnet in Figure \ref{fig:DeepMM:Series:UT}.
The loss function shares the same formula as equation \eqref{eq:loss:DeepMM} with $\mathcal{L}_{data}$, $\mathcal{L}_{op}$, and $\mathcal{L}_{G}$ are given by 
\begin{equation*}
\begin{aligned}
 &\mathcal{L}_{data} = \sum_{j= 1}^{n_{d}} \|U_{data}^j - U(x_j)\|^2 + \sum_{j= 1}^{n_{d}} \|T_{data}^j - T(x_j)\|^2,\\
 &\mathcal{L}_{op} = \sum_{j= 1}^{n_{op}} \|U^*(x_j) - U(x_j)\|^2 + \sum_{j= 1}^{n_{op}} \|T^*(x_j) - T(x_j)\|^2, \\
 &\mathcal{L}_{G} = \sum_{j = 1}^{n_G} \|\rho^* U^* (x_j) - Const.\|^2,
    \end{aligned}
\end{equation*}
where $\rho^* = \rho_{N_2}^* + \rho_{O_2}^*+ \rho_{N}^*+ \rho_{O}^*+ \rho_{NO}^*$.
In addition to the capacity of this configuration to predict all seven fields variables from limited data of just two inputs, the architecture also has the benefit that the inputs to $G_{U}$ and $G_{T}$ are `naturally' regularized because they are the outputs of the upstream DeepONets.  

\begin{figure}[http]
    \centering
    \subfigure[]{\label{DeepMM:P:D_N2}
    \includegraphics[width=0.31\textwidth]{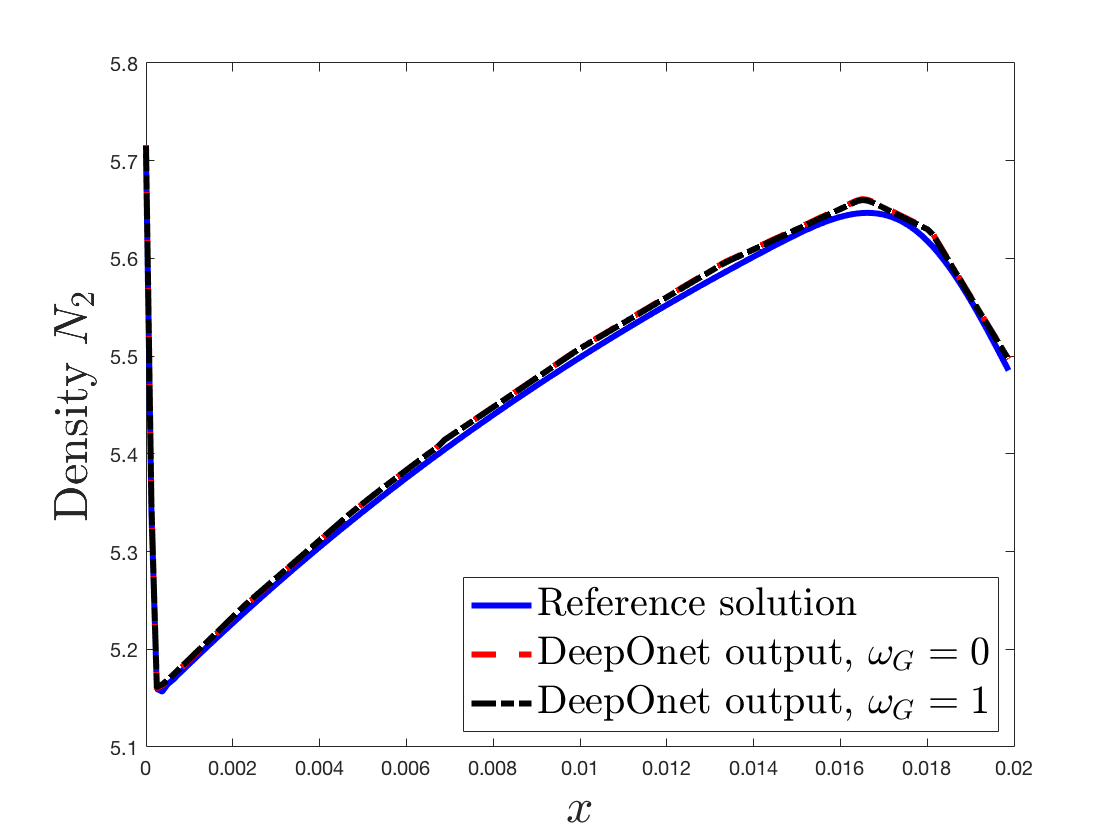}}
    \subfigure[]{\label{DeepMM:P:D_O2}
    \includegraphics[width=0.31\textwidth]{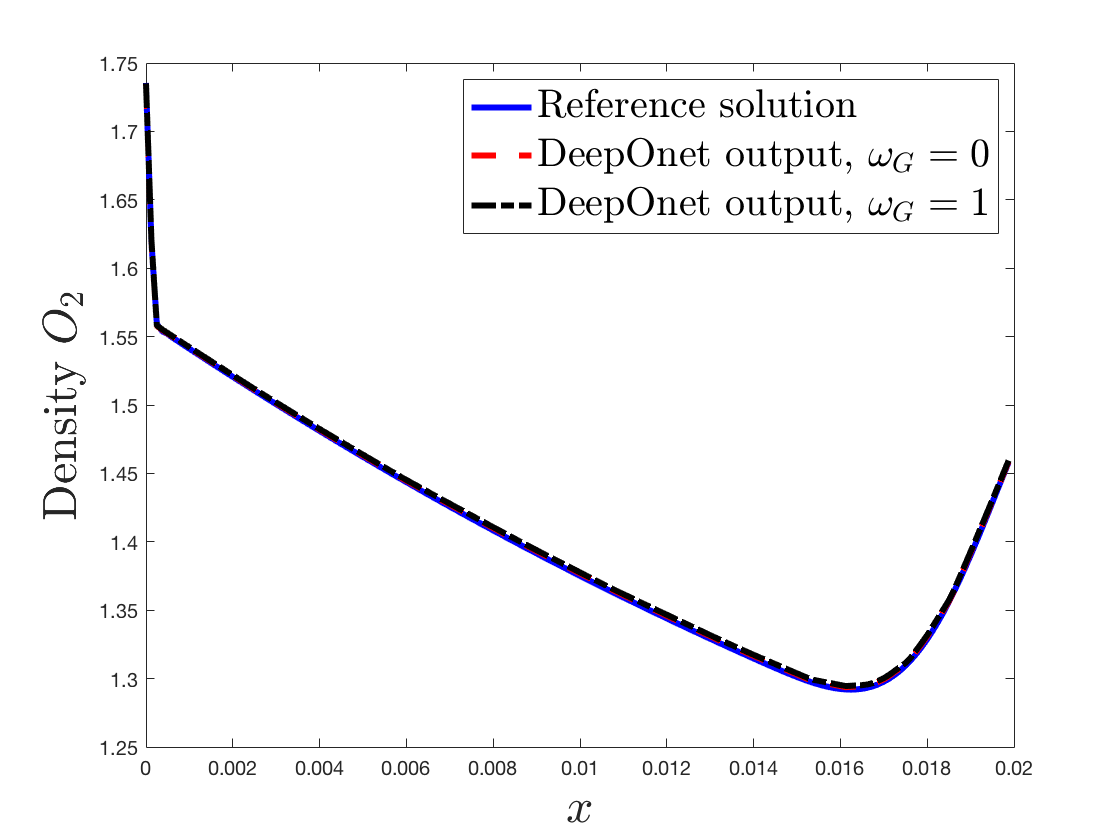}}
    \subfigure[]{\label{DeepMM:P:D_N}
    \includegraphics[width=0.31\textwidth]{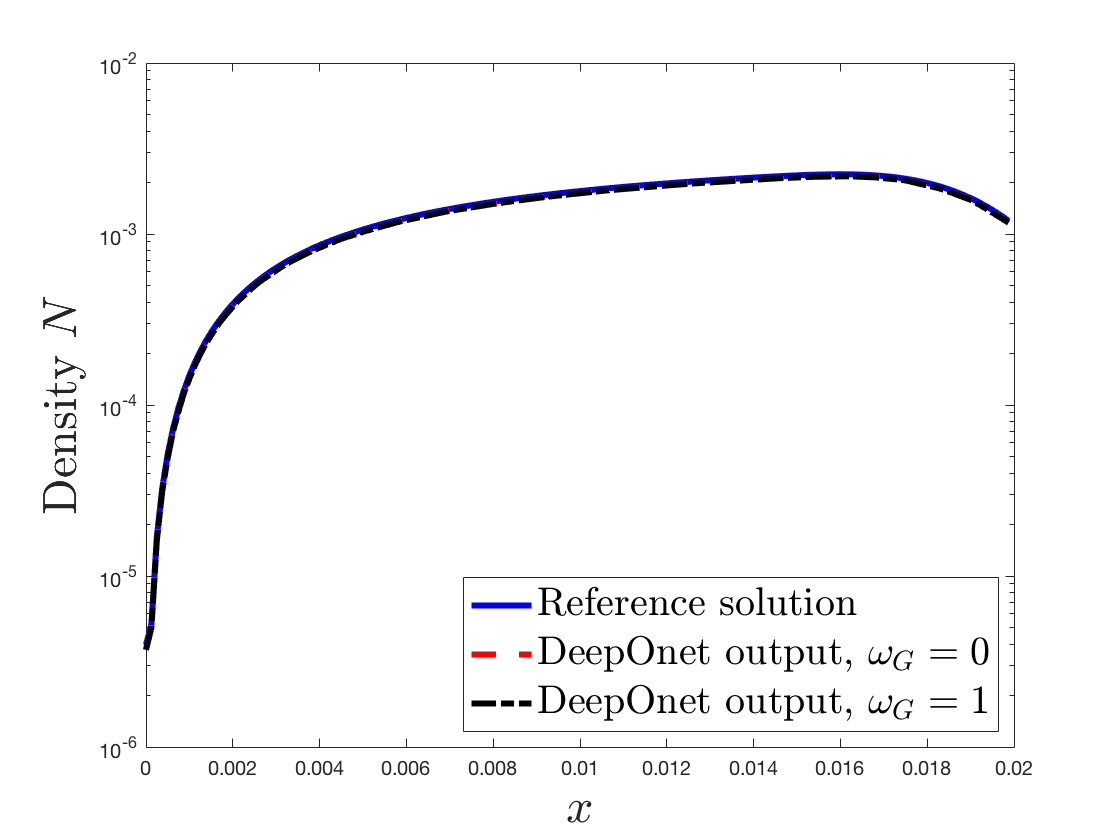}}
    \subfigure[]{\label{DeepMM:P:D_O}
    \includegraphics[width=0.31\textwidth]{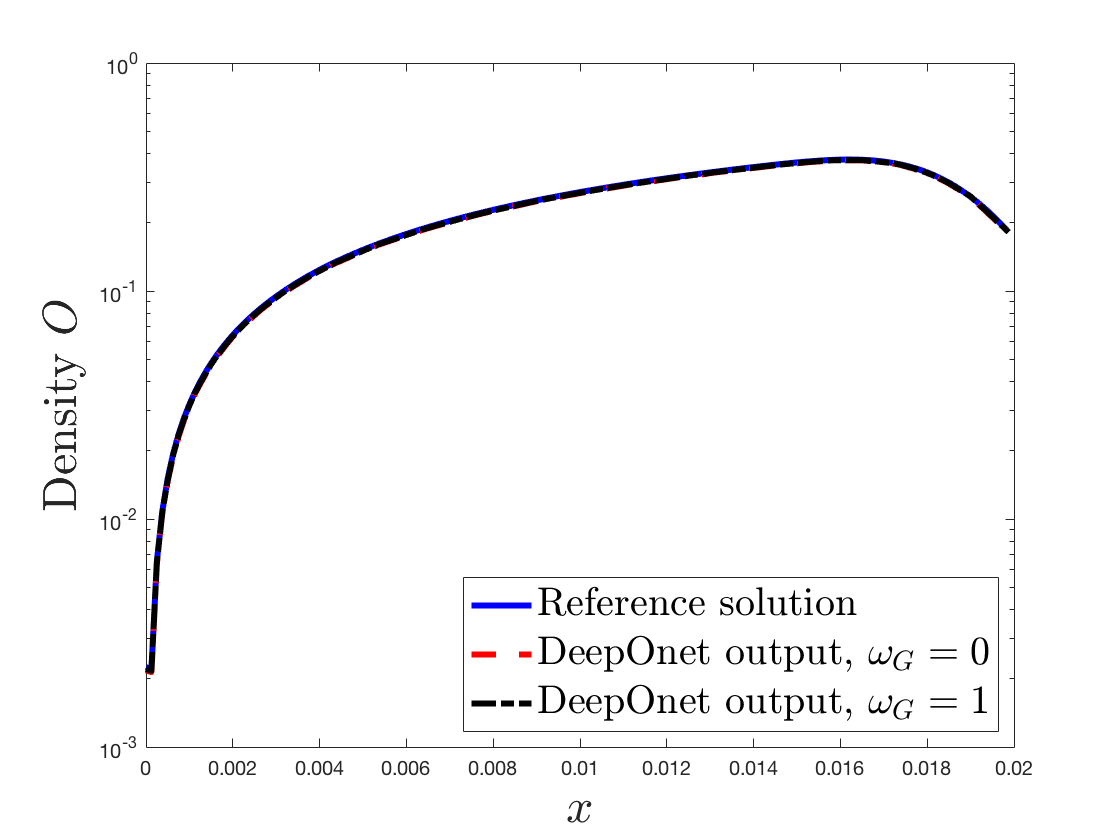}}
    \subfigure[]{\label{DeepMM:P:D_NO}
    \includegraphics[width=0.31\textwidth]{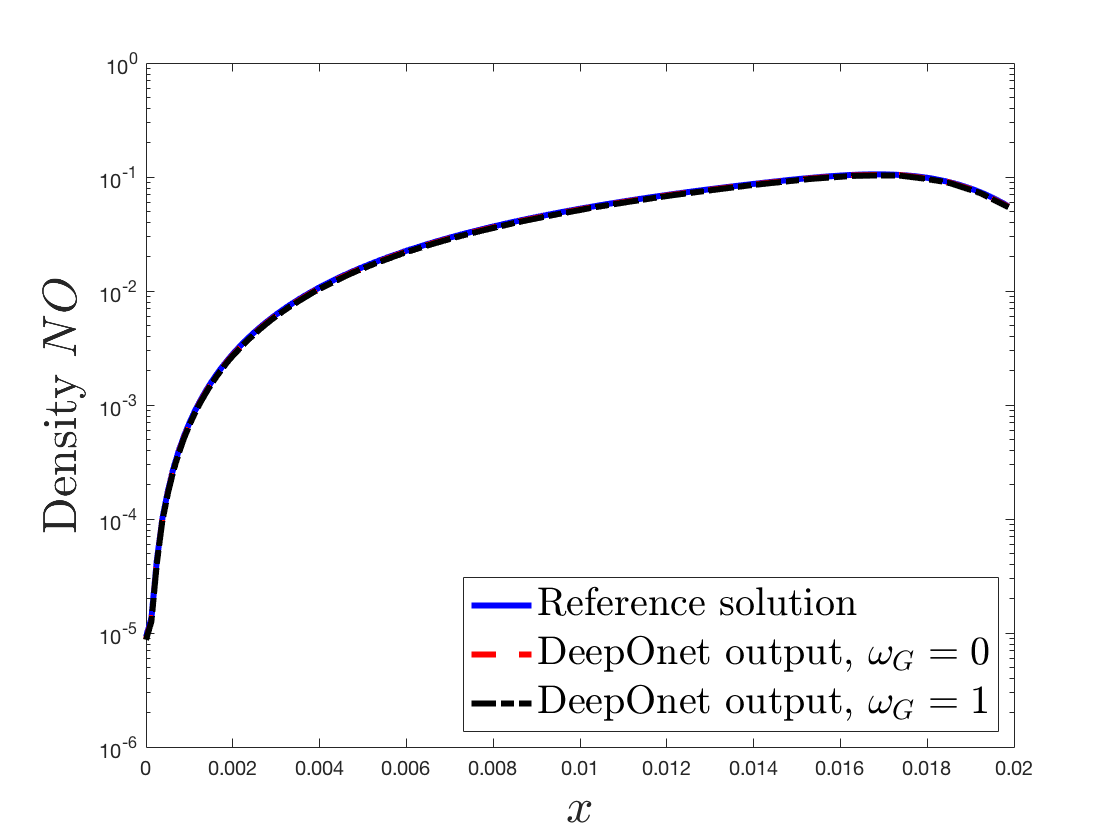}}
    \subfigure[]{\label{DeepMM:P:U}
    \includegraphics[width=0.31\textwidth]{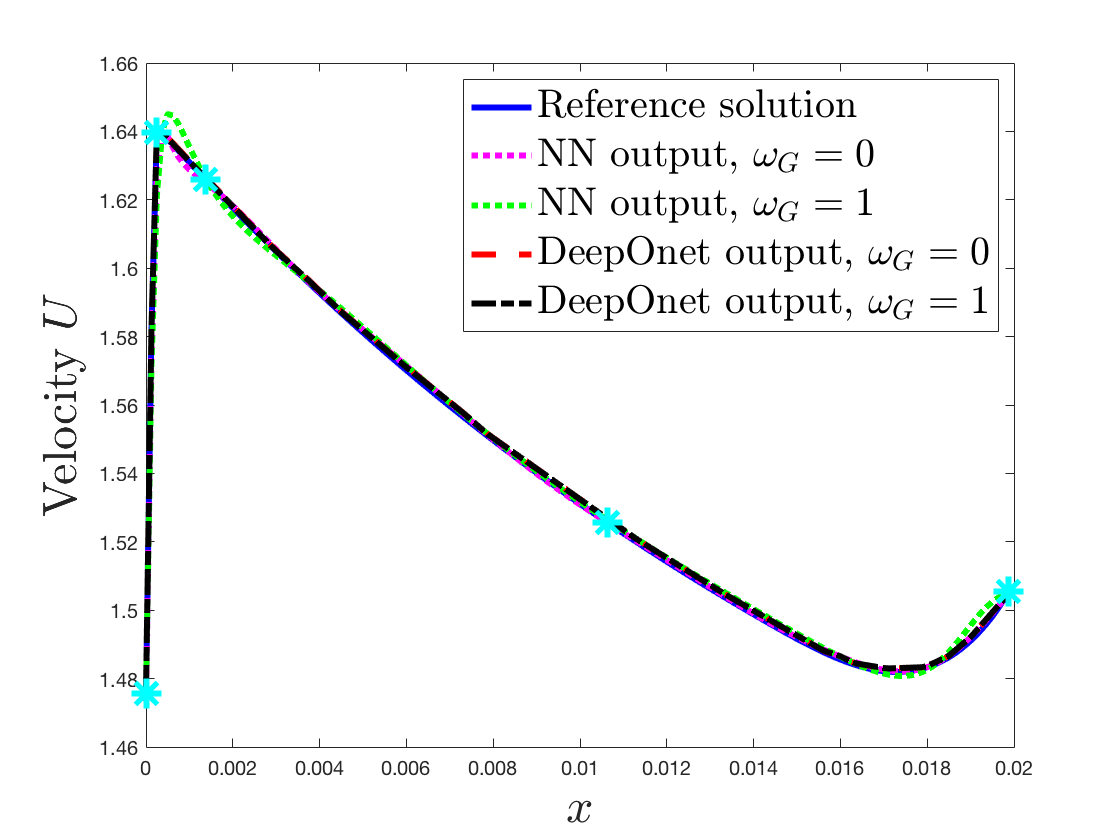}}
    \subfigure[]{\label{DeepMM:P:T}
    \includegraphics[width=0.31\textwidth]{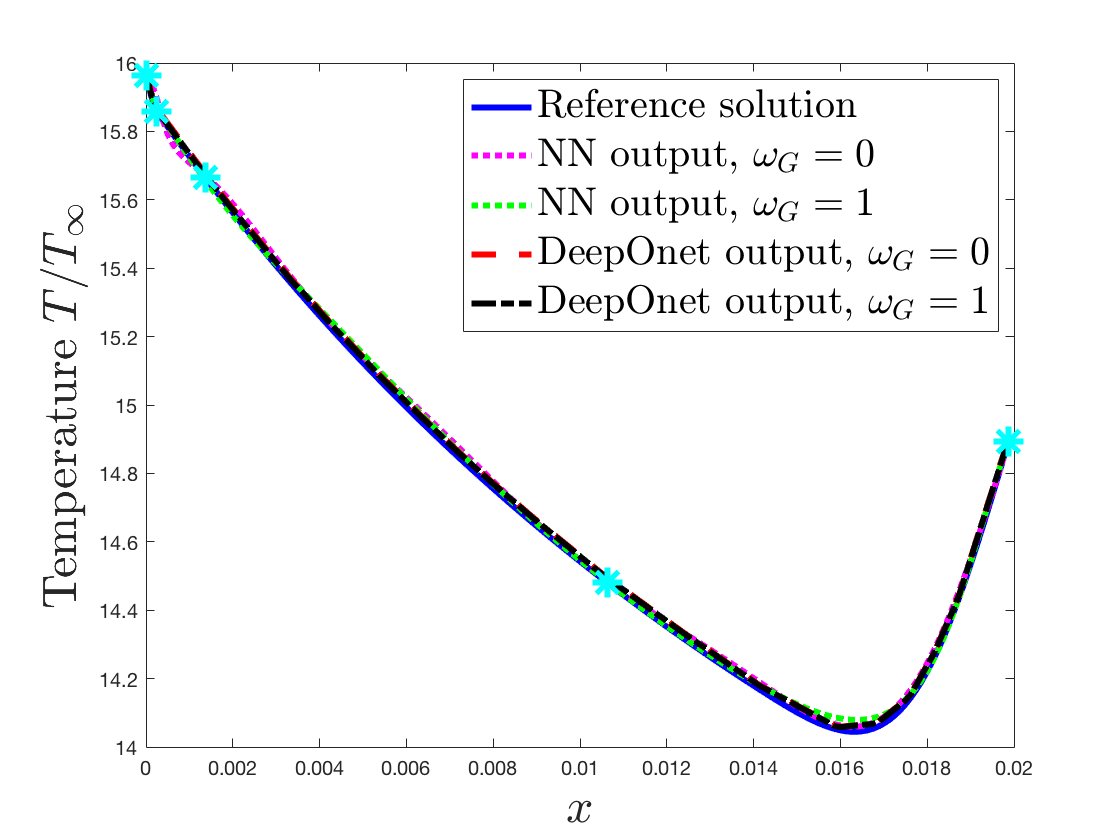}}
    \subfigure[]{\label{DeepMM:P:loss}
    \includegraphics[width=0.31\textwidth]{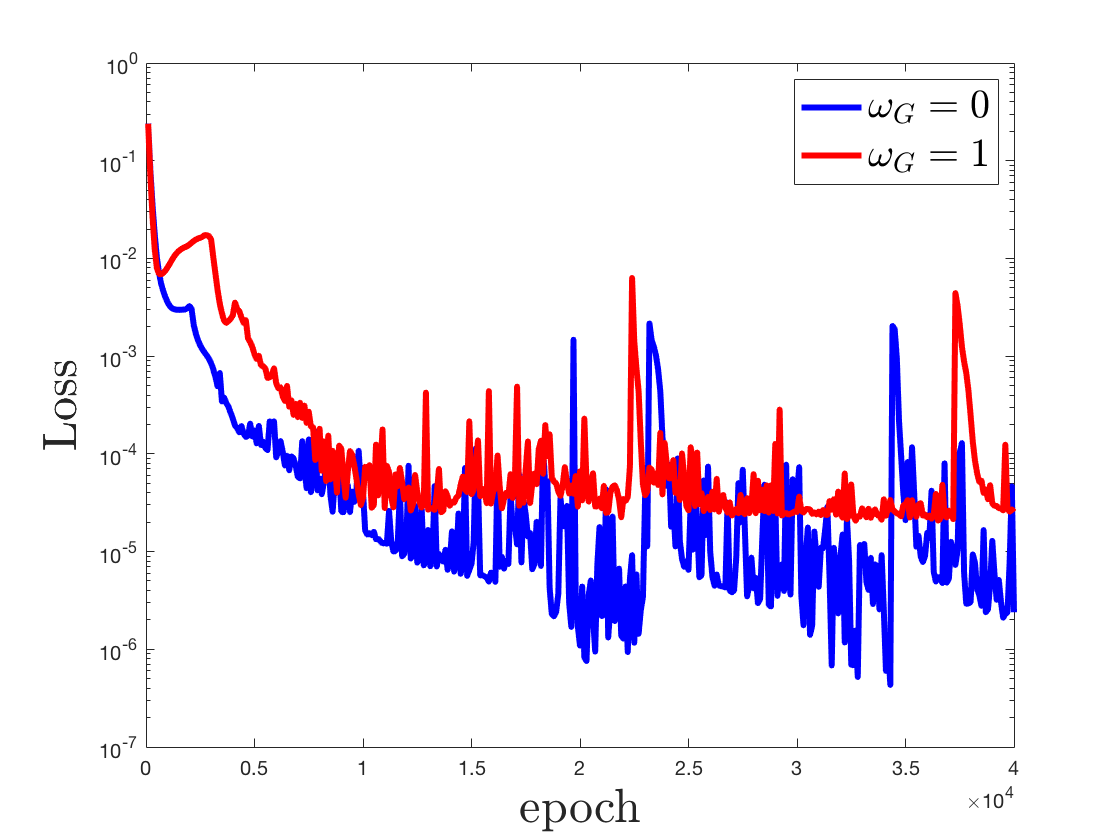}}
    \caption{(a)-(g): Predictions of all the variables (densities of the chemical species, the velocity and the temperature) using the series DeepM\&Mnet with ($\omega_G = 1$) or without ($\omega_G = 0$) global mass conservation constraint. $\omega_{R} = 0$. (h): Loss versus epoch.}
    \label{fig:DeepMM:S1}
\end{figure}

We use 5 data points for both the velocity and temperature, and train the network with the following parameters:
\begin{itemize}[topsep=0pt,itemsep=0pt,parsep=0pt]
    \item Hidden layers for the neural network ``NN": $6\times 50$.
    \item Activation function: $\tanh$;
    \item Learning rate: $6\times 10^{-4}$;
    \item Epochs: 400000.
\end{itemize}
The results for the series DeepM\&Mnet are shown in Figure \ref{fig:DeepMM:S1}.
Again, the predictions are in good agreement with the reference solutions. Also, the DeepONet outputs are always smooth and accurate, consistent with the notion that the pre-trained DeepONets regularize the predictions because the physics modeled by the operators are encoded during their training.
Similar to the parallel DeepM\&Mnet, we also show the mean square errors between the predictions of the DeepONet output and the reference data with and without  $L_2$ regularization or the global mass conservation for the series DeepM\&Mnet in Table \ref{tab:deepMM:S1:nui5}. 
We observe again that in the case of with the $L^2$ regularization, we obtain much more accurate predictions when using in addition the global conservation law.

\begin{table}[http]
\footnotesize\selectfont
    \centering
\begin{tabular}{|c|c|c|c|c|c|c|c|}
\hline
  &  $\rho_{N_2}$  &  $\rho_{O_2}$  & $\rho_{N}$  & $\rho_{O}$  & $\rho_{NO}$ & $U$  & $T$ \\
\hline
   $\omega_{G} = 0, \omega_{R} = 0$  & 6.31e-04 & 8.75e-05 & 1.77e-04 & 2.11e-06 & 1.24e-06 & 3.50e-07 & 1.18e-06 \\
   $\omega_{G} = 1, \omega_{R} = 0$  & 8.02e-04 & 3.92e-05 & 5.33e-04 & 1.98e-06 & 2.45e-06 & 3.29e-07 & 8.85e-07 \\
   $\omega_{G} = 0, \omega_{R} = 10^{-5}$  & 3.42e-03 & 1.32e-03 & 7.16e-04 & 2.99e-05 & 7.89e-06 & 2.42e-07 & 1.46e-06 \\
   $\omega_{G} = 1, \omega_{R} = 10^{-5}$  & 2.85e-04 & 1.14e-04 & 3.18e-05 & 1.25e-05 & 1.21e-06 & 1.63e-06 & 4.78e-06 \\
\hline
\end{tabular}
    \caption{The values of $\|Ref - Pred\|^2$ for the parallel DeepM\&Mnet using 5 data for each variable, where $Ref$ means the reference solutions and $Pred$ means the prediction of the DeepONet output in the DeepM\&Mnet architecture. $\omega_{G}$ is the weight for the global conservation term in the loss function. $\omega_R$ is the weight for the regularization term in the loss function.}
    \label{tab:deepMM:S1:nui5}
\end{table}

Furthermore, we observe that even without the regularization, the training processes for the parallel and series DeepM\&Mnets proposed in this subsection are stable. 
In \ref{sec:DeepMMnet:S2}, we illustrate the importance of regularization for another series DeepM\&Mnet, which does not employ the data of the velocity and temperature but the data of the densities of the chemical species.

\section{Conclusion}\label{sec:conclusion}
The simulation of hypersonic flow is a challenging multi-scale \& multi-physics (M\&M) problem, due to the combination of high Mach, the interaction with a shock leads to excessively high temperatures that can cause dissociation of the gas.  When the reaction rates are commensurate with the rates asscoiated with the flow itself, the dynamics of the dissociation chemistry and the flow are coupled and must be solved together.  These coupled physics lead to changes in the chemical composition of the gas, with densities spanning orders of magnitudes within very small regions, or steep boundary layers, behind the shock.  

We develop a new framework called DeepM\&Mnet to address M\&M problems in general and the hypersonics problem in particular.
We first presented the DeepONets, which serve as the building blocks of the DeepM\&Mnet for the model of the non-equilibrium chemistry that takes place behind a normal shock at high Mach numbers. We simulate the interplay of the flow velocity and temperature as well as the five chemical species whose densities span eight orders of magnitude downstream of the shock. 
Moreover, we test the case when the input is not within the input space, i.e., the Mach number is out of the training range $[8, 10]$. In this case, we cannot obtain accurate predictions, however, we resolve this issue by combining a few data and the pre-trained DeepONets with a simple supervised network.

To employ the pre-trained DeepONets as building blocks to form the DeepM\&Mnet, we develop two architectures, namely a parallel and a series DeepM\&Mnet. The parallel variant requires some data for all the variables while the series DeepM\&Mnet only requires some data for the velocity and the temperature. Moreover, we show that the framework can accommodate enforcing additional constraints such as global mass conservation, which, in addition to accuracy, is also stabilizing.  

We demonstrate in this work that DeepM\&Mnet may provide a new paradigm in computational science and engineering for complex M\&M problems. The codes are compact and easy to adopt in diverse domains as well as maintain or enhance further. The value of DeepM\&Mnet is best demonstrated in the context of assimilating scarce data from sensors in complex physics.  This problem of data assimilation has hindered simulation science for decades because it relies on repeated, or iterative, procedure with solutions of nonlinear coupled equations in the loop and knowledge from previous assimilation tasks are not effectively exploited.  In the context of DeepM\&Mnet, these previous solutions are encoded in the pre-training of DeepONet, and DeepM\&Mnets deliver a seamless integration of models and data all incorporated into a loss function. 
We develop DeepM\&M work specifically to address the data poor regimes. We have removed the tyranny of mesh generation, and with offline training strategies based on single field representations by the DeepONet, the trained DeepM\&Mnet can produce solutions in a fraction of a second. This totally eliminates the need for reduced order modeling that has been rather ineffective for nonlinear complex problems.

\section*{Acknowledgement}
The authors acknowledge support from DARPA/CompMods HR00112090062.

\appendix

\section{DeepONet predictions for the chemical species}\label{sec:apd:DeepOnet:s}

In this appendix, we demonstrate another DeepONet configuration to predict the densities $\rho_N$, $\rho_O$ and $\rho_{NO}$ using as inputs to the branch nets only the density $\rho_{N_2}$ and/or $\rho_{O_2}$.  For each of the target outputs, we build a separate DeepONet: $G_{\rho_{N}}$, $G_{\rho_{O}}$ and $G_{\rho_{NO}}$,
\begin{itemize}
    \item $G_{\rho_{N}}: \rho_{N_2} \to \rho_{N}$ uses $\rho_{N_2}$ as the input of the branch net, and predicts $\rho_{N}$.
    \item $G_{\rho_{O}}: \rho_{O_2} \to \rho_{O}$ uses $\rho_{O_2}$ as the input of the branch net, and predicts $\rho_{O}$.
    \item $G_{\rho_{NO}}: (\rho_{N_2}, \rho_{O_2}) \to \rho_{NO}$ uses both $\rho_{N_2}$ and $\rho_{O_2}$ as the input of the branch net, and predicts $\rho_{NO}$.
\end{itemize}
Note that the physics requires knowledge of the velocity and temperature, as well as the other species.  However, their influence is all encoded in the solution of the operator that each DeepONet is trained to predict.   

\begin{figure}[h]
\begin{center}
\includegraphics[scale=0.5,angle=0]{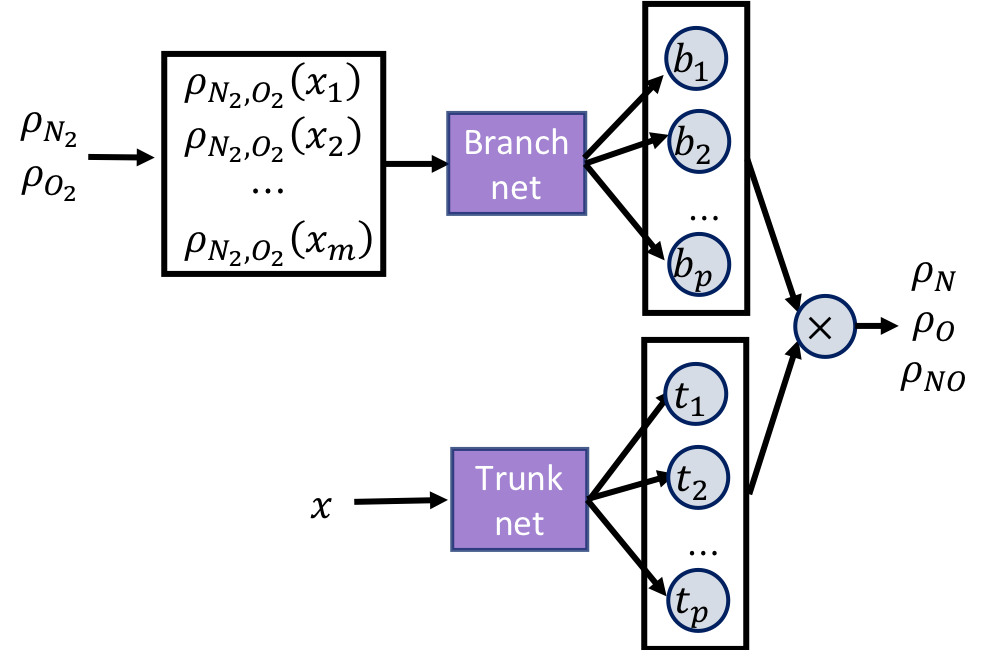}
\end{center}
\caption{Schematic of the DeepONets for the chemical species. Here we have three DeepONets: the first one uses $\rho_{N_2}$ to predict $\rho_N$, i.e., $G_{\rho_N}:~ \rho_{N_2}\rightarrow \rho_{N}$, the second one uses $\rho_{O_2}$ to predict $\rho_O$, i.e., $G_{\rho_O}:~ \rho_{O_2}\rightarrow \rho_{O}$, and the third uses $\rho_{N_2}$ and $\rho_{O_2}$ to predict $\rho_{NO}$, i.e., $G_{\rho_{NO}}:~ (\rho_{N_2}, \rho_{O_2}) \rightarrow \rho_{NO}$.
}\label{fig:DeepOnet:N2O2}
\end{figure}

For simplicity, we show the three DeepONets with one schematic in Figure \ref{fig:DeepOnet:N2O2}.
Note that we train the three DeepONets independently.
We also point out here that the data used for all densities are the logarithms of the original data.

The parameters used for the training are as follows: 
\begin{itemize}[topsep=0pt,itemsep=0pt,parsep=0pt]
    \item Hidden layers for both branch and trunk nets: $4\times 100$.
    \item Activation function: adaptive ReLU;
    \item Learning rate: $8\times 10^{-4}$;
    \item Epochs: 70000.
\end{itemize}

\begin{figure}[h]
    \centering
    \subfigure[]{\label{N2O2:D_N}
    \includegraphics[width=0.31\textwidth]{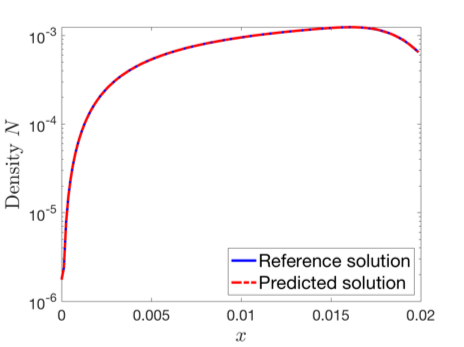}}
    \subfigure[]{\label{N2O2:D_O}
    \includegraphics[width=0.31\textwidth]{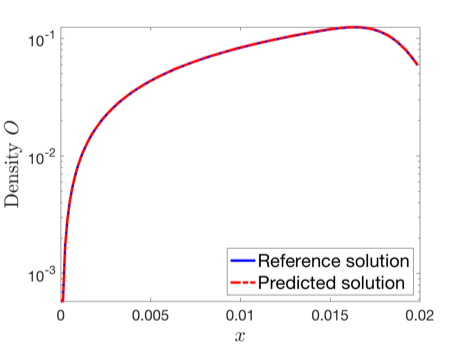}}
    \subfigure[]{\label{N2O2:D_NO}
    \includegraphics[width=0.31\textwidth]{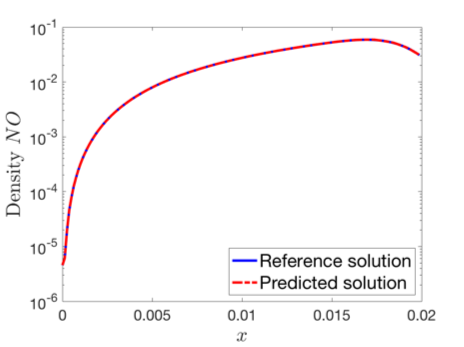}}
    \subfigure[]{\label{N2O2:loss_N}
    \includegraphics[width=0.31\textwidth]{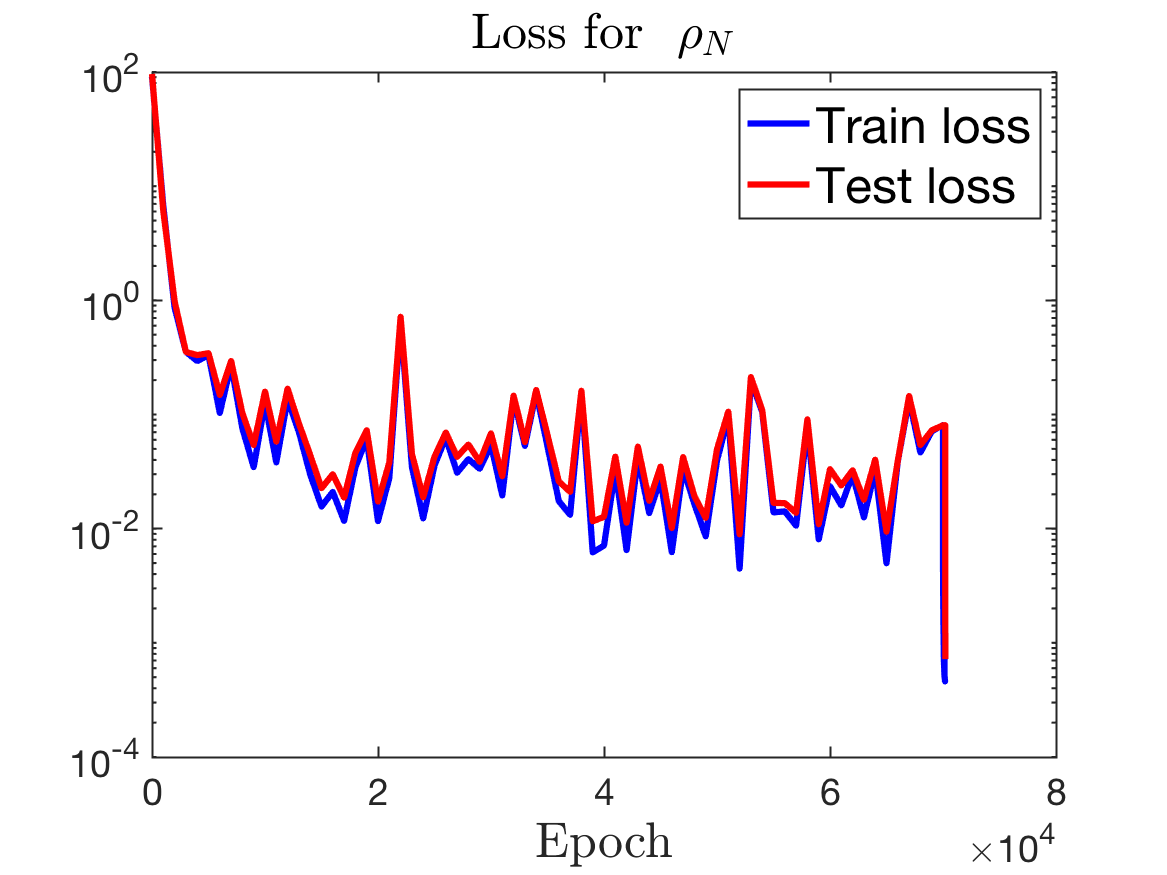}}
    \subfigure[]{\label{N2O2:loss_O}
    \includegraphics[width=0.31\textwidth]{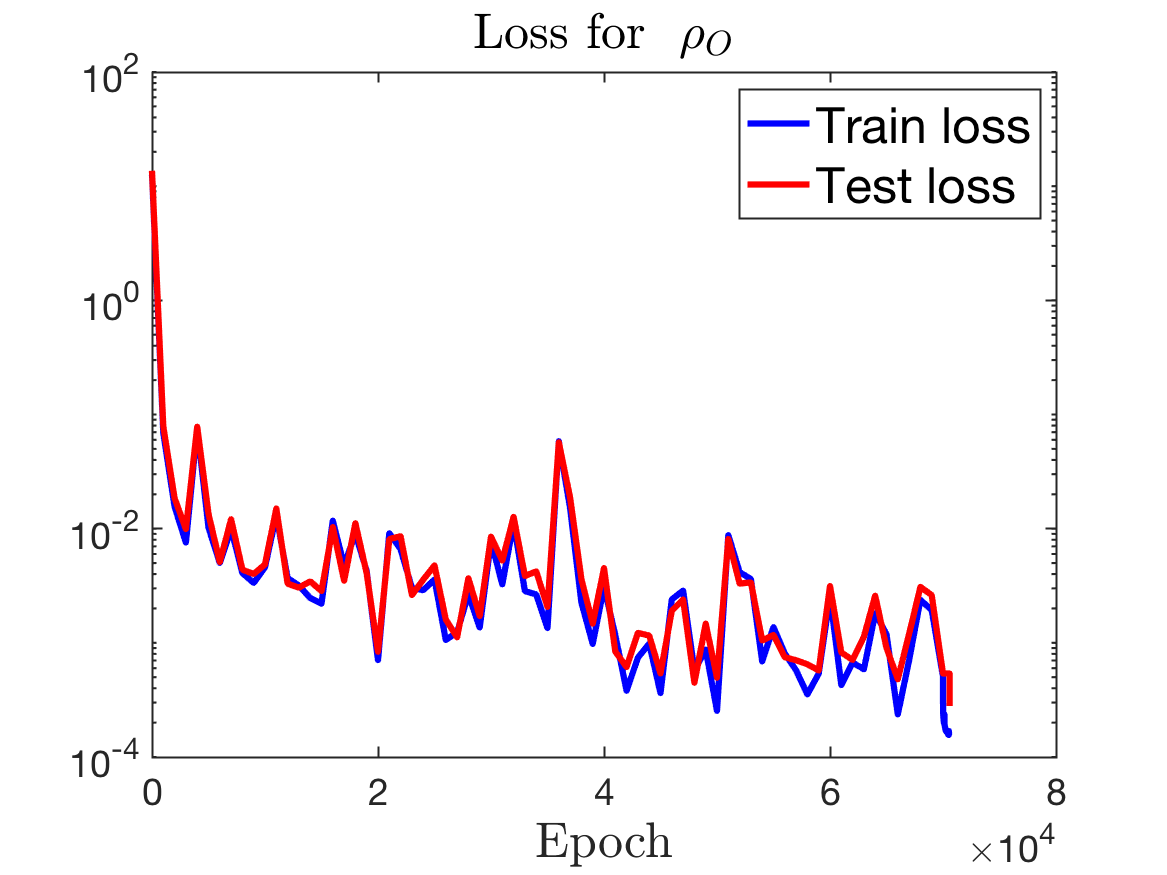}}
    \subfigure[]{\label{N2O2:loss_NO}
    \includegraphics[width=0.31\textwidth]{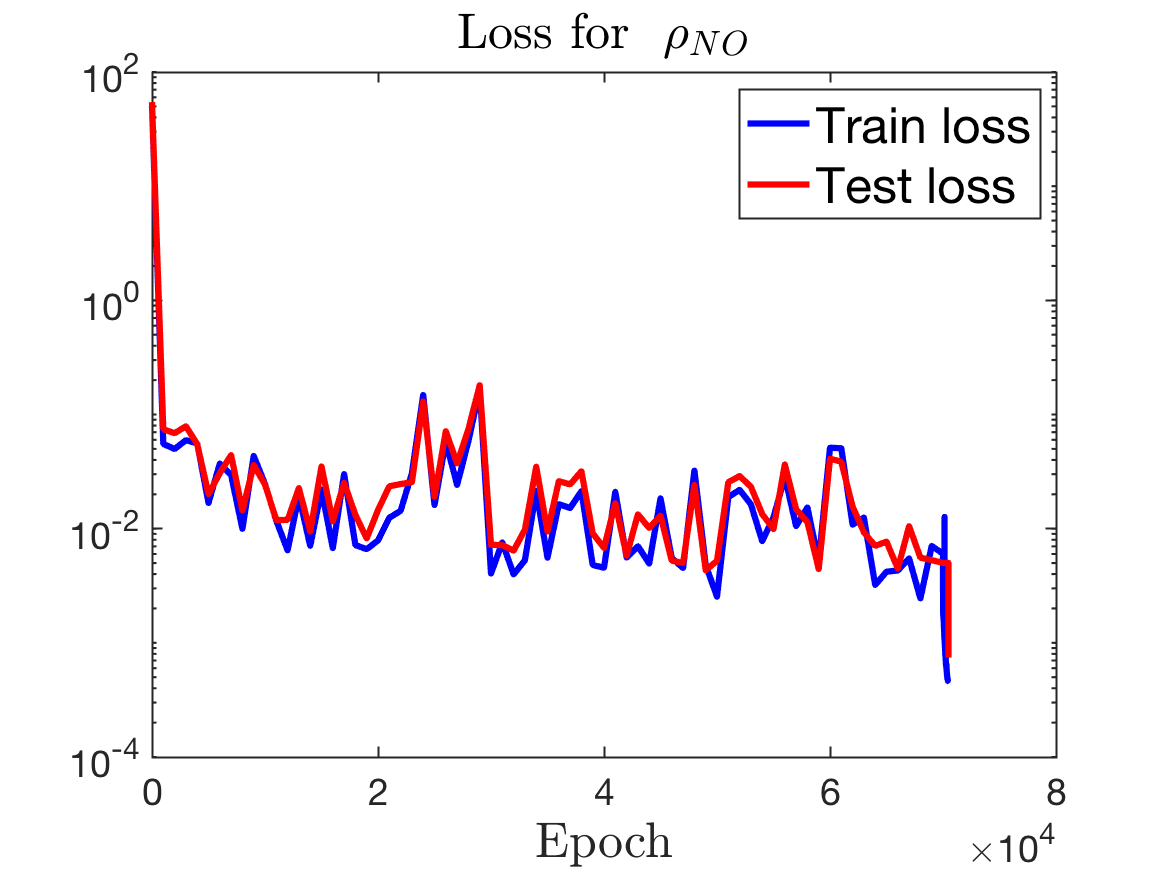}}
    \caption{(a)-(c): Predictions of $\rho_N,~~ \rho_O$, and $\rho_{NO}$ using the corresponding trained DeepONets. (d)-(f): The training losses and test losses for the three DeepONets, i.e., $G_{\rho_N}, ~~ G_{\rho_O}$, and $G_{\rho_{NO}}$.}
    \label{fig:N2O2:prediction}
\end{figure}

To demonstrate the capacity of DeepONets, we use trained DeepONets to predict new conditions that are not in the training data, and some prediction examples are shown in Figure \ref{fig:N2O2:prediction}. We show the predictions for the densities of the three species, i.e., $\rho_{N}$, $\rho_{O}$, and $\rho_{NO}$, in Figure  \ref{N2O2:D_N}-\ref{N2O2:D_NO} while we show the corresponding losses of training and testing in Figure  \ref{N2O2:loss_N}-\ref{N2O2:loss_NO}.  Our DeepONet predictions have good agreement with the reference solutions obtained from CFD. However, DeepONets have much less computational cost than CFD simulations, because the DeepONet prediction is only a forward-pass evaluation of neural networks. Quantitatively, the relative mean-squared error is about $10^{-5}$, while the speedup of DeepONet vs. CFD is about 100,000 X.

\section{Another type of series DeepM\&Mnet architecture}\label{sec:DeepMMnet:S2}
In this Appendix we consider another type of series DeepM\&Mnet architecture.
Similar to the DeepM\&Mnet considered in subsection \ref{sec:DeepMMnet:S}, we construct a DeepM\&Mnet by interchanging the roles of the densities and the flow.  This example will highlight the benefits of regularization in ensuring the stability of the data assimilation problem.  

\begin{figure}[http]
\begin{center}
\includegraphics[scale=0.65,angle=0]{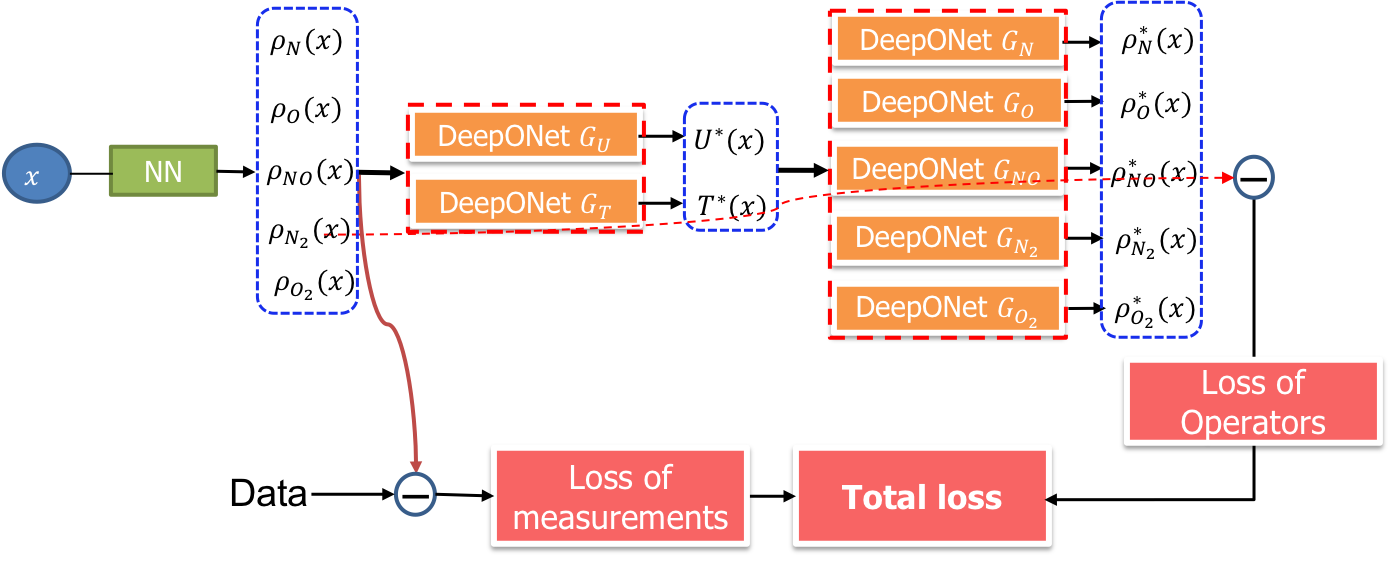}
\end{center}
\caption{Schematic of the series DeepM\&Mnet that employs some data of the densities of the chemical species.
}\label{fig:DeepMM:Series:D}
\end{figure}

We assume we have some data for the densities of the five chemical species, i.e., $[\rho_{k,data}^j], k = N_2, O_2, N,O,NO, ~j =1,2,\ldots, n_d$, where $n_d$ is the number of data. 
The series DeepM\&Mnet in this case is given as follows:
    \begin{enumerate}
        \item Similarly, we construct a neural network ``NN" (to be trained) by taking $x$ as the input and the densities of the five species $\rho_{N_2}, ~\rho_{O_2}, ~\rho_{N}, ~\rho_{O}, ~\rho_{NO}$ as the outputs, which would be fed as the input of the {\it trained} DeepONets $G_{U,T}$.
        \item We now feed the outputs of the DeepONets $G_{U,T}$, i.e., velocity $U^*(x)$ and temperature $T^*(x)$, as the inputs of the {\it trained} DeepONets $G_{\rho_{N_2}, \rho_{O_2}, \rho_{N}, \rho_{O}, \rho_{NO}}$ and output the densities of the five species $\rho_{N_2}^*, ~\rho_{O_2}^*, ~\rho_{N}^*, ~\rho_{O}^*, ~\rho_{NO}^*$. 
        \item Then, we define the total loss the sum of the mean square error between the outputs of the neural network ``NN", i.e., $\rho_{N_2}, ~\rho_{O_2}, ~\rho_{N}, ~\rho_{O}, ~\rho_{NO}$ and the outputs of the DeepONets $G_{\rho_{N_2}, \rho_{O_2}, \rho_{N}, \rho_{O}, \rho_{NO}}$, i.e., $\rho_{N_2}^*, ~\rho_{O_2}^*, ~\rho_{N}^*, ~\rho_{O}^*, ~\rho_{NO}^*$, and the loss of the measurements, i.e., the mean square error between the data and the the outputs of the neural network ``NN", i.e., $\rho_{N_2}, ~\rho_{O_2}, ~\rho_{N}, ~\rho_{O}, ~\rho_{NO}$. 
        \end{enumerate}
We show the schematic of this type of DeepM\&Mnet in Figure \ref{fig:DeepMM:Series:D}.
The loss function also has the same formula as the equation \eqref{eq:loss:DeepMM} with $\mathcal{L}_{data}$ and $\mathcal{L}_{op}$ are given by 
    \begin{equation}\label{eq:OpData:loss}
    \begin{aligned}
 &\mathcal{L}_{data} = \sum_{j= 1}^{n_{d}} \sum_{k\in \{N_2,O_2,N,O,NO\}}  \|\rho_{k,data}^j - \rho_{k}(x_j)\|^2,\\
 &\mathcal{L}_{op} = \sum_{j= 1}^{n_{op}} \sum_{k\in \{N_2,O_2,N,O,NO\}}  \|\rho_{k}^*(x_j) - \rho_{k}(x_j)\|^2, \\
 &\mathcal{L}_{G} = \sum_{j = 1}^{n_G} \|\rho U^* (x_j) - Const.\|^2,
    \end{aligned}
\end{equation}
where $\rho = \rho_{N_2} + \rho_{O_2}+ \rho_{N}+ \rho_{O}+ \rho_{NO}$.
We use 6 data points for the densities of all species and train the network with the same parameters used in subsection \ref{sec:DeepMMnet:S}.

The results are given in Table \ref{tab:deepMM:S2:nui6}, where we report the mean square errors between the predictions of the DeepONet output and the reference data with/without  $L_2$ regularization/global conservation. We observe that without regularization, even with the global conservation constraint, the training process is unstable. We obtain a stable training process with a strong regularization ($\omega_{R} = 10^{-4}$) or with a weak regularization ($\omega_{R} = 10^{-5}$) if additionally we use the global conservation constraint. Moreover, we observe from the table that the results obtained with a weak regularization ($\omega_{R} = 10^{-5}$) and global conservation have the best accuracy.

\begin{table}[http]
\footnotesize\selectfont
    \centering
\begin{tabular}{|c|c|c|c|c|c|c|c|}
\hline
  &  $\rho_{N_2}$  &  $\rho_{O_2}$  & $\rho_{N}$  & $\rho_{O}$  & $\rho_{NO}$ & $U$  & $T$ \\
\hline
   $\omega_{G} = 0, \omega_{R} = 0$  & unstable & unstable & unstable & unstable & unstable & unstable & unstable \\
   $\omega_{G} = 1, \omega_{R} = 0$  & unstable & unstable & unstable & unstable & unstable & unstable & unstable \\
   $\omega_{G} = 0, \omega_{R} = 10^{-5}$  & unstable & unstable & unstable & unstable & unstable & unstable & unstable \\
   $\omega_{G} = 1, \omega_{R} = 10^{-5}$  & {\bf 2.73e-04} & {\bf 4.56e-04} & {\bf 5.96e-04} & {\bf 4.63e-06} & {\bf 1.18e-05} & {\bf 7.25e-06} & {\bf 1.88e-05} \\
   $\omega_{G} = 0, \omega_{R} = 10^{-4}$  & 4.18e-04 & 2.09e-03 & 1.97e-03 & 3.51e-05 & 5.06e-05 & 4.37e-05 & 1.09e-04 \\
   $\omega_{G} = 1, \omega_{R} = 10^{-4}$  & 1.50e-03 & 1.75e-03 & 2.59e-03 & 1.15e-05 & 5.29e-05 & 2.11e-05 & 5.22e-05 \\
\hline
\end{tabular}
    \caption{The values of $\|Ref - Pred\|^2$ for the parallel DeepM\&Mnet using 6 data for each variable, where $Ref$ means the reference solutions and $Pred$ means the prediction of the DeepONet output in the DeepM\&Mnet architecture. $\omega_{G}$ is the weight for the global conservation term in the loss function. $\omega_R$ is the weight for the regularization term in the loss function. The result in {\bf bold} is the best compared to the other results in the table.}
    \label{tab:deepMM:S2:nui6}
\end{table}



\bibliographystyle{abbrvnat}
\bibliography{ref}

\begin{thebibliography}{37}
\providecommand{\natexlab}[1]{#1}
\providecommand{\url}[1]{\texttt{#1}}
\expandafter\ifx\csname urlstyle\endcsname\relax
  \providecommand{\doi}[1]{doi: #1}\else
  \providecommand{\doi}{doi: \begingroup \urlstyle{rm}\Url}\fi

\bibitem[Buchta and Zaki(2020)]{david_zaki_2019}
D.~A. Buchta and T.~A. Zaki.
\newblock Observation-infused simulations of high-speed boundary layer
  transition.
\newblock \emph{Journal of Fluid Mechanics}, page In review, 2020.

\bibitem[Chen and Chen(1993)]{chen1993approximations}
T.~Chen and H.~Chen.
\newblock Approximations of continuous functionals by neural networks with
  application to dynamic systems.
\newblock \emph{IEEE Transactions on Neural Networks}, 4\penalty0 (6):\penalty0
  910--918, 1993.

\bibitem[Chen and Chen(1995)]{chen1995universal}
T.~Chen and H.~Chen.
\newblock Universal approximation to nonlinear operators by neural networks
  with arbitrary activation functions and its application to dynamical systems.
\newblock \emph{IEEE Transactions on Neural Networks}, 6\penalty0 (4):\penalty0
  911--917, 1995.

\bibitem[Coussement et~al.(2012)Coussement, Gicquel, Caudal, Fiorina, and
  Degrez]{coussement2012three}
A.~Coussement, O.~Gicquel, J.~Caudal, B.~Fiorina, and G.~Degrez.
\newblock Three-dimensional boundary conditions for numerical simulations of
  reactive compressible flows with complex thermochemistry.
\newblock \emph{Journal of computational physics}, 231\penalty0 (17):\penalty0
  5571--5611, 2012.

\bibitem[Cybenko(1992)]{cybenko1992approximation}
G.~Cybenko.
\newblock Approximation by superpositions of a sigmoidal function.
\newblock \emph{Mathematics of Control, Signals and Systems}, 5\penalty0
  (4):\penalty0 455--455, 1992.

\bibitem[Day and Bell(2000)]{day2000numerical}
M.~S. Day and J.~B. Bell.
\newblock Numerical simulation of laminar reacting flows with complex
  chemistry.
\newblock \emph{Combustion Theory and Modelling}, 4\penalty0 (4):\penalty0
  535--556, 2000.

\bibitem[del {\'A}guila~Ferrandis et~al.(2019)del {\'A}guila~Ferrandis,
  Triantafyllou, Chryssostomidis, and Karniadakis]{del2019learning}
J.~del {\'A}guila~Ferrandis, M.~Triantafyllou, C.~Chryssostomidis, and
  G.~Karniadakis.
\newblock Learning functionals via {LSTM} neural networks for predicting vessel
  dynamics in extreme sea states.
\newblock \emph{arXiv preprint arXiv:1912.13382}, 2019.

\bibitem[Duan and Mart{\'\i}n(2009)]{duan2009procedure}
L.~Duan and M.~P. Mart{\'\i}n.
\newblock Procedure to validate direct numerical simulations of wall-bounded
  turbulence including finite-rate reactions.
\newblock \emph{AIAA journal}, 47\penalty0 (1):\penalty0 244--251, 2009.

\bibitem[Hilbert et~al.(2004)Hilbert, Tap, El-Rabii, and
  Th{\'e}venin]{hilbert2004impact}
R.~Hilbert, F.~Tap, H.~El-Rabii, and D.~Th{\'e}venin.
\newblock Impact of detailed chemistry and transport models on turbulent
  combustion simulations.
\newblock \emph{Progress in Energy and Combustion Science}, 30\penalty0
  (1):\penalty0 61--117, 2004.

\bibitem[Jahanbakhshi and Zaki(2019)]{jahanbakhshi_zaki_2019}
R.~Jahanbakhshi and T.~A. Zaki.
\newblock Nonlinearly most dangerous disturbance for high-speed boundary-layer
  transition.
\newblock \emph{Journal of Fluid Mechanics}, 876:\penalty0 87–121, 2019.
\newblock \doi{10.1017/jfm.2019.527}.

\bibitem[Karniadakis and Sherwin(2013)]{karniadakis2013spectral}
G.~Karniadakis and S.~Sherwin.
\newblock \emph{Spectral/hp element methods for computational fluid dynamics}.
\newblock Oxford University Press, Third edition, 2013.

\bibitem[Kitamura et~al.(2005)Kitamura, Men'shov, and
  Nakamura]{kitamura2005shock}
K.~Kitamura, I.~Men'shov, and Y.~Nakamura.
\newblock Shock/shock and shock/boundary-layer interactions in two-body
  configurations.
\newblock In \emph{35th AIAA fluid dynamics conference and exhibit}, page 4893,
  2005.

\bibitem[Liu et~al.(1994)Liu, Osher, Chan, et~al.]{liu1994weighted}
X.-D. Liu, S.~Osher, T.~Chan, et~al.
\newblock Weighted essentially non-oscillatory schemes.
\newblock \emph{Journal of Computational Physics}, 115\penalty0 (1):\penalty0
  200--212, 1994.

\bibitem[Lu et~al.(2019{\natexlab{a}})Lu, Jin, and Karniadakis]{lu2019deeponet}
L.~Lu, P.~Jin, and G.~E. Karniadakis.
\newblock {DeepONet}: Learning nonlinear operators for identifying differential
  equations based on the universal approximation theorem of operators.
\newblock \emph{arXiv preprint arXiv:1910.03193}, 2019{\natexlab{a}}.

\bibitem[Lu et~al.(2019{\natexlab{b}})Lu, Meng, Mao, and
  Karniadakis]{lu2019deepxde}
L.~Lu, X.~Meng, Z.~Mao, and G.~E. Karniadakis.
\newblock {DeepXDE}: A deep learning library for solving differential
  equations.
\newblock \emph{arXiv preprint arXiv:1907.04502}, 2019{\natexlab{b}}.

\bibitem[Magin and Degrez(2004)]{magin2004transport}
T.~E. Magin and G.~Degrez.
\newblock Transport algorithms for partially ionized and unmagnetized plasmas.
\newblock \emph{Journal of Computational Physics}, 198\penalty0 (2):\penalty0
  424--449, 2004.

\bibitem[Magin et~al.(2006{\natexlab{a}})Magin, Caillault, Bourdon, and
  Laux]{magin2006nonequilibrium}
T.~E. Magin, L.~Caillault, A.~Bourdon, and C.~Laux.
\newblock Nonequilibrium radiative heat flux modeling for the {H}uygens entry
  probe.
\newblock \emph{Journal of Geophysical Research: Planets}, 111\penalty0 (E7),
  2006{\natexlab{a}}.

\bibitem[Magin et~al.(2006{\natexlab{b}})Magin, Caillault, Bourdon, and
  Laux]{MaginCaillaultBourdonLaux2006}
T.~E. Magin, L.~Caillault, A.~Bourdon, and C.~O. Laux.
\newblock Nonequilibrium radiative heat flux modeling for the {H}uygens entry
  probe.
\newblock \emph{J. Geophys. Res}, 111\penalty0 (E07S12):\penalty0 1--11,
  2006{\natexlab{b}}.

\bibitem[Mao et~al.(2020)Mao, Jagtap, and Karniadakis]{mao2020physics}
Z.~Mao, A.~D. Jagtap, and G.~E. Karniadakis.
\newblock Physics-informed neural networks for high-speed flows.
\newblock \emph{Computer Methods in Applied Mechanics and Engineering},
  360:\penalty0 112789, 2020.

\bibitem[Marxen et~al.(2013)Marxen, Magin, Shaqfeh, and
  Iaccarino]{marxen2013method}
O.~Marxen, T.~E. Magin, E.~S. Shaqfeh, and G.~Iaccarino.
\newblock A method for the direct numerical simulation of hypersonic
  boundary-layer instability with finite-rate chemistry.
\newblock \emph{Journal of Computational Physics}, 255:\penalty0 572--589,
  2013.

\bibitem[Matheou et~al.(2008)Matheou, Pantano, and
  Dimotakis]{matheou2008verification}
G.~Matheou, C.~Pantano, and P.~E. Dimotakis.
\newblock Verification of a fluid-dynamics solver using correlations with
  linear stability results.
\newblock \emph{Journal of Computational Physics}, 227\penalty0 (11):\penalty0
  5385--5396, 2008.

\bibitem[Mons et~al.(2019)Mons, Wang, and Zaki]{mons2019kriging}
V.~Mons, Q.~Wang, and T.~A. Zaki.
\newblock Kriging-enhanced ensemble variational data assimilation for
  scalar-source identification in turbulent environments.
\newblock \emph{Journal of Computational Physics}, 398:\penalty0 108856, 2019.

\bibitem[{Nagarajan} et~al.(2003){Nagarajan}, {Lele}, and
  {Ferziger}]{Nagarajanetal03}
S.~{Nagarajan}, S.~K. {Lele}, and J.~H. {Ferziger}.
\newblock {A robust high-order method for large eddy simulation}.
\newblock \emph{J.\ Comput. Phys.}, 191:\penalty0 392--419, 2003.

\bibitem[Najm and Knio(2005)]{najm2005modeling}
H.~N. Najm and O.~M. Knio.
\newblock Modeling low {M}ach number reacting flow with detailed chemistry and
  transport.
\newblock \emph{Journal of Scientific Computing}, 25\penalty0 (1-2):\penalty0
  263, 2005.

\bibitem[Najm et~al.(1998)Najm, Wyckoff, and Knio]{najm1998semi}
H.~N. Najm, P.~S. Wyckoff, and O.~M. Knio.
\newblock A semi-implicit numerical scheme for reacting flow: {I}. stiff
  chemistry.
\newblock \emph{Journal of Computational Physics}, 143\penalty0 (2):\penalty0
  381--402, 1998.

\bibitem[Nicoud(2000)]{nicoud2000conservative}
F.~Nicoud.
\newblock Conservative high-order finite-difference schemes for low-{M}ach
  number flows.
\newblock \emph{Journal of Computational Physics}, 158\penalty0 (1):\penalty0
  71--97, 2000.

\bibitem[NVIDIA(2019)]{nvidiatalk}
NVIDIA.
\newblock {NVIDIA CEO} talk, {SC}’19, 44th min
  https://www.youtube.com/watch?v=69neepdejzu.
\newblock 2019.

\bibitem[Park and Zaki(2019)]{park_zaki_2019}
J.~Park and T.~A. Zaki.
\newblock Sensitivity of high-speed boundary-layer stability to base-flow
  distortion.
\newblock \emph{Journal of Fluid Mechanics}, 859:\penalty0 476–515, 2019.
\newblock \doi{10.1017/jfm.2018.819}.

\bibitem[Prakash et~al.(2011)Prakash, Parsons, Wang, and
  Zhong]{prakash2011high}
A.~Prakash, N.~Parsons, X.~Wang, and X.~Zhong.
\newblock High-order shock-fitting methods for direct numerical simulation of
  hypersonic flow with chemical and thermal nonequilibrium.
\newblock \emph{Journal of Computational Physics}, 230\penalty0 (23):\penalty0
  8474--8507, 2011.

\bibitem[Raissi et~al.(2019)Raissi, Perdikaris, and
  Karniadakis]{raissi2019physics}
M.~Raissi, P.~Perdikaris, and G.~E. Karniadakis.
\newblock Physics-informed neural networks: {A} deep learning framework for
  solving forward and inverse problems involving nonlinear partial differential
  equations.
\newblock \emph{Journal of Computational Physics}, 378:\penalty0 686--707,
  2019.

\bibitem[Raissi et~al.(2020)Raissi, Yazdani, and Karniadakis]{raissi2020hidden}
M.~Raissi, A.~Yazdani, and G.~E. Karniadakis.
\newblock Hidden fluid mechanics: {L}earning velocity and pressure fields from
  flow visualizations.
\newblock \emph{Science}, 367\penalty0 (6481):\penalty0 1026--1030, 2020.

\bibitem[{Wang} et~al.(2019){Wang}, {Wang}, and {Zaki}]{WangM_2019}
M.~{Wang}, Q.~{Wang}, and T.~A. {Zaki}.
\newblock {Discrete adjoint of fractional-step incompressible {N}avier-{S}tokes
  solver in curvilinear coordinates and application to data assimilation}.
\newblock \emph{Journal of Computational Physics}, 396:\penalty0 427--450,
  2019.
\newblock \doi{10.1016/j.jcp.2019.06.065}.

\bibitem[Wang et~al.(2019)Wang, Hasegawa, and Zaki]{WangQ_2019}
Q.~Wang, Y.~Hasegawa, and T.~A. Zaki.
\newblock Spatial reconstruction of steady scalar sources from remote
  measurements in turbulent flow.
\newblock \emph{Journal of Fluid Mechanics}, 870:\penalty0 316–352, 2019.
\newblock \doi{10.1017/jfm.2019.241}.

\bibitem[Wang et~al.(2011)Wang, Yee, Sj{\"o}green, Magin, and
  Shu]{Wangetal2011}
W.~Wang, H.~C. Yee, B.~Sj{\"o}green, T.~Magin, and C.-W. Shu.
\newblock {Construction of low dissipative high-order well-balanced filter
  schemes for non-equilibrium flows}.
\newblock \emph{Journal of Computational Physics}, 230\penalty0 (11):\penalty0
  4316--4335, May 2011.

\bibitem[Zanus et~al.(2020)Zanus, Mir{\'o}, and Pinna]{zanus2020parabolized}
L.~Zanus, F.~M. Mir{\'o}, and F.~Pinna.
\newblock Parabolized stability analysis of chemically reacting boundary-layer
  flows in equilibrium conditions.
\newblock \emph{Proceedings of the Institution of Mechanical Engineers, Part G:
  Journal of Aerospace Engineering}, 234\penalty0 (1):\penalty0 79--95, 2020.

\bibitem[Zhang and Shu(2011)]{ZhangShu2011JCP}
X.~Zhang and C.-W. Shu.
\newblock Positivity-preserving high order discontinuous {G}alerkin schemes for
  compressible {E}uler equations with source terms.
\newblock \emph{Journal of Computational Physics}, 230\penalty0 (4):\penalty0
  1238--1248, 2011.

\bibitem[Zhong(1996)]{zhong1996additive}
X.~Zhong.
\newblock Additive semi-implicit {R}unge--{K}utta methods for computing
  high-speed nonequilibrium reactive flows.
\newblock \emph{Journal of Computational Physics}, 128\penalty0 (1):\penalty0
  19--31, 1996.

\end{thebibliography}
\end{document}